\documentclass[a4paper,11pt]{article}
\pdfoutput=1
\usepackage{graphicx,rotating,hyperref,slashed,amsmath,xcolor,amssymb,amsfonts,colortbl,cite, subcaption,float}
\makeatletter  
\usepackage{graphics} 

\usepackage{graphicx} %Loading the package 
\graphicspath{{plots/}}%here the directory where it looks for the figures

\hypersetup{colorlinks,bookmarksopen,bookmarksnumbered,
linkcolor=blus,pdfstartview=FitH,urlcolor=rossos,citecolor=verde}
\numberwithin{equation}{section}

\hypersetup{%
    ,urlcolor=rossos
    ,citecolor=rossos
    ,linkcolor=rossos
    }

\AtBeginDocument{% optionally set colors to your liking
  \hypersetup{
    urlcolor=verdes,
    citecolor=verdes,
    linkcolor=verdes,
  }%
}

\allowdisplaybreaks

\def\lsim{\mathrel{\rlap{\lower3pt\hbox{\hskip0pt$\sim$}}
   \raise1pt\hbox{$<$}}}         %less than or approx. symbol
\def\gsim{\mathrel{\rlap{\lower4pt\hbox{\hskip1pt$\sim$}}
   \raise1pt\hbox{$>$}}}         %greater than or approx. symbol

 \newcommand{\sfootnote}[1]{} 
\definecolor{bluc}{cmyk}{1,1,0,0.1}
\definecolor{rossoCP3}{cmyk}{0,.88,.77,.40}
\definecolor{rosso}{cmyk}{0,1,1,0.4}
\definecolor{rossos}{cmyk}{0,1,1,0.55}
\definecolor{rossoc}{cmyk}{0,1,1,0.2}
\definecolor{verdes}{cmyk}{0.92,0,0.59,0.4}

\hypersetup{colorlinks, bookmarksopen, bookmarksnumbered,
citecolor=verdes, linkcolor=bluc, pdfstartview=FitH, urlcolor=rossos}

\newcommand{\mio}[1]{}

\definecolor{Gray}{gray}{0.95}

\usepackage{multicol}
\usepackage{color}
\definecolor{rosso}{cmyk}{0,1,1,0.4}
\definecolor{rossos}{cmyk}{0,1,1,0.55}
\definecolor{rossoc}{cmyk}{0,1,1,0.2}
\definecolor{blu}{cmyk}{1,1,0,0.3}
\definecolor{blus}{cmyk}{1,1,0,0.6}
\definecolor{bluc}{cmyk}{1,1,0,0.1}
\definecolor{verde}{cmyk}{0.92,0,0.59,0.25}
\definecolor{verdec}{cmyk}{0.92,0,0.59,0.15}
\definecolor{verdes}{cmyk}{0.92,0,0.59,0.4}

\def\circa#1{\,\raise.3ex\hbox{$#1$\kern-.75em\lower1ex\hbox{$\sim$}}\,}

\newcommand{\beq}{\begin{equation}}
\newcommand{\eeq}{\end{equation}}

\newcommand{\bea}{\begin{eqnarray}}
\newcommand{\eea}{\end{eqnarray}}
\newcommand{\be}{\begin{equation}}
\newcommand{\ee}{\end{equation}}
\newfam\rsfsfam
\def\mathscr#1{{\fam\rsfsfam\relax#1}}

\def\circa#1{\,\raise.3ex\hbox{$#1$\kern-.75em\lower1ex\hbox{$\sim$}}\,}
\makeatletter

\def\hhref#1{\href{http://arxiv.org/abs/#1}{arXiv:#1}} % in bibliography

\newcommand{\doi}[1]{\href{http://dx.doi.org/#1}{[doi]}}

\def\hhref#1{\href{http://arxiv.org/abs/#1}{arXiv:#1}} 
 
\def\art{\@ifnextchar[{\eart}{\oart}}
\def\eart[#1]#2#3#4#5#6{{\rm #2}, {\em #3 \bf #4} {\rm (#6) #5} ({\em #1})}

\def\article{\@ifnextchar[{\earticle}{\oarticle}}
\def\oarticle#1#2#3#4#5#6{{\rm #1}, {\em ``#6''}, {\rm #2 #3 (#5) #4}}
\def\earticle[#1]#2#3#4#5#6#7{{\rm #2}, {\em ``#7''}, {\rm #3 #4 (#6) #5}  [\hhref{#1}]}
\def\hepart[#1]#2{{\rm #2, \em#1}}
\def\heparticle[#1]#2#3{#2, {\em ``#3''} [\hhref{#1}]}

\newcounter{alphaequation}[equation]
\def\thealphaequation{\theequation\hbox to
0.6em{\hfil\alph{alphaequation}\hfil}}
\def\eqnsystem#1{
\def\@eqnnum{{\rm (\thealphaequation)}}
\def\@@eqncr{\let\@tempa\relax \ifcase\@eqcnt \def\@tempa{& & &} \or
  \def\@tempa{& &}\or \def\@tempa{&}\fi\@tempa
  \if@eqnsw\@eqnnum\refstepcounter{alphaequation}\fi
\global\@eqnswtrue\global\@eqcnt=0\cr}
\refstepcounter{equation} \let\@currentlabel\theequation \def\@tempb{#1}
\ifx\@tempb\empty\else\label{#1}\fi
\refstepcounter{alphaequation}
\let\@currentlabel\thealphaequation
\global\@eqnswtrue\global\@eqcnt=0 \tabskip\@centering\let\\=\@eqncr
$$\halign to \displaywidth\bgroup \@eqnsel\hskip\@centering
$\displaystyle\tabskip\z@{##}$&\global\@eqcnt\@ne
\hskip2\arraycolsep\hfil${##}$\hfil& \global\@eqcnt\tw@\hskip2\arraycolsep
$\displaystyle\tabskip\z@{##}$\hfil
\tabskip\@centering&\llap{##}\tabskip\z@\cr}
\def\endeqnsystem{\@@eqncr\egroup$$\global\@ignoretrue} \makeatother

\definecolor{fiorentina}{rgb}{.5,0,.5}

\begin{document}
%%%%%%%%%%%%%%%%%%%%%%%%%%%%%%%%%%%%%%%%%%%%%%%%%%%%%%%%%%%%%%%%%%%%%%%%%%%%%%%%%%%%%%%%%%%%%%%%%%%%%%%%%%%%%%%%%%%%%%%%%%%%%%%%%%%%%%%%%%%%%%%%%%%%%%%%%%%%%%%%%%%%%%%%%%%%%%%%%%%%%%%%%%%%%%%%%%%%%%%%%%%%%%%%%%%%%%%%%%%%%%%%%%%%%%%%%%%%%%%%%%%%%%%%%%%%%%%%%%%%%%%%%%%%%%%%%%%%%%%%%%%%%%%%%%%%%%%%%%%%%%%%%%%%%%%%%%%%%%%%%%%%%%%%%%%%%%%%%%%%%%%%%%%%%%%%%%%%%%%%

%\vspace{1truecm}
% \begin{center}
%\boldmath

%%%%%%%%%%%%%%%%%%%%%%%%%%%%%%%%%%%%%%%%%%%%%%%%%%%%%%%%%%%%

%%%%%%%%%%%%%%%%%%%%%%%%%%%%%%%%%%%%%%%%%%%%%%%%%%%%%%%%%%%%

\vspace{0.8cm}
\begin{center}

{\fontsize{19}{28}\selectfont  \sffamily \bfseries {
Exploring cosmological gravitational wave backgrounds \\ \vskip0.2cm  through the synergy of LISA and ET
}}
\end{center}
\vspace{0.2cm}

\begin{center}
{\fontsize{12}{30}\selectfont  
Alisha Marriott-Best$^{a}$ \footnote{\texttt{2347066.at.swansea.ac.uk}}, Debika
Chowdhury$^{b}$ \footnote{\texttt{debika.chowdhury.at.iiap.res.in}},  
Anish Ghoshal$^{c}$ \footnote{\texttt{anish.ghoshal.at.fuw.edu.pl}}
, Gianmassimo Tasinato$^{a, d}$ \footnote{\texttt{g.tasinato2208.at.gmail.com}}
} 
\end{center}

\begin{center}

\vskip 8pt
\textsl{$^{a}$ Physics Department, Swansea University, SA2 8PP, United Kingdom}\\
\textsl{$^{b}$}
Indian Institute of Astrophysics, II Block, Koramangala, Bengaluru 560034, India
\\
\textsl{$^{c}$}
Institute of Theoretical Physics, Faculty of Physics, ul. Pasteura 5, 02-093 Warsaw, Poland
\\ 
\textsl{$^{d}$ Dipartimento di Fisica e Astronomia, Universit\`a di Bologna,\\
 INFN, Sezione di Bologna, I.S. FLAG, viale B. Pichat 6/2, 40127 Bologna,   Italy}
\vskip 7pt

\end{center}

\begin{abstract}
\noindent
The gravitational wave (GW) interferometers LISA and ET are expected to be functional in the next decade(s),  possibly around the same time. They will operate over different frequency ranges, with similar integrated sensitivities to the amplitude of a stochastic GW  background (SGWB).  We investigate the synergies between these two detectors, in terms of a multi-band detection of a cosmological SGWB characterised by  a large amplitude, and a broad frequency  spectrum. We develop the notion of integrated sensitivity and propose a novel signal-to-noise (SNR) optimal for characterization of the geometrical properties of the interferometer systems of LISA and ET operating simultaneously. By investigating various examples of SGWBs, such as those arising from cosmological phase transition, cosmic string, primordial inflation, we show that LISA and  ET operating together  will have the opportunity to  assess  more effectively   the characteristics  of the  GW  spectrum produced by the same cosmological source, but at separate frequency scales. Moreover,  the two experiments in tandem can be sensitive to features of early universe cosmic expansion before big-bang nucleosynthesis (BBN), which affects the SGWB  frequency profile, and which would  not be possible to detect otherwise, since two different frequency ranges correspond to two different pre-BBN (or post-inflationary) epochs. Besides considering the GW spectrum, we additionally undertake a preliminary study of the sensitivity of LISA and ET to  soft limits of higher order tensor correlation functions.  Given that these experiments operate at different frequency bands, their synergy constitutes an ideal direct probe of squeezed limits of higher order GW correlators, which can not be measured operating with a single  instrument only. 
\end{abstract}

%%%%%%
\section{Introduction }
%%%%%%

{The detection of gravitational waves (GWs) from astrophysical sources by the LIGO-Virgo collaboration in 2015~\cite{Abbott:2017xzu} opened up a new window into GW astronomy. For cosmology, upcoming upgrades of LIGO-Virgo \cite{Aasi:2014mqd} and proposed future detectors such as LISA~\cite{Audley:2017drz}, BBO-DECIGO~\cite{Yagi:2011wg}, the 
Einstein Telescope~(ET)~\cite{Punturo:2010zz,Hild:2010id},
and Cosmic Explorer~(CE)~\cite{Evans:2016mbw} will also open up a new possible observational window into the early universe. Unlike photons, the gravitons (primordial GWs) that were produced in the early Universe can propagate freely throughout cosmic history and therefore would constitute ideal messengers of the history of the Universe \cite{Allen:1996vm, Caprini:2018mtu,Simakachorn:2022yjy}.}

In fact,
the recent hints of detection of a 
 stochastic gravitational wave background (SGWB)
 in the nano-Hertz regime 
 by Pulsar
Timing Array collaborations \cite{NANOGrav:2023gor,Reardon:2023gzh,Xu:2023wog,EPTA:2023fyk} initiated the era of experimental characterization of the SGWB. Still, much has to be done to distinguish between different sources of SGWB, astrophysical or
cosmological (see e.g. \cite{Christensen:2018iqi,Renzini:2022alw} for recent topical 
reviews on SGWB sources and 
 detection techniques).

The next generation  of gravitational wave (GW) detectors promises to    improve upon the current 
experimental sensitivity to SGWB in frequency ranges between milli-Hz and deca-Hz,
 much higher
than the nano-Hz regime probed by pulsar timing arrays. These higher frequency regimes are more
suitable for detecting GWs produced by various early universe cosmic sources, such as those arising from phase 
transitions, cosmic strings,
cosmological inflation etc.  Early universe scenarios can lead to a SGWB with intriguing properties such as a rich
frequency profile, chirality, non-Gaussianity, all of which are important
  to accurately characterise for future targets (see e.g. \cite{LISACosmologyWorkingGroup:2022jok} for  a comprehensive
  recent 
review). 
 It is essential to develop  tools to detect and better characterise the SGWB in a frequency range which can be tested with high sensitivity by future experiments,  say between
 $10^{-5}\le f/{\rm Hz} \le 10^{2}$. 
In this work, we explore such a possibility of studying the early-universe sources of SGWB spanning this frequency range by  
exploiting  synergies 
between the Laser Interferometer Space Antenna (LISA) \cite{LISA:2017pwj,LISACosmologyWorkingGroup:2022jok,Colpi:2024xhw} and the Einstein Telescope (ET) \cite{Punturo:2010zz,Hild:2010id,Sider:2022ntn,Maggiore:2019uih} experiments, for
a multi-band detection of the SGWB.
  Both experiments are planned to take data in the next decade, and
will have similar sensitivities to the amplitude of SGWB. Hence
it is interesting to inquire what we can gain from detecting a SGWB with both the experimental facilities. 

  While LISA  will have
  its maximal sensitivity  for frequencies $f$ in the milli-Hertz (mHz) regime, ET will be more sensitive to signals in the deca-Hertz
 range. If the SGWB has a broad enough frequency profile and 
 a sufficiently large amplitude,
  it will be advantageous to have both the experiments detecting its features in different frequency ranges.  
   We can then measure the properties
   of the GW source more accurately, and  study aspects of early-universe
   cosmology which cannot be probed by each single experiment.

In the context of beyond the Standard Model (BSM) of particle physics there are several concrete predictions of SGWBs over multi-band frequency ranges as we will discuss below. Firstly, the very well understood temperature anisotropies in the Cosmic Microwave Background Radiation (CMBR) superimposed on the perfectly smooth background implies that the  universe at the very beginning has undergone an accelerated expansion, a phenomenon also known as the cosmic inflation~\cite{Guth:1980zm, Linde:1981mu, Albrecht:1982wi, Akrami:2018odb}. 
However, the history of the primordial universe post inflation (plausibly after temperature $T\leq 10^{14}~\rm GeV$ ) and before the beginning of Big Bang Nucleosynthesis (BBN), that is the at temperature above the SM plasma temperature of $T\gtrsim 1~\rm MeV$, remains unconstrained by any observational data at the moment. The standard assumption that the pre-BBN universe is filled with radiation and becomes radiation-dominated after the end of the inflationary phase, is often  challenged by open problems in the Standard Model (SM) of particle physics, e.g., the microscopic origin of dark matter (DM), the explanation for the observed matter-antimatter asymmetry, the flavor puzzle, or the ultraviolet SM Higgs field dynamics, the Strong CP problem (see e.g. Ref.~\cite{Espinosa:2015qea,Gouttenoire:2022gwi} for a review). Introducing new BSM physics which resolves these puzzles of modern particle physics and cosmology often is associated with new energy scales (other than the electroweak or Planck scales) and on many occasions new degrees of freedom (like new particles) or interactions which sometimes generate deviations from the standard radiation domination era before the onset of BBN.
   
   We start our work with section \ref{sec_corr_exp} explaining why and how a detection
  of the SGWB in synergy  between LISA and ET can improve the signal-to-noise ratio (SNR)
   on the measurements of parameters characterizing a SGWB with a broad frequency spectrum. 
   We then move on to section \ref{sec_theory} to discuss and analyse several early universe
   scenarios that are able to produce  GWs spanning over a broad frequency range. By means of a Fisher
   analysis, we quantitatively demonstrate how a detection of GW with the
two experiments together can  help us to measure specific model parameters. Section \ref{sec_sencur}
discusses the notion of integrated sensitivity curves, which offer a simple
visual aid to demonstrate the advantages of the synergy between the two experiments with regard to detecting SGWB with certain
frequency shapes. Cosmological SGWB can be characterised
by non-Gaussian features, which motivate the study 
of $n$-point correlation functions going beyond the GW
power spectrum and energy density.
Given that LISA and ET operate in different frequency ranges {which corresponding to different energy scales of tensor Fourier modes},
in section \ref{sec_nonG} we address  the problem  of
the detectability of soft limits
of $n$-point correlation functions, discussing the response
function of the LISA-ET system to such observables.  We conclude in section \ref{sec_concl}. A technical appendix complements our arguments. 
 We work with natural units $c=\hbar=1$. We fix the $h$ in
 the Hubble parameter as $h=0.67$.

%%%%%%

%%%%%%
\section{Synergies between LISA and Einstein Telescope  }
\label{sec_corr_exp}
%%%%%%

The aim of this section is to start discussing in practical terms
the possibility of making a synergetic detection of a SGWB
with the Laser Interferometer Space Antenna (LISA) and the Einstein Telescope (ET) instruments.
  In the next section we  {will} describe
theoretical motivations to do so. 

\subsubsection*{Gravitational waves
and their detection}

Gravitational waves (GWs) are associated with spin-2 fluctuations $h_{ij}$ of the Minkowski metric:
\be
d s^2\,=\,-d t^2+\left( \delta_{ij}+h_{ij}(t, \vec x)\right)\,d x^i d x^j
\,.
\ee
We decompose $h_{ij}$
 into Fourier  modes as
\be
\label{fouhij}
h_{ij}(t,\vec x)\,=\,\sum_{\lambda}\,\int_{-\infty}^{+\infty} d f\,\int d^2 \hat n\,{ e}^{-2 \pi i f \,\hat n \vec x}\,e^{2 \pi i f t}\,
{\bf e}_{ij}^\lambda (\hat n)\,h_{\lambda}(f, \hat n)\,,
\ee
imposing the condition
\be
\label{relcoa}
h_\lambda (-f, \hat n)\,=\,h^*_\lambda (f, \hat n),
\ee
  which ensures that $h_{ij}(t, \vec x)$ is a real function.
  The quantities
   $f$, $\hat n$, and $\lambda$ denote respectively the GW frequency, direction, and  polarization
   ($\lambda \,= \,+,\times$).    In the previous expressions,
 we have decomposed
  the GW momentum as $\vec k \,=\,2 \pi f\,\hat n$, with $f$ being the GW frequency, and $\hat
  n$ being its direction. 
  We  assume that the polarization tensors ${\bf e}_{ij}^{\lambda}$ are real
quantities. We adopt  a $(+,\times)$ basis, and use
the normalization: $\sum_{ij}{\bf e}_{ij}^{\lambda} {\bf e}_{ij}^{\lambda'} \,=\,2 \delta^{\lambda \lambda'}$.

 We assume in this section that the SGWB is isotropic, stationary, and Gaussian.  The GW energy density  is expressed
 in terms of the function
 $\Omega_{\rm GW}(f)$, defined starting from the two-point function for GW Fourier modes (see e.g. \cite{Maggiore:1999vm}). It
 is defined as 
 \be
 \label{def_ogw}
\Omega_{\rm GW}(f)\,=\,\left(\frac{4 \pi^2}{3\,H_0^2} \right)\,f^3\,I(f),
\ee
where the GW intensity $I(f)$  is given by
\be
\label{defoi}
\left\langle \left(h^{\lambda} (f,\hat n)\right)^*\,h^{\lambda'}(f',\hat n')\right\rangle
\,=\,\frac{\delta^{\lambda \lambda'}}{2}\,\frac{\delta(\hat n-\hat n')}{4\pi}\,\delta(f-f')\,I(f)\,.
\ee

 \smallskip
We shall now discuss how the interferometers LISA and ET respond to the presence of GWs. Their behaviour
depends on the so-called response functions, and on the sources of noise which affect
a possible GW detection.  
For the case of LISA, this topic
 is explained in a  clear and pedagogical way in \cite{Smith:2019wny},
which we briefly review here (see also \cite{Romano:2016dpx} for a more systematic discussion). 
We extend their analysis 
to include ET\footnote{For the case of ET, the arguments  leading to the definition of sensitivity curves are formally very similar, and we refer the reader
e.g. to \cite{Maggiore:2019uih,Mentasti:2020yyd} for  transparent discussions.} and the synergy between the two detectors.
We  mainly use the notation and conventions of  \cite{Smith:2019wny}, adapting them to the present context.
 
 We assume that both the LISA and the ET instruments have shapes corresponding to equilateral 
 triangles~\footnote{
 This is certain for LISA, and possible  for
ET -- see \cite{Branchesi:2023mws} for a discussion of various  possible ET configurations. 
}.
 The GW is  detected as an effect of the time
difference between signals measured at different vertices of the triangular interferometer.   We shall denote the three  vertices of a   triangle  (it can
 be the LISA or the ET instrument) with the combination of letters $(abc)$. Let us consider the vertex $a$
 as reference. The instrument measures the phase difference $\Phi$
 \be
 \Phi_{a_{bc}}\,=\,\Delta \varphi_{a_{bc}}+n_{a_{bc}}
 \ee
  of the GW signals travelling along the arms $(ab)$ and $(ac)$, plus
  the contribution $n$ of noise. In what follows, we  {will} neglect the time dependence
 of the positions of the detectors.
 
The interferometer response and the   GW signal contribution can then be expanded in Fourier modes as
  \be
 \Delta \varphi_{a_{bc}}(t)\,=\,
 \int_{-\infty}^{+\infty} df\,e^{2 \pi i f t}\,\Delta \tilde \varphi_{a_{bc}}(f),
 \ee
 where the signal Fourier mode $\Delta \tilde \varphi$ is given by {the}  combination of the spin-2 mode $ h^{\lambda} (f,\hat n)$
 and the interferometer response  $F_{a_{bc}}^\lambda$,
 as contained in the following definition
 \be
\Delta \tilde \varphi_{a_{bc}}(f)
\,=\,\sum_\lambda\,\int d^2n\,
 h^{\lambda} (f,\hat n)\,F_{a_{bc}}^\lambda (f, \hat n)
 \,.
 \ee
 The quantity $F_{a_{bc}}^\lambda$ is expressed as
 \bea
 F_{a_{bc}}^\lambda (f, \hat n)&=&\frac{e^{-2 \pi i \,f \,\hat n\cdot \vec x_a}}{2}
\,{\bf e}_{ij}^{\lambda}(\hat n)\,\left[ {\cal F}^{ij}(\hat \ell_{ab}\cdot \hat n,f)- {\cal F}^{ij}(\hat \ell_{ac}\cdot \hat n,f)
\right],
\eea
with the
unit vector $\hat \ell$ corresponding to the  direction of the detector arm.
The geometry of the detector enters into the functions $ {\cal F}^{ij}$. Their expressions
depend on the type
of interferometer one considers -- space-based
 (LISA) or ground-based (ET). For the case of LISA, they read
\bea
 {\cal F}_{\rm LISA}^{ij}(\hat \ell \cdot \hat n,f)&=&\frac{\hat \ell^i \hat \ell^j}{2}
e^{-i f (3+\hat \ell \cdot \hat n)/(2 f_\star)}\,
{\rm sinc}\left( \frac{f}{2 f_\star}(1-\hat \ell\cdot \hat n) \right)
\nonumber
\\
&+&
\frac{\hat \ell^i \hat \ell^j}{2}\,e^{-i f (1+\hat \ell \cdot \hat n)/(2 f_\star)}\,
{\rm sinc}\left( \frac{f}{2 f_\star}(1+\hat \ell\cdot \hat n) \right),
 \eea
 where
 the pivot scale $f_\star=1/(2 \pi L)$ -- with $L$ being
 the length of the interferometer arms -- is of the order of the 
milli-Hz frequencies probed by LISA.
In contrast, the expression for ${\cal F}$ is much simpler for a ground-based detector such as
 ET, and corresponds to the following low-frequency limit of the previous equation:
 \bea
  {\cal F}_{\rm ET}^{ij}(\hat \ell \cdot \hat n,f)&=&{\hat \ell^i \hat \ell^j}\,.
 \eea
 
 Starting from the above formulae,
   we can measure the phase difference
of signals travelling between the arms $(ab)$ and $(ac)$, and correlate signals measured at different vertices by  computing  their
two-point functions. 
They depend on the GW intensity $I(f)$ (see eq \eqref{defoi}), weighted by the instrument response to the GW signal, and on possible noise sources. The
signal two-point function in  Fourier space reads
\be
\label{tostwpofu}
\langle \Phi_{a_{bc}}(f) \Phi_{x_{yz}}(f')
 \rangle\,=\,\frac{\delta(f-f')}{2}
\left[ R_{a_{bc},\,x_{yz}} (f)\,I(f)+N_{a_{bc},\,x_{yz}} (f)
\right],
 \ee
 with $N_{a_{bc},\,x_{yz}}$ being the correlated noise. The signal response functions are
 \be
 \label{resccor}
 R_{a_{bc},\,x_{yz}} (f)\,=\,\int \frac{d^2 \hat n}{4 \pi}
 \,\left[F_{a_{bc}}^+ (f, \hat n)
 F_{x_{yz}}^+ (-f, \hat n)
 +
 F_{a_{bc}}^\times (f, \hat n)
 F_{x_{yz}}^\times (-f, \hat n)
 \right]\,.
 \ee

 At a given frequency $f$,
 we assume that there is neither noise correlation nor
 contaminations between the two detectors LISA and ET  -- correlated
 noise is present only between arms of the same interferometer --  hence the functions
 $N_{a_{bc},\,x_{yz}} $ are zero for correlations among vertices of two different experiments.
In this section, moreover, we do not take into account   correlations of the signal intensity $I(f)$ among 
arms of the  two different interferometers. In fact, the latter  
have best sensitivities in  different  frequency ranges, hence  we expect that at a given frequency $f$ the signal intensity can be at best probed by one individual experiment only. Namely, we
consider
 only signal  correlations {to exist} between
the arms of each
interferometer. 
 Under these hypotheses, 
   the phase covariance of the correlated signals among the different vertices
 of the two equilateral triangles (LISA and ET) results in a block-diagonal
 $6\times6$ matrix:
 
 \be
 \begin{pmatrix}
 C_1 & C_2 & C_2&0 &0&0\\
  C_2 & C_1 & C_2&0 &0&0\\
    C_2 & C_2 & C_1&0 &0&0
    \\
    0 & 0 & 0& C_3 &C_4&C_4
\\
    0 & 0 & 0& C_4 &C_3&C_4
    \\
        0 & 0 & 0& C_4 &C_4&C_3
 \end{pmatrix}
 \ee
 
\begin{figure}[t!]
	%\centering
	\noindent\includegraphics[width=0.5\linewidth]{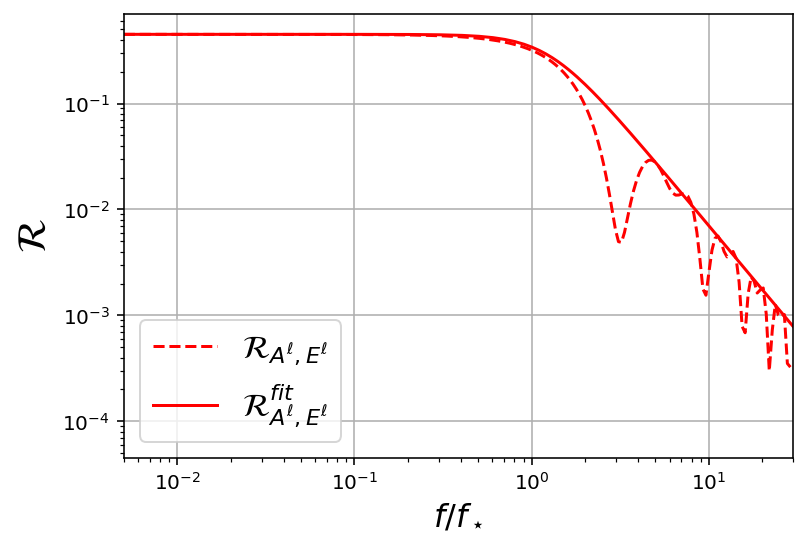}
\hfill
 \includegraphics[width=0.5
\linewidth]{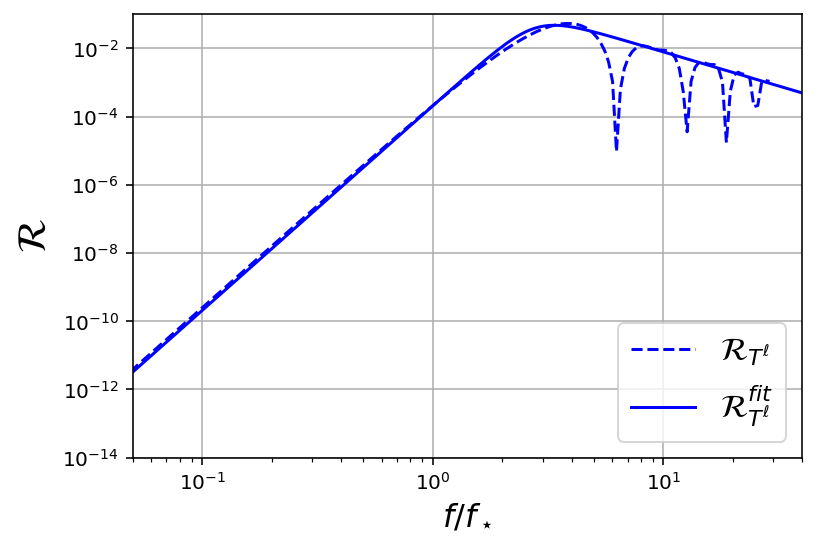}
	\caption{\small \it The numerical LISA response functions for A,E ({\bf
	left panel}) 
	and T ({\bf right panel}) orthogonal channels, (dashed lines) as 
	well as the corresponding analytical fits of eqs \eqref{ra_fit}, \eqref{rt_fit} (continuous
 lines).}
	\label{fig:resp1}
\end{figure}

 The upper block corresponds to the LISA and the lower block to the ET 
  equilateral triangle. The quantities $$C_i= S_i+N_i\,,$$ are combinations of (possible) GW signal ($S_i$) and instrumental noise ($N_i$)
  at each detector.
 This matrix can easily be diagonalized,
  leading to the definition of six orthogonal channels.
   In analogy with the names  traditionally assigned to the LISA channels, these are called
 $\left(A^{\ell}, E^\ell, T^\ell, A^e, E^e, T^e\right)
 $. They are given by
 \bea
 C_{A^{\ell}}&=&C_{E^{\ell}}\,=\,C_1-C_2,
 \\
  C_{T^{\ell}}&=&C_1+2 C_2,
 \\
  C_{A^{e}}&=&C_{E^{e}}\,=\,C_3-C_4,
\\
 C_{T^{e}}&=&C_3+2 C_4.
 \eea 
 Starting from these considerations, we can obtain
the response functions for the diagonal channels, in our
approximation of static set-up~\footnote{The results are obtained by performing the integrations
 in eq \eqref{resccor}. The integrals
contain the relative positions of the interferometer vertices.  
 For definiteness, extending  \cite{Smith:2019wny}, 
  we set the positions of the LISA interferometer vertices as  (with $L$ being the LISA arm length)
  \be
  \label{dirLIar}
  \vec x_A\,=\,\{0,0,0\} \hskip0.5cm,\hskip0.5cm \vec x_B\,=\,L\,\{1/2,\sqrt{3}/2,0\} \hskip0.5cm,\hskip0.5cm \vec x_C\,=\,L\,\{-1/2,\sqrt{3}/2,0\}\,.
  \ee 
For simplicity  we choose the same arm directions for the triangle forming  ET, which of course has much shorter arm lengths. As mentioned above, in this work we do not consider
effects of the relative motion between LISA and ET detectors.
  \label{foot_position}
}. 
In the case of LISA, the response functions depend on the frequency; in
the small-frequency limit they can be expressed as 
\bea
\label{ra2p}
R_{A^\ell}&=&R_{E^\ell}\,=\,\frac{9}{20}-\frac{169}{1120} \left(\frac{f}{f_\star} \right)^2+{\cal O}\left({f}/{f_\star} \right)^4,
\\
\label{rt2p}
R_{T^\ell}&=&\frac{1}{4032} \left(\frac{f}{f_\star} \right)^6+{\cal O}\left({f}/{f_\star} \right)^8.
\eea
The complete frequency dependence of the response functions $R_{A^\ell,\,E^\ell}$ and  $R_{T^\ell}$ 
for LISA  can be easily obtained numerically, as explained in \cite{Smith:2019wny} (see Fig \ref{fig:resp1}).
 Suitable
analytical approximations for these two quantities  
are
\bea
\label{ra_fit}
R^{\rm fit}_{A^\ell,\,E^\ell}(f)&=&\frac{9}{20} \left( 1+\left(
\frac{f}{1.25\,f_\star}
\right)^3 \right)^{-2/3},
\\
\label{rt_fit}
R^{\rm fit}_{T^\ell}(f)&=&\frac{1}{10} \left(\frac{f}{2.8 \,f_\star} \right)^6 \left(1+\left(\frac{f}{2.8 \,f_\star} \right)^6  \right)^{-4/3},
\eea

\noindent
which
 are also represented in Fig \ref{fig:resp1}. 
  The $T$-channels $T^\ell$ and $T^e$ are either weakly sensitive or not sensitive at all to the GW signal (and the sensitivity in any case vanishes in the small-frequency limit). For the
ground-based interferometer ET, the response functions
are independent from frequency, and are proportional
to the zero-frequency limit of eqs \eqref{ra2p}, \eqref{rt2p}: see
e.g. \cite{Mentasti:2020yyd} for details.

\subsubsection*{The optimal
signal-to-noise ratio}

After the  characterization of the geometrical
properties of the interferometer system  as described above, we investigate the optimal signal-to-noise ratio (SNR)
for detecting a SGWB with the two instruments LISA and ET
working 
together. We assume that the two detectors take data
approximately for the same amount of time $T$ (although
not necessarily simultaneously, exploiting  the stationarity
of the SGWB). 
 The optimal signal-to-noise ratio (SNR) for measuring  a SGWB with LISA and ET in synergy 
 is  built using techniques based  on  Wiener filtering, combining  information obtained from the independent  channels $A^{\ell, e}$ and $E^{\ell,e}$. We follow 
 \cite{Smith:2019wny} (see also \cite{Romano:2016dpx}), extending it to the general case where we work with two instruments together.  
   Working in the weak-signal limit, denoting with 
 $S_i$ the signal on each independent channel, $N_i$ the noise, and $Q_i$ the filter,
 the signal-to-noise ratio (SNR)  in
  Fourier space reads 
 \bea\label{ans_snrf}
 {\rm SNR}\,=\,\sqrt{\frac{T}{2}}
 \,\frac{\sum_i \int_{-\infty}^{\infty}\,df\,S_i(f) Q_i(f)}{\sqrt{
 \sum_i \int_{-\infty}^{\infty}\,df\,N^2_i(f) Q^2_i(f)
 }}\,,
 \eea  
 where the sums are over the four channels  $A^{\ell, e}$ and $E^{\ell,e}$ which are most sensitive to the signal. 
 As mentioned above, the quantity $T$ indicates 
  the duration of the measurements, which we consider to be comparable in the 
  two experiments.
  We now wish to determine the optimal filter which maximizes eq \eqref{ans_snrf}. We
define a positive definite inner product
\be
( P_i,  Q_i)\,=\,\sum_i \int_{-\infty}^\infty\,d f\,P_i(f) Q_i(f)\,N_i^2(f),
\ee
 which acts on the four vectors $(P_i)$, where $i=A^{\ell, e},E^{\ell,e}$. The SNR can then be expressed
 as
 \be
 {\rm SNR}\,=\,\sqrt{\frac{T}{2}}
 \,\frac{(S_i/N_i^2,Q_i)}{\sqrt{(Q_i, Q_i)}},
 \ee
 and the filter that maximizes the previous expression, up to an overall factor, is $Q_i=S_i/N_i^2$. 
 The optimal SNR is (we now integrate over positive frequencies only)
 \be
 \label{exp_snr}
 {\rm SNR}\,=\,\left[ T\,\sum_{i=A^{\ell, e},E^{\ell,e}}\,
 \int_0^{+\infty}d f\,\frac{S_i^2(f)}{N_i^2(f)}
 \right]^{1/2}
\,. \ee
We can now decompose the integrand in the previous formula  as
\bea
\sum_{i=A^{\ell, e},E^{\ell,e}}
\,\frac{S_i^2(f)}{N_i^2(f)}&=&\left[ 
\left(
\frac{{ R}_{A^{\ell}}(f)}{N_{A^\ell}(f)}\right)^2+
\left(
\frac{{ R}_{E^{\ell}}(f)}{N_{E^\ell}(f)}\right)^2
\right]\,I^2(f)
+
\left[ 
\left(
\frac{{ R}_{A^{e}}(f)}{N_{A^e}(f)}\right)^2+
\left(
\frac{{ R}_{E^{e}}(f)}{N_{E^e}(f)}\right)^2
\right]\,I^2(f).
\nonumber\\
\eea
 It is convenient to assemble the above result as
 (recall that $\Omega_{\rm GW}$ is defined
 in eq \eqref{def_ogw})
 \bea
 \sum_{i=A^{\ell, e},E^{\ell,e}}
\,\frac{S_i^2(f)}{N_i^2(f)}&=&\frac{\Omega_{\rm GW}^2(f)}{\Sigma_{\rm LISA}^2(f)}
+\frac{\Omega_{\rm GW}^2(f)}{\Sigma_{\rm ET}^2(f)},
 \eea
 with
 \bea
 \Sigma_{\rm LISA}(f)&=&\left(\frac{4 \pi^2}{3\,H_0^2} \right)\,f^3\,\left[ 
 \left(
\frac{{ R}_{A^{\ell}}(f)}{N_{A^\ell}(f)}\right)^2+
\left(
\frac{{ R}_{E^{\ell}}(f)}{N_{E^\ell}(f)}\right)^2
 \right]^{-1/2},
 \eea
 and analogously for ET. 

%%%%%%%%%%%%%%%%%%%%%%%%%%%%%%%%
\begin{figure}[t!]
	\centering
	\includegraphics[width=0.4\linewidth]{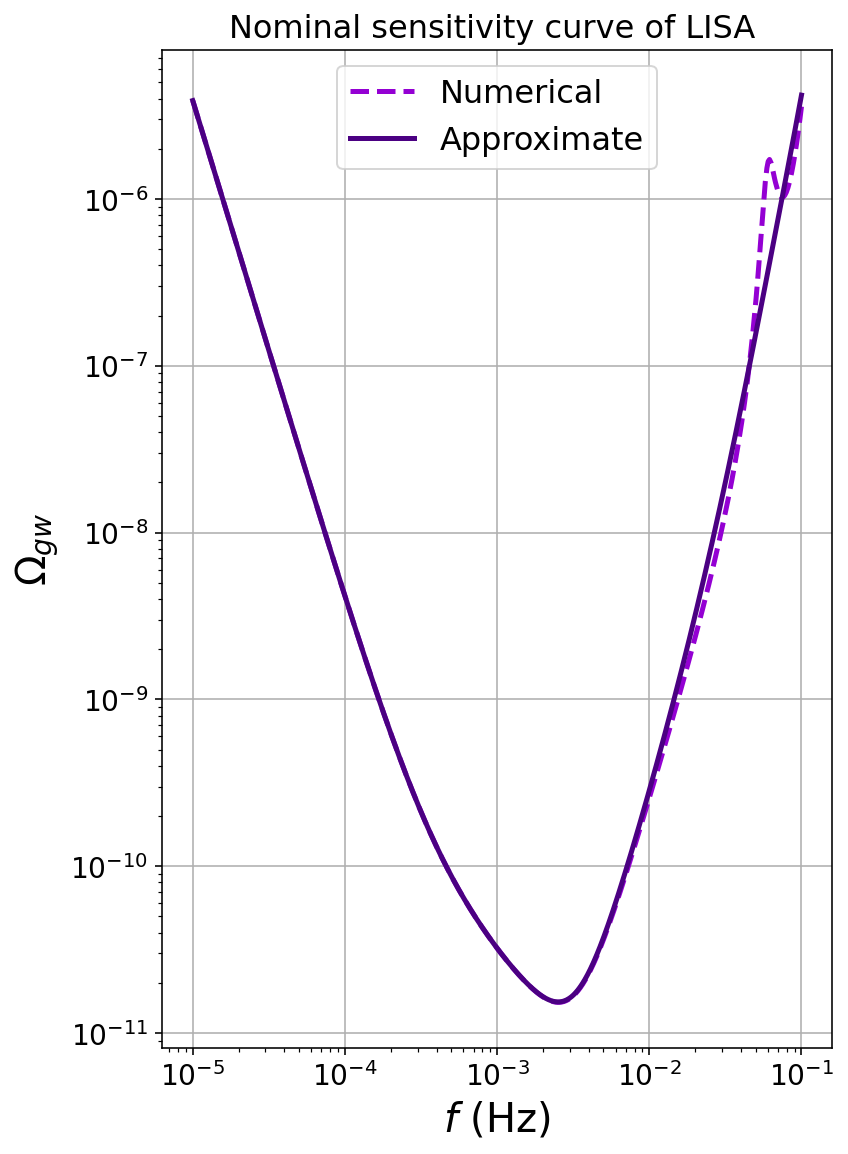}
% \hfill
 \includegraphics[width=0.4\linewidth]{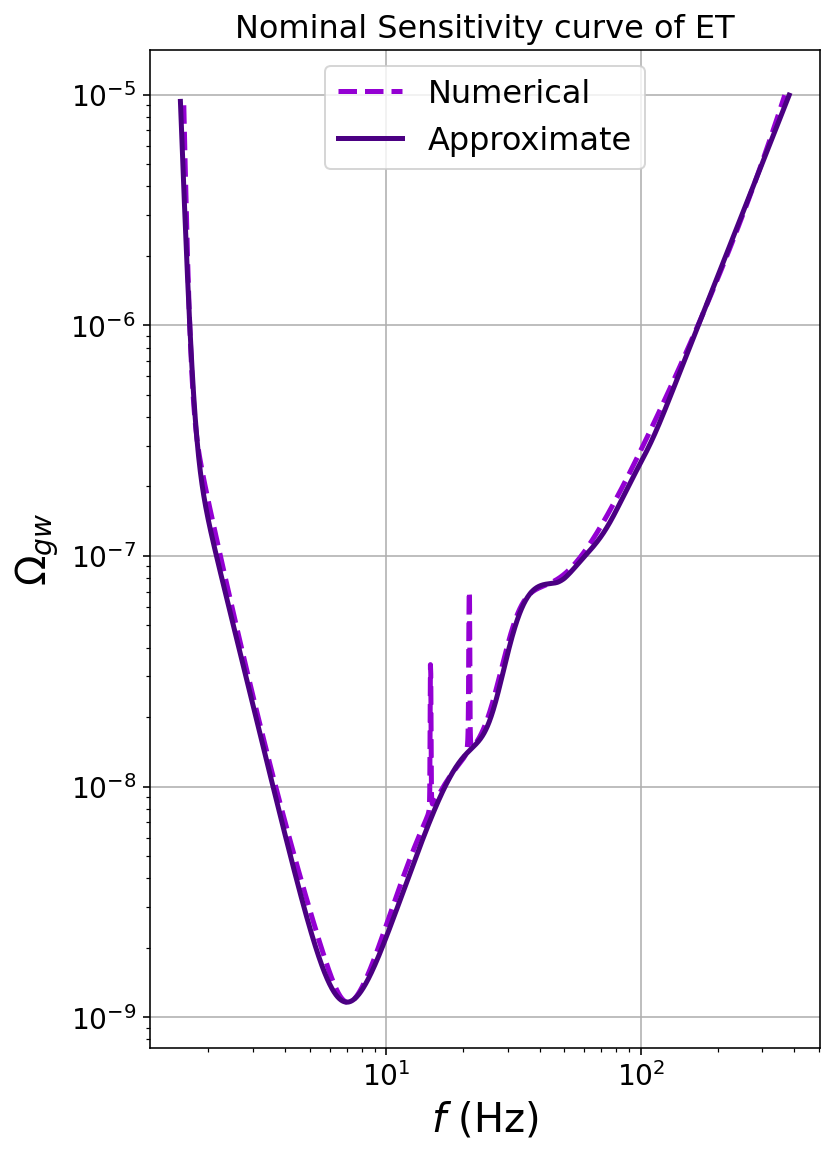}
	\caption{\small \it Nominal sensitivity curves for LISA  and ET. For the latter, we represent the so-called
	ET-D curve.  The approximate 
analytical fit for ET is discussed in Appendix \ref{app_fit}. Recall we take $h=0.67$.} 
	\label{fig:ETLISAnoms}
\end{figure}
%%%%%%%%%%%%%%%%%%%%%%%%%%%%%%%%%%%%%%%%

 \smallskip
The result
depends on the instrument response ${ R}$ to the signal, and on the noise curve ${ N}$ for each independent channel.
 The functions $\Sigma_{\rm LISA}(f)$
and $\Sigma_{\rm ET}(f)$ are called nominal sensitivity curves -- see
\cite{Moore:2014lga} for a general discussion -- and we represent them in Fig \ref{fig:ETLISAnoms} for the two experiments under consideration.
Besides the numerically evaluated sensitivity curves, we
also represent analytical approximations for the curves in Fig \ref{fig:ETLISAnoms}. The analytical fit we
use for LISA is obtained from \cite{Flauger:2020qyi},   while the one for ET is a new result of
the present work, and
we discuss it in Appendix \ref{app_fit}.
For each experiment,  the function  $\Sigma$ encapsulates a
 weighted  combination of response functions and noise on each channel, and
 is  useful for visually understanding the sensitivity
 of the instruments.  For an extended
 discussion on sensitivity curves, see section \ref{sec_sencur},
 where we will also discuss the more refined notion
 of {\it integrated} sensitivity curves
 in this context. To summarize, the square
 of the total SNR is the sum of the squares of the individual SNRs:
\bea
\label{snr_tog}
{\rm SNR}_{\rm tot}&=&\sqrt{T\,\int_0^{\infty}\,df\,\left[ 
\frac{\Omega^2_{\rm GW}(f)}{\Sigma_{\rm LISA}^2(f)}
+\frac{\Omega^2_{\rm GW}(f)}{\Sigma_{\rm ET}^2(f)}
\right]}
%\\
%&=&
\,=\,
\sqrt{{\rm SNR}_{\rm LISA}^2+{\rm SNR}_{\rm ET}^2},
\eea

a formula which will be used in what follows.  This expression demonstrates
that, by working in synergy, the two detectors can reach higher values of SNR than each experiment operating individually. 

\smallskip
Interestingly, the very same results can be obtained in terms of a likelihood
function associated with a measurement of $\Omega_{\rm GW}$
 carried out by the two experiments together. This method is also useful
 for applications to Fisher {matrix} forecasts. 
 We assume  the following structure for the Gaussian likelihood: 
 \be
 \label{def_like}
\ln {\cal L}\,=\,{\rm const}-\frac{1}{2}\int_0^\infty\,d f \,d f'\, \left(\hat{\Omega}_{\rm GW}(f)
-{\Omega}_{\rm GW}^{\rm th}(f)\right)\,C^{-1}(f,f')\,
\left(\hat{\Omega}_{\rm GW}(f')
-{\Omega}_{\rm GW}^{\rm th}(f')\right),\
\ee
where $\hat{\Omega}_{\rm GW}$ is the measured value, and ${\Omega}_{\rm GW}^{\rm th}$ is the theoretical 
 {prediction from various sources} for the quantity ${\Omega}_{\rm GW}$  we wish to test. The inverse of the
covariance matrix 
 corresponding to the GW measurement  by the two experiments together is 
\be
\label{def_cofun}
C^{-1}(f,f')\,=\,T\,\delta(f-f')\,\left( \frac{1}{\Sigma_{\rm LISA}^2(f)}
+
 \frac{1}{\Sigma_{\rm ET}^2(f)}
\right).
\ee
 Considering
$\hat{\Omega}_{\rm GW}(f)$ to be the quantity to measure, we can compute the following quantity corresponding
to a continuous version of  the Fisher matrix:
\bea
{\bf F}(f,f')&=&-\frac{\delta^2 \ln {\cal L} }{\delta {\Omega}_{\rm GW}^{\rm th}(f)\,
\delta {\Omega}_{\rm GW}^{\rm th}(f')}
\\
&=&C^{-1}(f,f').
\eea
The optimal SNR can then be
computed in terms of a convolution integral 
\be
{\rm SNR}_{\rm opt}^2\,=\,\int_0^{+\infty}\, df d f'\,{\Omega}_{\rm GW}^{\rm th}(f) {\Omega}_{\rm GW}^{\rm th}(f')\,
{\bf F}(f,f')\,.
\ee
Substituting the inverse covariance function \eqref{def_cofun}, this result coincides with eq \eqref{snr_tog}. 

\smallskip

 The concept of Fisher matrices, of course, can be used more directly to make forecasts
 on the prospective error bars associated
 with measured  quantities,
  see e.g. \cite{Tegmark:1996bz,Coe:2009xf}. 
 Suppose we are interested in measuring
 the components of a parameter vector $\Theta_i$, with $i$ being an index running
 over the number of model parameters we are interested in. 
  The corresponding Fisher matrix
  is
  \bea
{\bf F}_{ij}&=&-\frac{\delta^2 \ln {\cal L} }{\delta \Theta_i\,
\delta \Theta_j
},
  \eea
  where we consider the quantity given in eq \eqref{def_like} as likelihood function.
  Then the errors on the measurements of  $\Theta_i$ are at least 
\be
\Delta \Theta_i\,=\,\sqrt{({\bf F})^{-1}_{ii}}\,.
\ee
This is a standard formula that we  utilise in later sections. 
After discussing how to use the two instruments together 
to measure SGWB signals, in the next section we  {will} provide
motivations and examples of {broad multiband} SGWB sources, which can benefit
from a joint detection by LISA and ET.

%%%%%%
%%%%%%
\section{Examples of  SGWB  with broad frequency profiles}
\label{sec_theory}
%%%%%%

In this section we  discuss examples  of cosmological SGWB sources, which are  able to produce a broad GW signal with a sizeable  amplitude spanning several decades in frequency.
We  consider, in succession, GW sources from  first order cosmological phase transitions, cosmic
strings, and primordial inflation.
We are interested in SGWB 
 spectra enhanced  within the  broad frequency band
 \be
 \label{int_range}
{ {\cal B}}_{\rm tot}\,=\, 10^{-5}\le f/{\rm Hz}\le 445\,.
 \ee 
 The lower part of the interval \eqref{int_range} corresponds to the region of maximal
 sensitivity 
of LISA (the milli-Hz), while the upper part corresponds to the region (the deca-Hz) where 
  ET is more sensitive to a GW signal (see both panels of Fig \ref{fig:ETLISAnoms}).  
  We intend to
 demonstrate that  important physical information about the GW source and the universe's
 evolution history can be extracted by
 measuring   in synergy the SGWB within the broad frequency interval \eqref{int_range}.

We are  not interested though  in measuring
 the finer details of the frequency dependence of the GW spectrum
 (for methods to do so, see e.g. \cite{Caprini:2019pxz,Flauger:2020qyi}). Instead, we wish to 
 characterise  the overall frequency profile of the spectrum, including the properties 
 of a SGWB  which extends all the way between the lower and the upper
 regions of the frequency  band in eq \eqref{int_range}. 
 A  GW spectrum is  particularly interesting for us  if it has a structure {\it evolving in frequency} throughout the entire interval \eqref{int_range}. In such a  case, 
 the synergy between the two experiments LISA and ET can be especially  useful for better characterizing the signal and extracting its physical properties, compared to measurements made with a single instrument (LISA {\it or} ET). We explore this topic quantitatively by means of Fisher
 forecasts on the detectability of the properties of the SGWB shape.

\medskip

A useful SGWB template  to keep in mind for the GW energy density of eq \eqref{def_ogw},
with the properties we need, is the so-called 
{\it broken power law} (BPL) function 
which well describes, at least up to first approximation,  GW spectra produced by several early universe phenomena \cite{Kuroyanagi:2018csn}. 
 This template applies well to GW spectra from
phase transitions
 and cosmic strings (see sections \ref{sec_pt} and \ref{sec_cs}). We
 adopt the
 frequency shape parametrization of \cite{Caprini:2024hue}:
 \begin{equation}
  \label{BPLt}
    \Omega_{\rm GW}(f) = \Omega_\star\, \left(\frac{f}{f_{\star}}\right)^{n_1}\, \left[\frac12+\frac12 \left(\frac{f}{f_{\star}}\right)^{\sigma}\right]^{\frac{n_2-n_1}{\sigma}}. 
\,
 \end{equation}
      The quantities $f_\star$  and $\Omega_\star$ in eq \eqref{BPLt} control the position of the break and the amplitude of the spectrum around the break.  The quantities $n_{1,2}$ are related
      to the spectral indexes before and after the break, while
        $\sigma$ controls the smoothness of the break -- the smaller  $\sigma$ is, the smoother is the transition.  If the break occurs somewhere the middle 
     of band \eqref{int_range}, it will be interesting to detect it with the two experiments
     in synergy for the possibility
     of measuring both the indexes 
     $n_1$ and $n_2$.

      If $n_1$ and $n_2$ have opposite sign,  the break position $f_{\rm break}$ and the  corresponding value of $\Omega_{\rm GW}^{\rm break}$
   are given by
    \bea
    \label{eq_break}
    f_{\rm break}&=&
    \left(-{n_1}/{n_2} \right)^{{1}/{\sigma}}\,f_\star,
    \\
    \Omega_{\rm GW}^{\rm break}&=&{\Omega_\star}
    \,\left[ \frac{\left(-n_2/n_1 \right)^{n_1/(n_1-n_2)}}{2}
    +\frac{\left(-n_1/n_2 \right)^{n_2/(n_2-n_1)}}{2}
    \right]^{(n_2-n_1)/\sigma}.
    \eea 

Currently, we have an indirect bound on the amplitude of a cosmological SGWB signal in the frequency band \eqref{int_range}, because its amplitude should not exceed the big-bang nucleosynthesis (BBN) bound $\Omega_{\rm GW}\le 1.7\times 10^{-6}$ \cite{Pagano:2015hma}. Moreover,
 at the frequency scales of ground-based interferometers -- around deci-Hertz -- the Ligo-Virgo-Kagra
 collaboration currently
 sets the upper  bound $\Omega_{\rm GW}\le 6\times10^{-8}$ at the reference ground-based frequency of $25$ 
 Hz \cite{LIGOScientific:2019vic}, for a  flat GW spectrum. In this work, we  consider the BBN bound as reference
   for the maximal amplitude of the SGWB even when studying GW sources active after BBN.

After these preliminary considerations, we can start looking at concrete early universe sources of GW. We do not plan
to be exhaustive, but to discuss selected examples of sources 
which lead to a broad GW spectrum, and 
whose detection
would benefit from  synergies between LISA and ET. We focus
on the theoretical aspects of the discussion, and also present 
Fisher estimates on the capability of the two instruments
together to better detect properties of the SGWB profile.
 Although the  template \eqref{BPLt} is simple and general
enough to accommodate several early universe sources -- as discussed
in sections \ref{sec_pt} and \ref{sec_cs}  -- for  selected cases
in the context of inflation
we   go beyond the profile of eq \eqref{BPLt}, and we consider a different broad  ansatz for $\Omega_{\rm GW}$ 
--  the so-called log-normal profile 
--
to better describe the frequency dependence
of the SGWB (see section \ref{sec_ci}).

 %%%%%%
 %%%%%%
 %%%%%%
 \subsection{Cosmological Phase transitions}
  \label{sec_pt}
 %%%%%%
\begin{figure}[t!]
    \centering
    \includegraphics[width=0.31\linewidth]{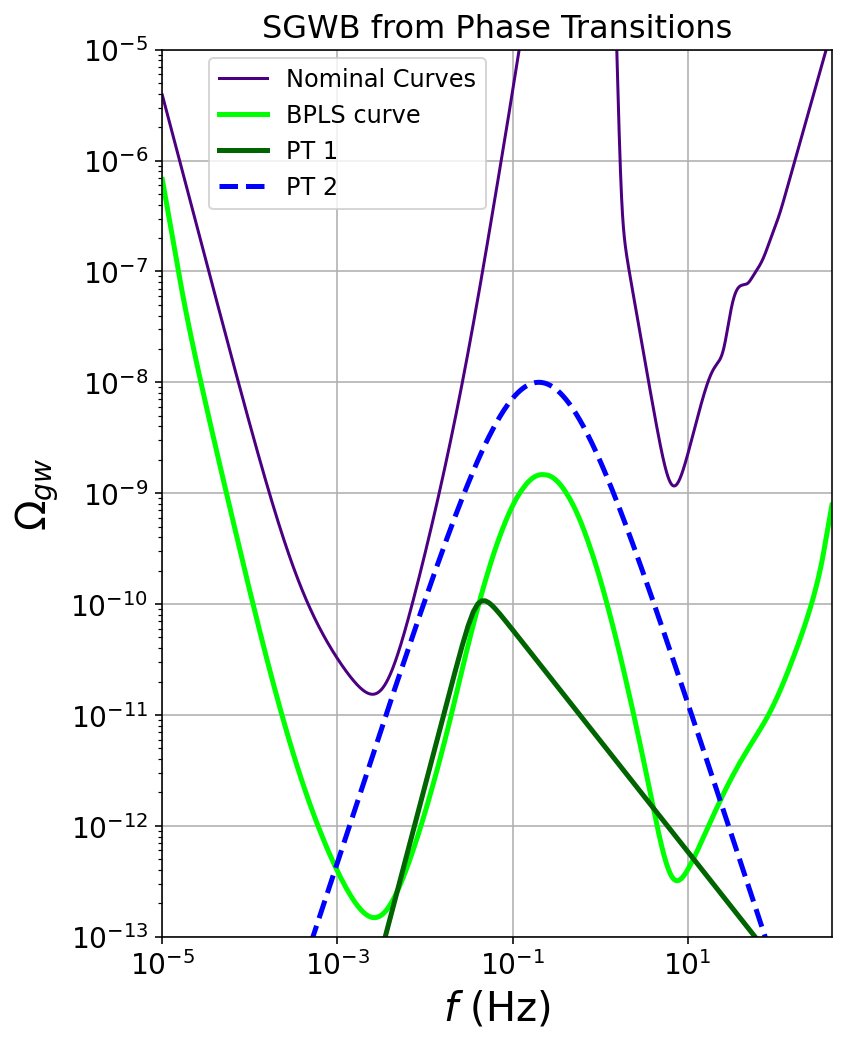}
    \caption{\it 
\small   Examples of SGWB from phase transitions (PT),
    see section \ref{sec_pt}. The purple curves correspond to the LISA and ET nominal curves, the green curve corresponds to the integrated sensitivity curve for Broken Powerlaw scenarios discussed in Section \ref{sec_sencur}, the blue and green lines represent the benchmark scenarios in Table \ref{tab:valuesPT}.}
    \label{fig:phaseonly}
\end{figure}

Several well-motivated models
of particle physics predict the existence of scalar
sectors beyond the Standard Model, whose potentials
are  characterised by  local minima.
The energy release of  strong
first order phase transitions (PTs) 
between different  vacua produces a stochastic background of GW. The detection  of such a background
would provide invaluable information on   physics beyond the Standard Model, and on the early cosmic evolution of our universe.
  We refer to \cite{Caprini:2015zlo,Caprini:2019egz,LISACosmologyWorkingGroup:2022jok,Caprini:2024hue} for complete
discussions and reviews on PT and their consequences
for different aspects of  GW production. 
  There are essentially three mechanisms \footnote{Recently it was pointed out in Ref.\cite{Jinno:2022fom}, a fourth source, namely particle production from bubbles sources GW during first order PT.} for GW production:  collisions among bubbles of different vacua \cite{Kosowsky:1991ua,Kosowsky:1992vn,Kamionkowski:1993fg}, sound waves in the primordial plasma \cite{Hindmarsh:2013xza,Hindmarsh:2015qta}, and turbulent motion \cite{Kosowsky:2001xp,Dolgov:2002ra,Caprini:2009yp}. 
The resulting shape in frequency  of the SGWB spectrum has a characteristic peak structure associated with
 the duration and properties of the PT responsible for the GW emission. The SGWB from strong first order PT  increases 
 from small towards large frequencies, reaches a maximum associated with the Hubble size during the PT in early universe, and then decreases in amplitude.

According
to the recent discussion in \cite{Caprini:2024hue}, in the first case of bubble collisions we can expect a SGWB with a broken power-law profile as in eq \eqref{BPLt}, while the other
two cases are better described by a double broken power law template. At first approximation --  since as mentioned earlier  we are not interested in fine details of the SGWB frequency dependence, but only on its overall
 structure within the broad interval \eqref{int_range} -- we do not take into account the differences
between
the latter and the former template. We  consider the BPL profile in eq \eqref{BPLt} as describing reasonably well the overall frequency dependence of a  SGWB from first order PT, 
which can in principle span over the frequency band  \eqref{int_range}. 
In this context,
the parameters in the template   \eqref{BPLt} depend on the GW production mechanisms, as well
as on the particle physics models sourcing the PT 
in the first place (see the recent work \cite{Caprini:2024hue} for a more detailed
analysis).

\subsubsection*{Position and height of the peak}

 For the case of bubble collisions, 
 many studies over the years have clarified the role of bubble dynamics and surrounding relativistic
fluid shells for GW production (see e.g. \cite{Caprini:2019egz} which contains a complete review). The dynamics of fluid shells might be important,  requiring to go beyond the so-called thin-shell approximation (see \cite{Ghosh:2023aum,Guo:2024gmu} for latest developments). 
In  the limit of strong phase  transitions, the inverse duration of the transition, denoted as  $\beta/H_\star$ (normalized against the Hubble parameter at the transition epoch), and the temperature  $T_*$ at transition are related to the BPL amplitude $\Omega_\star$ and break position $f_\star$
by the formulae \cite{Caprini:2024hue}
\begin{eqnarray}
\Omega_\star&\simeq&\frac{2\,H_*^2}{ 10^6\,\beta^2} 
\hskip0.7cm ; \hskip0.7cm
%\\
\frac{f_*}{\rm Hz}\,\simeq\,
\frac{1}{10^{8}\,\sqrt{\Omega_\star}}
\,\left( 
\frac{T_*}{100\,{\rm GeV}}
\right).
 \end{eqnarray}
  We refer to \cite{Caprini:2024hue} for details. 
Hence, by tuning appropriately the transition
temperature and its duration, the position of the break (see eq \eqref{eq_break}) might be placed freely within the band \eqref{int_range}.

A precise measurement of 
the value of $f_{\rm break}$ informs us of when 
the PT occurs during cosmological history, and the time-scale 
of its duration.
As an explicit example,  let us assume $n_1=-n_2\ge 0$ (so that $f_{\rm break}= f_\star$) and a  high SGWB amplitude  $\Omega_\star \,=\,10^{-6}$. The transition temperature corresponds to the electroweak value -- $T_* = 10^{2}$ GeV -- for a break in the  LISA band at $f_\star =  10^{-5}$ Hz, the lower
extremum of the interval
 \eqref{int_range}. On the other hand, we find an intermediate scale of $T_* = 10^{9}$ GeV for a break within the  ET band at $f_\star = 5 \times 10^{2}$ Hz, the upper extremum  of the interval
 \eqref{int_range} (see the recent discussion \cite{Caprini:2024ofd}). 
Such intermediate-case PTs are very well motivated from scenarios of BSM involving axion physics with classic Peccei-Quinn symmetry breaking scales around $10^{9} - 10^{11}$ GeV \cite{Dev:2019njv, DelleRose:2019pgi,VonHarling:2019rgb,Conaci:2024tlc}. Interestingly, for this first order PT, the energy scale of new physics the ground-based detectors are  sensitive to, roughly coincides with the lowest possible energy scale at which the Peccei-Quinn (PQ) symmetry $U(1)_{\rm PQ}$ has to be broken in QCD axion models which also address the strong CP problem of the SM~\cite{Peccei:1977hh, Peccei:1977ur, Weinberg:1977ma,Wilczek:1977pj}. The involved axion scalar field is a viable cold dark matter (DM) candidate~\cite{Preskill:1982cy, Abbott:1982af,Dine:1982ah}, and even more generally, axion-like particles (ALPs) are well-motivated, since they are naturally present as pseudo Nambu-Goldstone bosons in many BSM extensions with a spontaneously broken global $U(1)$ symmetry, e.g. in several string theory avatars~\cite{Svrcek:2006yi, Arvanitaki:2009fg, Marsh:2015xka}. 
 Yet another motivation for intermediate scale PT comes from neutrino mass generation also known as seesaw mechanism connected to the scale of baryogenesis via leptogenesis, see e.g.  \cite{Dasgupta:2022isg,Huang:2022vkf,Chun:2023ezg,Azatov:2021irb,Borah:2022cdx}. In fact, 
 while standard thermal leptogenesis is a simple and elegant mechanism, it requires  a small window of right-handed neutrino
masses in the high energy regime $10^{9} - 10^{11}$ GeV. Hence, GW detectors 
could probe these energy ranges which can not be  
probed by accelerator experiments 
(see  the analysis \cite{Badger:2022nwo} in terms of existing GW data).

Moreover, 
 it is also possible to push $T_\star$
to high values considering non-minimal Higgs
scenarios, or scalar setups belonging to dark 
sectors beyond the Standard Model (see e.g. \cite{Schwaller:2015tja}, and also see \cite{Grojean:2006bp} for an early, complete analysis of
the possibility of tuning the scale
of the transition to intermediate values and its consequences
for interferometer detections and physics beyond
the standard model).
The possibility of detecting a break in the spectrum
somewhere within the entire range \eqref{int_range} would be an
important opportunity to study the physics of PTs  and probe
high energy physics beyond the electroweak scale.

\subsubsection*{The spectral indexes, and our
benchmark scenarios}

\renewcommand{\arraystretch}{1.25}
\begin{table}[h!]
\begin{center}
\begin{tabular}{| c | c | c | c | c | c | }
%\hline
\hline
 {\rm }& \cellcolor[gray]{0.9} $\Omega_\star$  &\cellcolor[gray]{0.9}$n_1$&\cellcolor[gray]{0.9}$n_2$ &\cellcolor[gray]{0.9}$\sigma$&\cellcolor[gray]{0.9}$f_\star$ \\
\hline
\cellcolor[gray]{0.9} {\rm PT1} & $1\times10^{-10}$ & $3$ & $-1$ & $7.2$ & $0.04$ 
\\
\hline
\cellcolor[gray]{0.9} {\rm PT2} & $1\times10^{-8}$ & $2.4$ & $-2.4$ & $1.2$ & $0.2$
\\
\hline
\end{tabular}
\caption{\it Benchmark values for the scenarios corresponding to cosmological  phase transitions. \label{tab:valuesPT} }
\end{center}	
\end{table} 
Interestingly, the values of the tilts $n_{1,2}$ and of the smoothing quantity $\sigma$ depend  more specifically on the PT scenario and GW source under consideration. Measuring  both indexes $n_1$ and $n_2$
accurately
is then essential to reconstruct the details of the physics leading to the PT: such a measurement
can be achieved by the synergy of LISA and ET, as we are going to demonstrate. We analyze
two benchmark scenarios, PT1 and PT2, as summarized in Table \ref{tab:valuesPT}. 
For uncorrelated  primordial sources, 
one finds a slope $n_1=3$  
in the deep infrared (see e.g. \cite{Caprini:2009fx} for a detailed analysis). But more generally, the slope depends sensitively on the GW production.
We first consider  a benchmark scenario  PT1,  with spectral index $n_1=3$  in the  infrared; in the UV, we consider $n_2=-1$ as predicted  in  scenarios  where GWs are  produced by sound waves of the bubble surrounding plasma, or by effects of turbulent behaviour in the fluid. For the second benchmark scenario, PT2, 
we consider the model 
  of \cite{Lewicki:2022pdb}
in the context of highly relativistic fluid shell dynamics, which 
finds $n_1\,=\,-n_2\,\simeq\,2.4$
and $\sigma\,\simeq\,1.2$. 
 These two scenarios are presented in the left panel of Fig \ref{fig:phaseonly}.

\begin{figure}[H]        \includegraphics[width=0.41\textwidth]{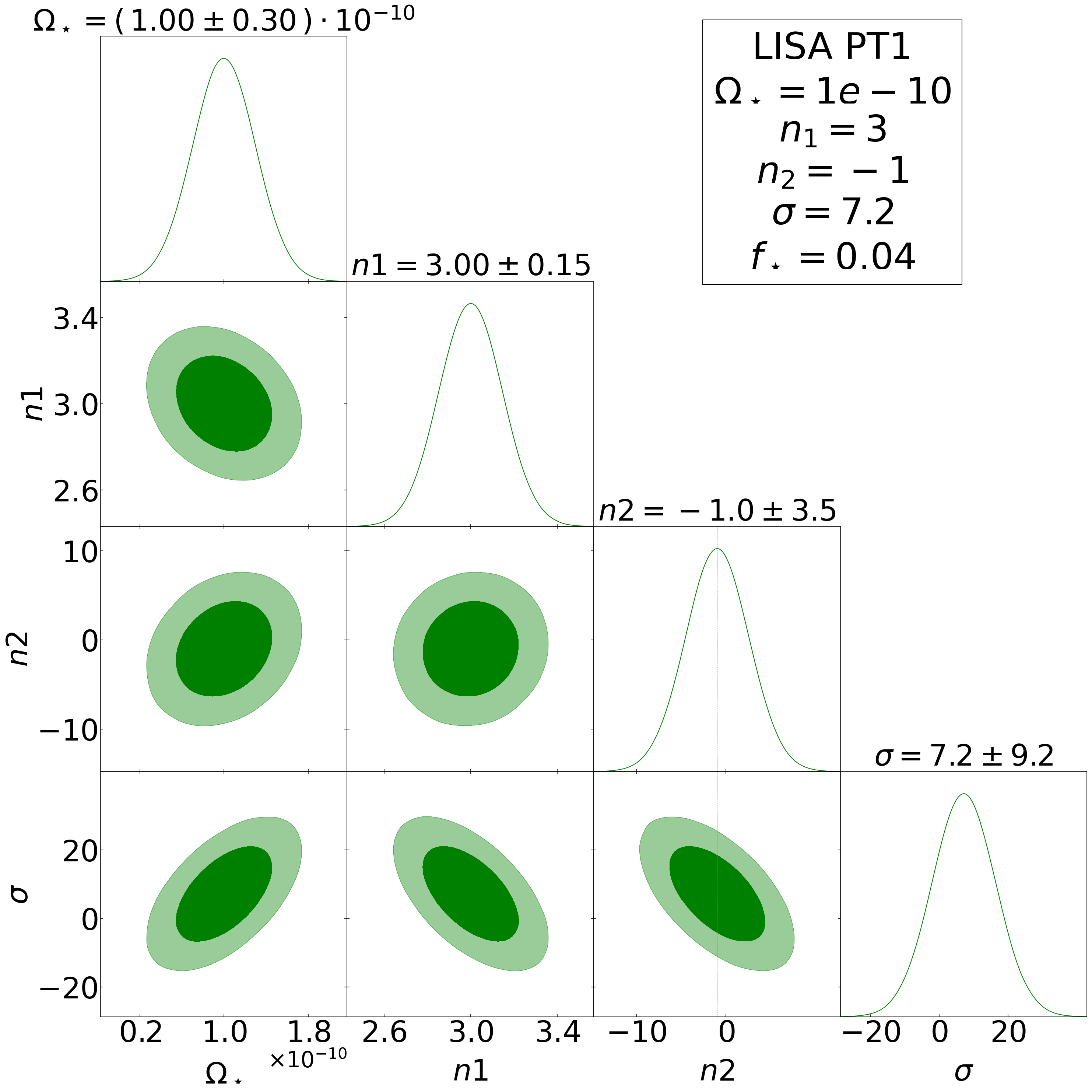}   \includegraphics[width=0.41\textwidth]{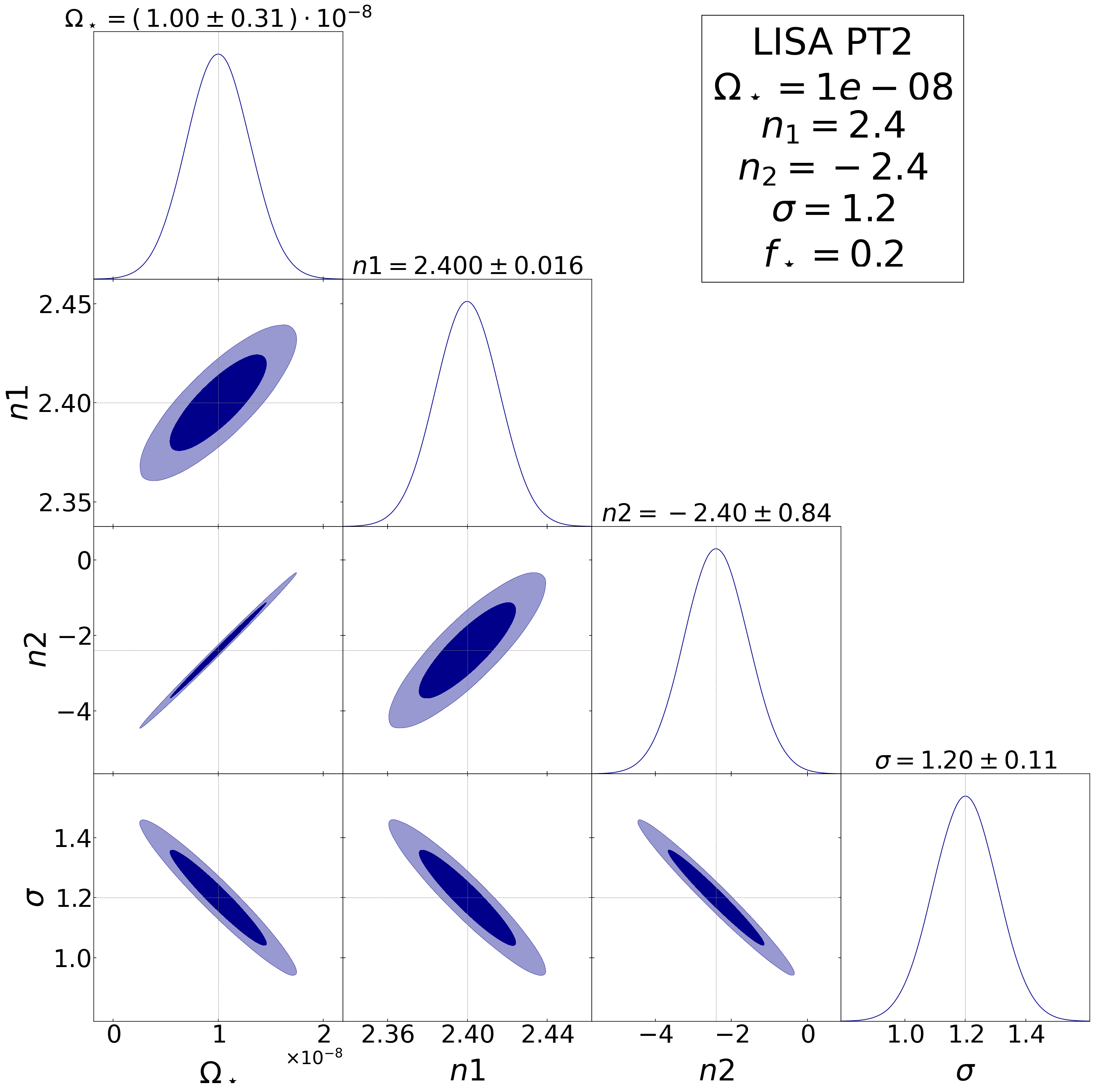}\\        
 \includegraphics[width=0.41\textwidth]{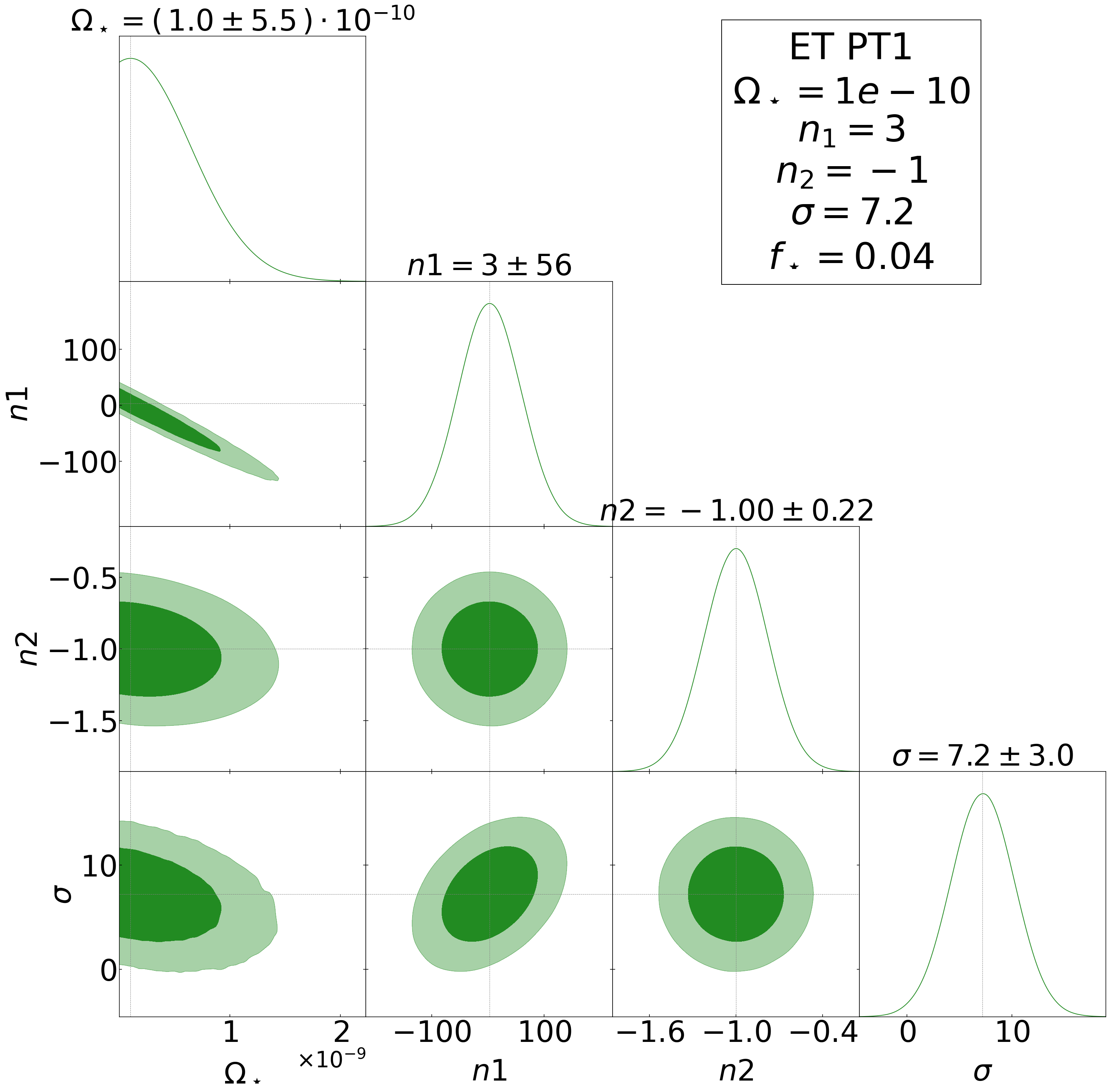}
 \includegraphics[width=0.41\textwidth]{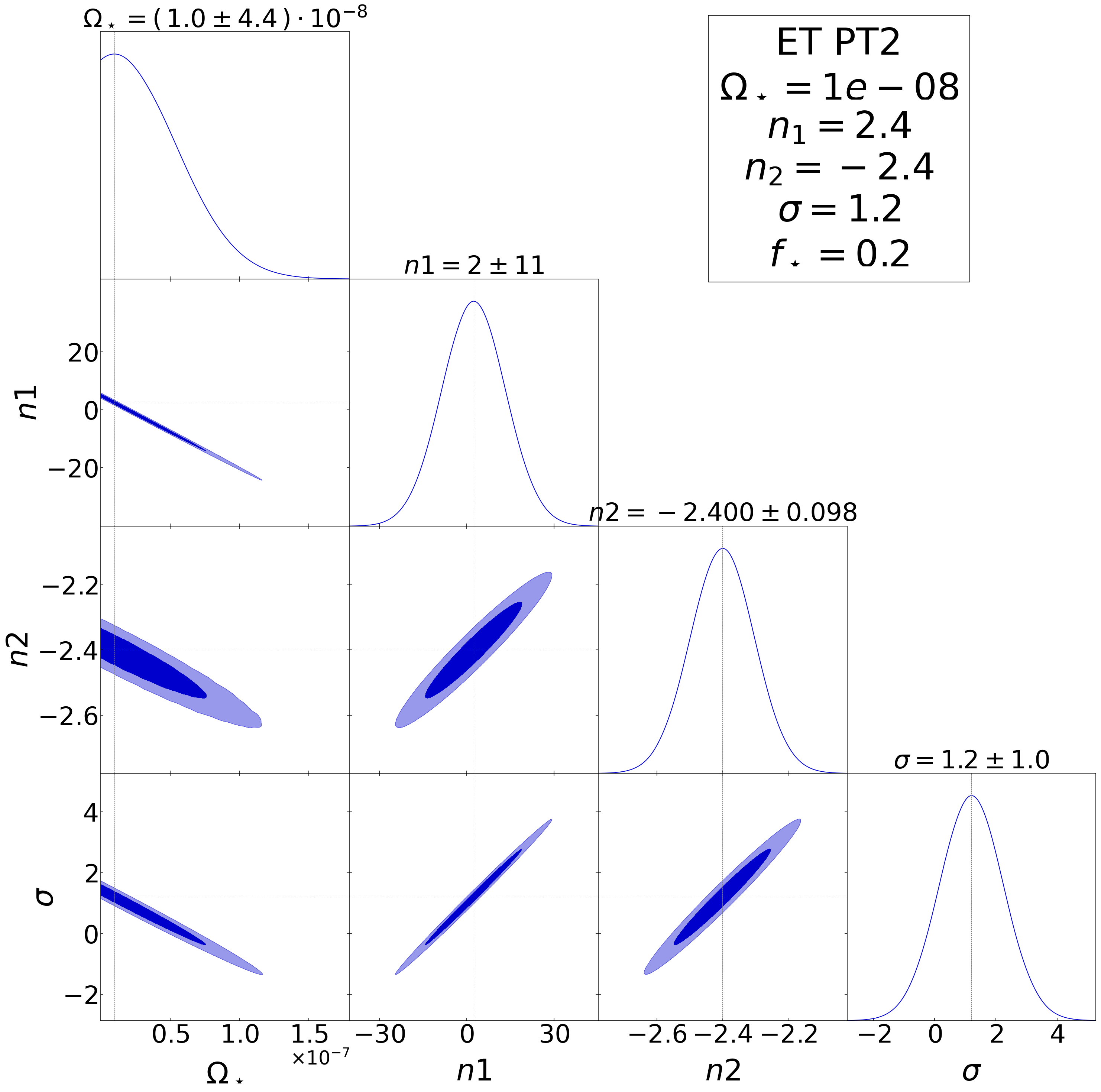}\\
  \includegraphics[width=0.41\textwidth]{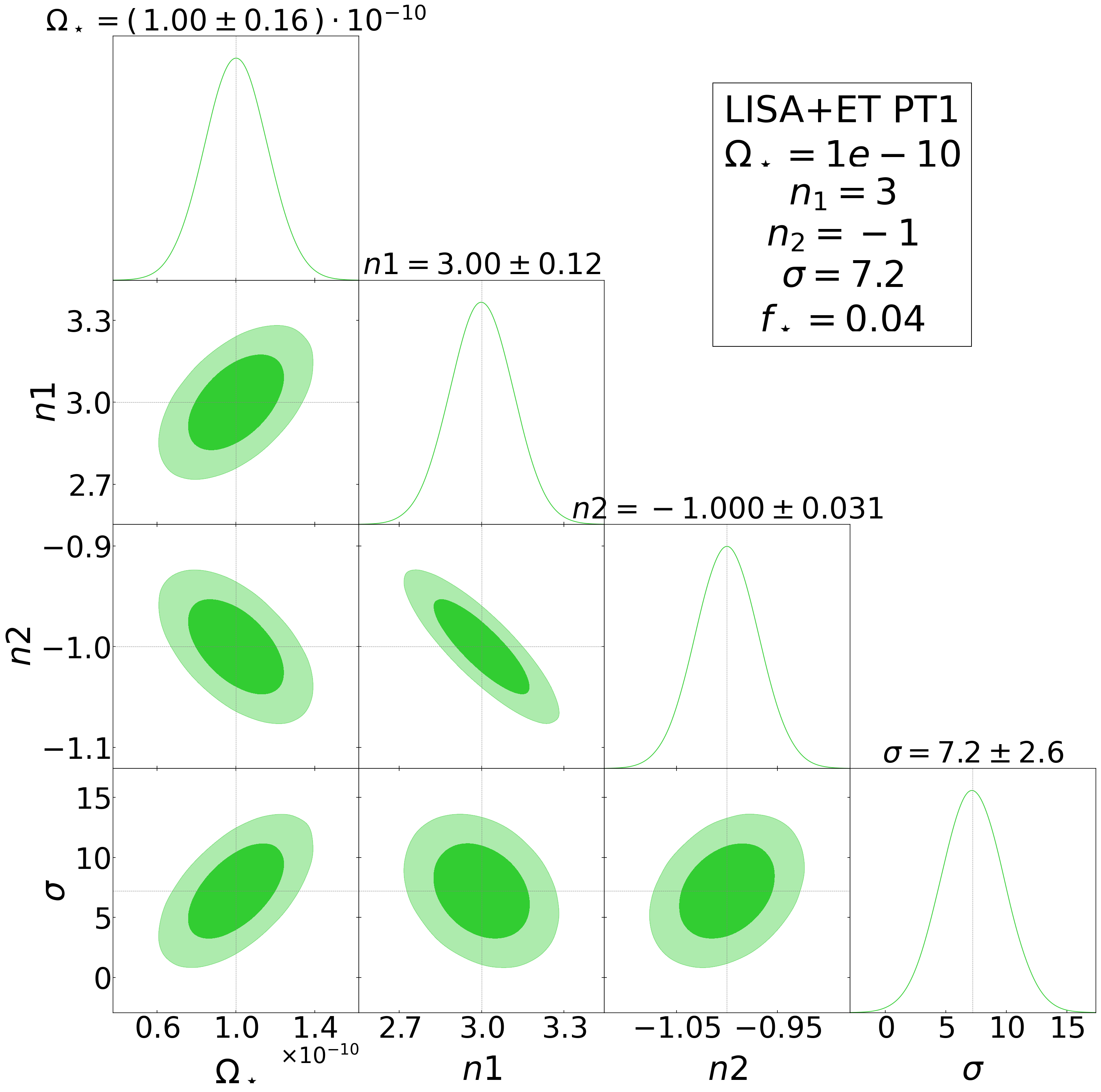}\includegraphics[width=0.41\textwidth]{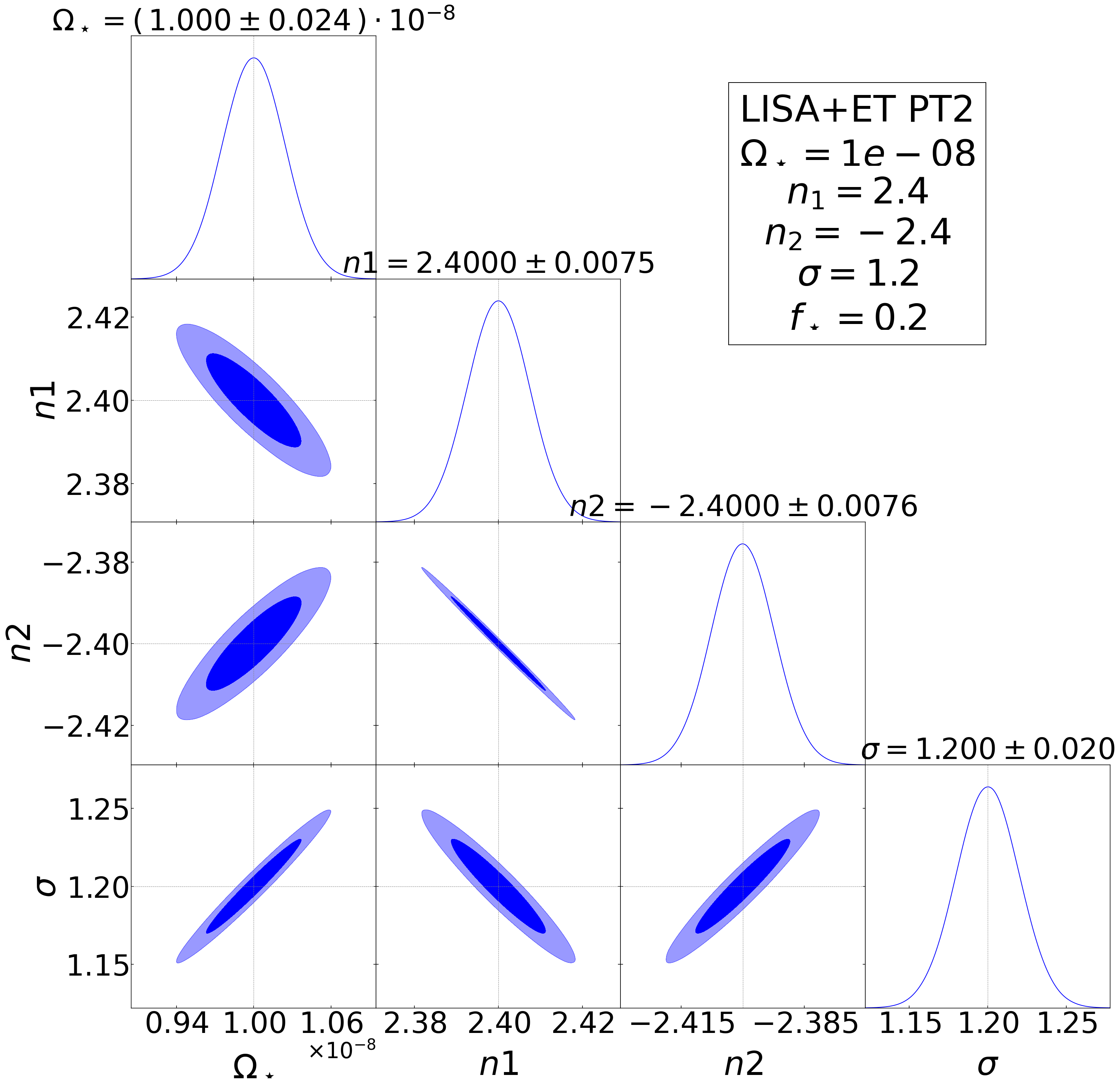}
    \caption{ \it \small Fisher forecasts for the phase transition benchmark scenarios
    PT1 and PT2, summarized in Table \ref{tab:valuesPT}.}
\label{fig:Fishphasel}
\end{figure}

\subsubsection*{What can we learn about PT 
by LISA and ET in synergy?}

In both of the aforementioned benchmark models, the position
of the break of the broken power-law spectrum is located somewhere in the middle
between LISA and ET bands, see Fig \ref{fig:phaseonly}. 
 The plots
   suggests that a measurement in synergy between LISA and ET would allow us to get important information on the position of the break and the value of the spectral tilts, and thereby  on particle physics models
leading to a first order PT.
 Notice that, importantly, the signal profile lies well below the nominal sensitivity curve of both
 experiments. 
  Nevertheless,  it can be detected by integrating over frequencies: recall the expression for the SNR (eq \eqref{exp_snr}). 
 Such an integration allows one to acquire sufficiently high values of SNR even if the signal
 lies well below the nominal sensitivity curves of the experiment. In fact, this property suggests the definition of broken power law sensitivity curve, as depicted with light green colour in Figures \ref{fig:phaseonly} and \ref{fig:CS1}, left and middle panel: the GW signals lie well above such a  curve. We will reconsider
 this topic in section \ref{sec_sencur},
 in the context
 of frequency-integrated
 sensitivity curves.

%%%%%%%%%%%%%%%%%%%
In fact,  we can carry out a Fisher analysis using a likelihood whose structure is
given in eq \eqref{def_like}, and assuming
the BPL ansatz \eqref{BPLt} with the aforementioned two sets of  benchmark values for the parameters, summarised
in Table \ref{tab:valuesPT}. 
 The benchmark values of $\Omega_\star$ are selected in a way 
such as to show how the two instruments {\it together} can achieve good accuracy 
in the measurements of the template parameters. 
  The results 
are shown in Fig \ref{fig:Fishphasel}. From now
on, we will present Fisher plots obtained  using the
 \verb|GetDist| package \cite{Lewis:2019xzd}.

 For both the scenarios PT1 and PT2 as shown in Fig \ref{fig:Fishphasel},  each of the two experiments -- LISA and ET -- can measure
with good accuracy  {\it only one} of the two spectral indexes $n_1$ or $n_2$. The two experiments in synergy, though, can
measure {\it both} these quantities well, with an
accuracy of at least $10$ percent. Apart from the spectral
indexes, the parameter $\sigma$
controlling the degree of smoothness of the transition can also be
measured accurately by the synergy
of the two experiments.
 This implies
that, by working with LISA and ET together, we can obtain much richer information 
on the physics of PT occurring at high temperature
scales. 
Additionally, for both scenarios, the  
ET experiment by itself cannot accurately measure  the amplitude of the SGWB. Only  
in synergy with LISA can it do so,  measuring  with a $10$  percent accuracy all the parameters characterizing our benchmark models.  Moreover, the synergy of the two experiments can help
in alleviating degeneracies in the parameter
measurements -- see for example the measurement
of $\sigma$ and $\Omega_\star$ by ET only in the second
row of Fig
\ref{fig:Fishphasel}.

The  correlations, shown by the ellipses, are an indication of how the parameters co-vary according to the constraints on LISA or ET. Let us first consider the case of PT1. We can estimate that the break in the signal occurs at a frequency of approximately $0.046$ Hz, which is within the LISA band. This break is farther within the LISA band than in the case of PT2, wherein the break occurs around $0.2$ Hz. Therefore, ET can not effectively constrain $\sigma$ or $n_1$  on its own. Moreover, due to the low amplitude of the signal and the resultant low SNR, even LISA is unable to constrain the quantity $\sigma$ well. ET mainly gathers its SNR from the part of the signal after the break. If this part of the signal becomes steeper -- that is, if $n_2$ becomes a larger negative number -- then in order to compensate for this fact and accumulate adequate SNR the amplitude of the signal should  become larger. This behaviour is manifest in the correlation between $n_2$ and $\Omega_\star$ for ET. Considering the combination LISA+ET, if the slope $n_2$ alone become steeper, the total SNR decreases. Due to the low overall amplitude of the signal, making $n_2$ very steep could result in the total SNR falling below the value of $5$,  the minimum SNR value we  consider in this work for the detectability of a signal. Hence the $\Omega_\star$  has to increase, for allowing  for the total SNR to cross such a threshold value.

In order to elucidate this point further, we consider the case of PT2 also. As mentioned earlier, in this case the break in the signal occurs within the LISA band, around $0.2$ Hz. Due to the relatively higher amplitude of the signal compared to PT1, LISA by itself can constrain $n_1$ and $\sigma$ quite well. However, ET can not constrain sufficiently well either of these parameters  on its own. Considering the case of  ET only, if the part of the signal after the break becomes steeper -- that is, if $n_2$ becomes a large negative number -- then the signal amplitude has to become larger in order to compensate for this fact and accumulate adequate SNR. This is what we learn in the correlation between $n_2$ and $\Omega_\star$ for ET. Such behaviour is similar to what we find for PT1. However, when considering the combination LISA+ET -- which includes adequate SNR from the LISA band -- by increasing $\Omega_\star$ we would increase the total combined SNR to very large values, even though the SNR for ET alone might remain the same. Therefore, the correlation pattern is reversed. 
\renewcommand{\arraystretch}{1.25}
\begin{table}[h!]
\begin{center}
\begin{tabular}{| c | c | c | c | }
%\hline
\hline
 {\rm }& \cellcolor[gray]{0.9} LISA  &\cellcolor[gray]{0.9}ET &\cellcolor[gray]{0.9}LISA$+$ET \\
\hline
\cellcolor[gray]{0.9} {\rm PT1 SNR} & $11$ & $13$ & $17$ \\
\hline
\cellcolor[gray]{0.9} {\rm PT2 SNR} & $528$ & $544$& $758$\\
\hline
\end{tabular}
\caption{\it Calculated SNR values for the scenarios corresponding to cosmological  phase transitions. \label{tab:valuesPTSNR} }
\end{center}	
\end{table} 
We see that the SNR for PT$2$ is much higher with respect to PT$1$. The values are $42$ to $48$ times larger for PT$2$ than for PT$1$.
In conclusion,
this analysis demonstrates
quantitatively, by means of the Fisher plots of Fig \ref{fig:Fishphasel}, the advantages of measuring
the profile of SGWB over a broad interval, for reconstructing the physics of the PT and the details
of the frequency profile around the peak. 

 %%%%%%
 %%%%%%
 %%%%%%
 %%%%%%
 \subsection{ Cosmic strings}
 \label{sec_cs}
 %%%%%%
\begin{figure}[t!]
    \centering
\includegraphics[width=0.31\linewidth]{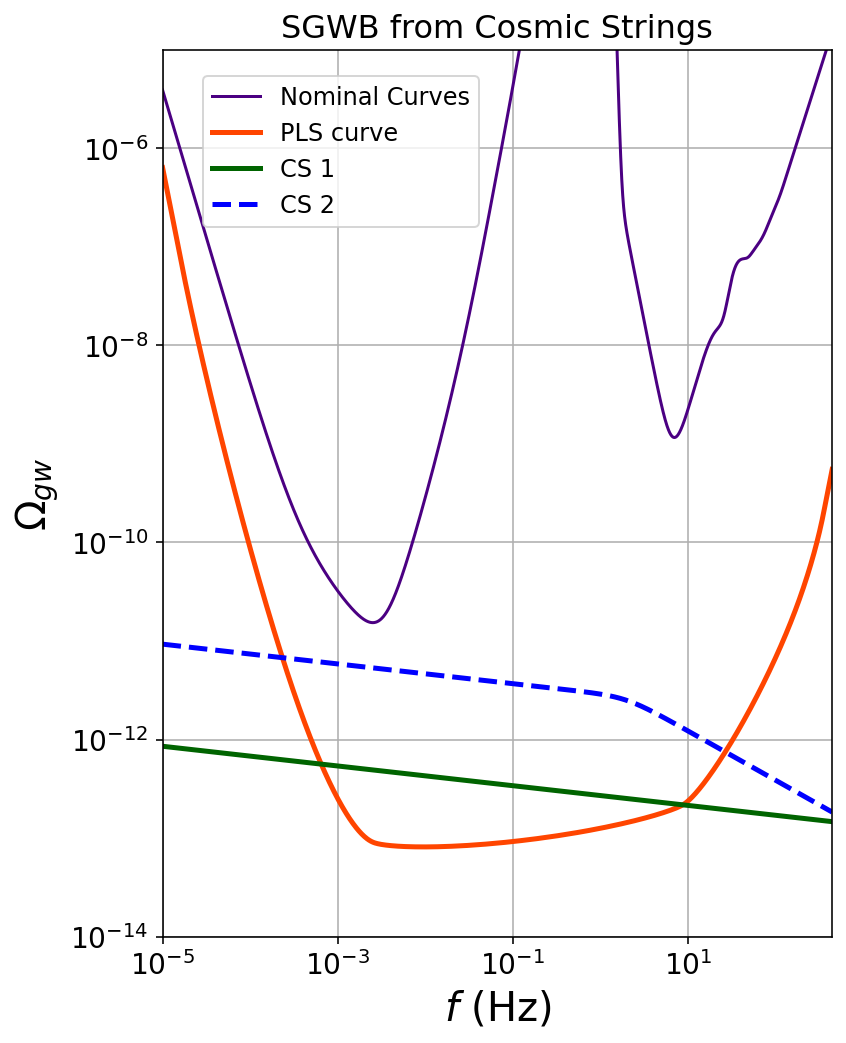}
    \caption{\it 
\small     Examples of SGWB from cosmic strings (CS), see section \ref{sec_cs}. 
The purple curves correspond to the LISA and ET nominal curves, the green curve corresponds to the integrated sensitivity curve for Broken Powerlaw scenarios discussed in Section \ref{sec_sencur}, and the blue and green lines
 represent the benchmark scenarios in Table \ref{tab:valuesCS}
    \label{fig:CS1}.}
\end{figure}
Another opportunity for determining the pre-BBN cosmic history of the universe is
associated with the detection of GW sourced by a network of cosmic strings. Cosmic strings are basically one-dimensional objects produced by the spontaneous breaking of a $U(1)$ symmetry in the early universe~\cite{Nielsen:1973cs, Kibble:1976sj}, or sometimes considered as fundamental objects, for instance, in superstring theory~\cite{Copeland:2003bj, Dvali:2003zj, Polchinski:2004ia, Jackson:2004zg, Tye:2005fn}. The essential feature in the GW emmittted by the cosmic strings is that they are sources of very long-standing over the entire history of the evolution of the universe \cite{Vilenkin:1981bx,Vilenkin:1981zs,Vachaspati:1984gt,  Hindmarsh:1994re, Vilenkin:2000jqa,Damour:2000wa}. Let us try to understand why: after the formation of the network of cosmic strings, it assumes a constant fraction of the total energy budget of the universe, this is very popularly known as the scaling regime \cite{Albrecht:1984xv, Bennett:1987vf, Allen:1990tv,Martins:2000cs, Figueroa:2012kw, Martins:2016ois}. Consequently as long as the strings exist in the universe it will keep on emitting GW during and this happens through most of the universe history. Since frequency of cosmic sources of GW represents time in the early universe (higher frequency means earlier time) this generates a GW spectrum spanning many orders of magnitude in frequencies. Therefore a possible measurement of the GW spectrum from high to lower frequencies will  determine the universe expansion rate from early to later times by investigating the features on the cosmic string GW spectrum \cite{Cui:2017ufi, Cui:2018rwi,Gouttenoire:2019kij,Gouttenoire:2019rtn,Blasi:2020mfx,Ghoshal:2023sfa}. A detailed study of the impact of various pre-BBN cosmological epochs on the GW spectrum emitted from local and global cosmic strings was carried out in Refs. \cite{Ghoshal:2023sfa} which clearly predicts the multi-band frequency spectrum of the GW detectors. In
fact, the SGWB from cosmic strings
 can be conveniently studied
in synergy between LISA and ET. 
(See
e.g. \cite{Auclair:2019wcv,Blanco-Pillado:2024aca} and
references therein for
a recent assessment 
in the context
of LISA physics.) 

The process of string loop formation, evolution, and decay into GWs is
quite complex. It is usually studied numerically, although 
accurate semi-analytical fits for the frequency shape of the SGWB can be determined (see e.g. the recent account \cite{Blanco-Pillado:2024aca}).
The SGWB characteristics depend on the string tension $(G\,\mu)$, normalized against the gravitational constant $G$, and on the loop size $\alpha$, normalized against its time of formation.
Typically, the SGWB frequency profile initially increases during the first phase of the decay of string loops into GWs, up to a maximum at the frequency 
\cite{Kuroyanagi:2018csn}
\be
f_{\rm max}\simeq 3 \times 10^{-8}\,\left( \frac{G \mu}{10^{-11}}\right)^{-1}\,{\rm Hz}
\,.
\ee
The value of the quantity $G \mu$ is quite model dependent,
but $f_{\rm max}$
usually occurs  at frequencies well below the band \eqref{int_range} we are interested in.  For example, 
for a specific model of loop distribution, the Ligo-Virgo-Kagra collaboration sets a bound ${G \mu}\le4 \times 10^{-15}$ 
\cite{LIGOScientific:2021nrg}.
 Then, at larger frequencies, the SGWB  becomes nearly constant, or slightly decaying with an approximately
constant slope. We refer the reader to \cite{Kuroyanagi:2018csn}
for a more complete discussion and references therein. As
a consequence, if future measurements favour a SGWB
entering from the left side of the band \eqref{int_range} with a negative
slope, they will provide circumstantial
evidence for a cosmic string origin of the signal. 
{Going beyond the discussions of  local cosmic strings  as above, there are several additional instances of CS sources as:  meta-stable cosmic strings \cite{Buchmuller:2023aus}, global cosmic strings \cite{Chang:2021afa} or cosmic superstrings \cite{Sousa:2016ggw}, current-carrying  \cite{Auclair:2022ylu}  and superconducting strings \cite{Rybak:2022sbo}.  Various other topological defects like monopoles and textures can interact with cosmic strings \cite{Dunsky:2021tih}: a separate dedicated analysis would be required since for each of them carries features of top of the standard flat spectrum, which  can be tested and searched for in the broad band interval of
eq \eqref{int_range}.

Motivated by
the previous considerations we shall now discuss two benchmark models. In our first cosmic
string benchmark
scenario -- CS1 -- we assume a constant power-law profile in 
the frequency band \ref{int_range}, with a spectral index
 $n_1=-0.1$.
  If the amplitude of $\Omega_{\rm GW}$, proportional to ${(G \mu)^2}/{H_0^2}$, is sufficiently large, 
then
such a  power law profile can be probed with  both the LISA and ET instruments. We present this
case in Fig \ref{fig:CS1}. Note that the constant slope lies well below
the nominal sensitivity curves. Nevertheless, as mentioned above,
 it can be detected with
sufficient SNR by integrating
over frequencies. For this
reason we represent
in the same plot in light green the corresponding broken-power-law
sensisitivity curve: more on this in the next section.

\renewcommand{\arraystretch}{1.25}
\begin{table}[h!]
\begin{center}
\begin{tabular}{| c | c | c | c | c | c | }
%\hline
\hline
 {\rm }& \cellcolor[gray]{0.9} $\Omega_\star$  &\cellcolor[gray]{0.9}$n_1$&\cellcolor[gray]{0.9}$n_2$ &\cellcolor[gray]{0.9}$\sigma$&\cellcolor[gray]{0.9}$f_\star$ \\
\hline
\cellcolor[gray]{0.9} {\rm CS1} & $4\times10^{-13}$ & $-0.1$ & $-0.1$ & $-$ & $0.02$ 
\\
\hline
\cellcolor[gray]{0.9} {\rm CS2} & $2.5\times10^{-12}$ & $-0.1$ & $-\frac{1}{2}$ & $3$ & $2$
\\
\hline
\end{tabular}
\caption{\it Benchmark  values  for each cosmic string scenario. \label{tab:valuesCS} }
\end{center}	
\end{table}

Let us also consider another interesting possibility offered 
by  the synergy of LISA with ET: an accurate test
of  the early expansion of our universe.
By measuring the frequency profile of the spectrum 
we can  probe (or constrain) early epochs of non-standard cosmic expansion,  preceding big bang nucleosynthesis. Early matter domination eras, kination domination, or the early presence of extra degrees of freedom beyond the
Standard Model can modify the string network evolution, and the corresponding dynamics of GW production  (see
e.g. \cite{Allahverdi:2020bys} and references therein for a comprehensive review, and \cite{Ghoshal:2023sfa} for a dedicated analysis).
  Non-standard early cosmological epochs  lead  to sudden changes, as breaks and features  in slope at frequency
 \footnote{Here $T_{\rm rd}$ is the universe temperature at the transition between non-standard evolution and  radiation domination. We do not take into account in this formula possible effects of  extra degrees of freedom beyond the Standard Model active at early epochs. }
\cite{Kuroyanagi:2018csn} 
\be
f_{\rm break}\,\simeq\,
\left(9\times 10^{-3}\, {\rm Hz}\right)\,\left(
\frac{T_{\rm rd}}{\rm GeV}
\right) \left(\frac{10^{-12}}{\alpha\,G\,\mu} \right)^{1/2}.
\ee
By making appropriate choices of the string
network properties, the break position
can 
occur within the interval \eqref{int_range}.
Right after the break, the slope of the spectrum changes to a slope depending on $\omega$ -- the equation of state during the non-standard cosmological expansion. For $\omega \ge 1/4$, the tilt
$n_2$ is given by:
$
n_2\,=\,-2 {\left(3 \omega-1 \right)}/{\left(3 \omega+1\right)}
$ \cite{Blanco-Pillado:2024aca}. Hence, knowledge of the spectral tilts $n_{1,2}$ and of the break position $f_{\rm break}$
offers us crucial information not only on the cosmic
string properties, but also on the evolution of the universe
prior to BBN. In Fig \ref{fig:CS1}, we show an explicit example of this phenomenon 
for the benchmark scenario dubbed CS2 in Table \ref{tab:valuesCS}, where we have
chosen $\omega=5/9$.

\bigskip
Other than continuous symmetries which when broken leads to GW (as discussed above), domain walls (DWs)~\cite{Vilenkin:1984ib} are topological defects, are formed when a discrete symmetry in some BSM scenario is broken after inflation. As well studied, during the scaling regime when the DW network evolves and expands along with its surroundings, the energy density stored is $\rho_{\rm DW}=c\, \sigma H$~\cite{Hiramatsu:2013qaa, Kibble:1976sj}, where $\sigma$ is the surface tension of the wall and $c = \mathcal{O}(1)$ is a scaling parameter. DWs keep on emitting GWs until they annihilate at a temperature given by $T = T_{\rm ann}$~\cite{Gelmini:1988sf,Coulson:1995nv,Larsson:1996sp,Preskill:1991kd}. The peak frequency of the resulting GW spectrum from DW annihilation tells us about the horizon size at the time of DW annihilation, $f_{\rm peak} = f_H(T_{\rm ann})$. Another important feature is that the frequencies $f\gg f_{\rm peak}$ the amplitude of the GW spectrum scales as $f^{-1}$. Studying closely the approximation for  the GW spectrum at the formation time $T=T_{\rm ann}$ are shown in Ref.~\cite{Hiramatsu:2013qaa,Ferreira:2022zzo} from which it can be understood the GW spectrum depends on 
on $\alpha_* \equiv \rho_{\rm DW}(T_{\rm ann})/\rho_r(T_{\rm ann})$ which is the energy density in the domain walls relative to the radiation energy density $\rho_r$ of the universe at the time of DW annihilation. 
The microscopic physics parameters of the DW model are the relative energy density in DWs, $\alpha_*$, and the temperature at which they annihilate, $T_{\rm ann}$. Depending upon if the DWs annihilate completely into dark radiation or into visible sector radiation (SM radiation), the energy density can be constrained as the equivalent number of neutrino species~\cite{Ferreira:2022zzo} $\Delta N_{\rm eff}$ which is constrained by BBN ($\Delta N_{\rm eff} < 0.33$)~\cite{Fields:2019pfx} and CMB ($\Delta N_{\rm eff} < 0.3$)~\cite{Planck:2018vyg,Ramberg:2022irf}.

 We
point out that the symmetry breaking scale  probed
by GW in this context  is different than in  PT scenarios, due to the different microscopic physics involved, see comparative analysis in Ref. \cite{Ellis:2023oxs}. 
\begin{figure}[H]
\centering
        \includegraphics[width=0.39\textwidth]{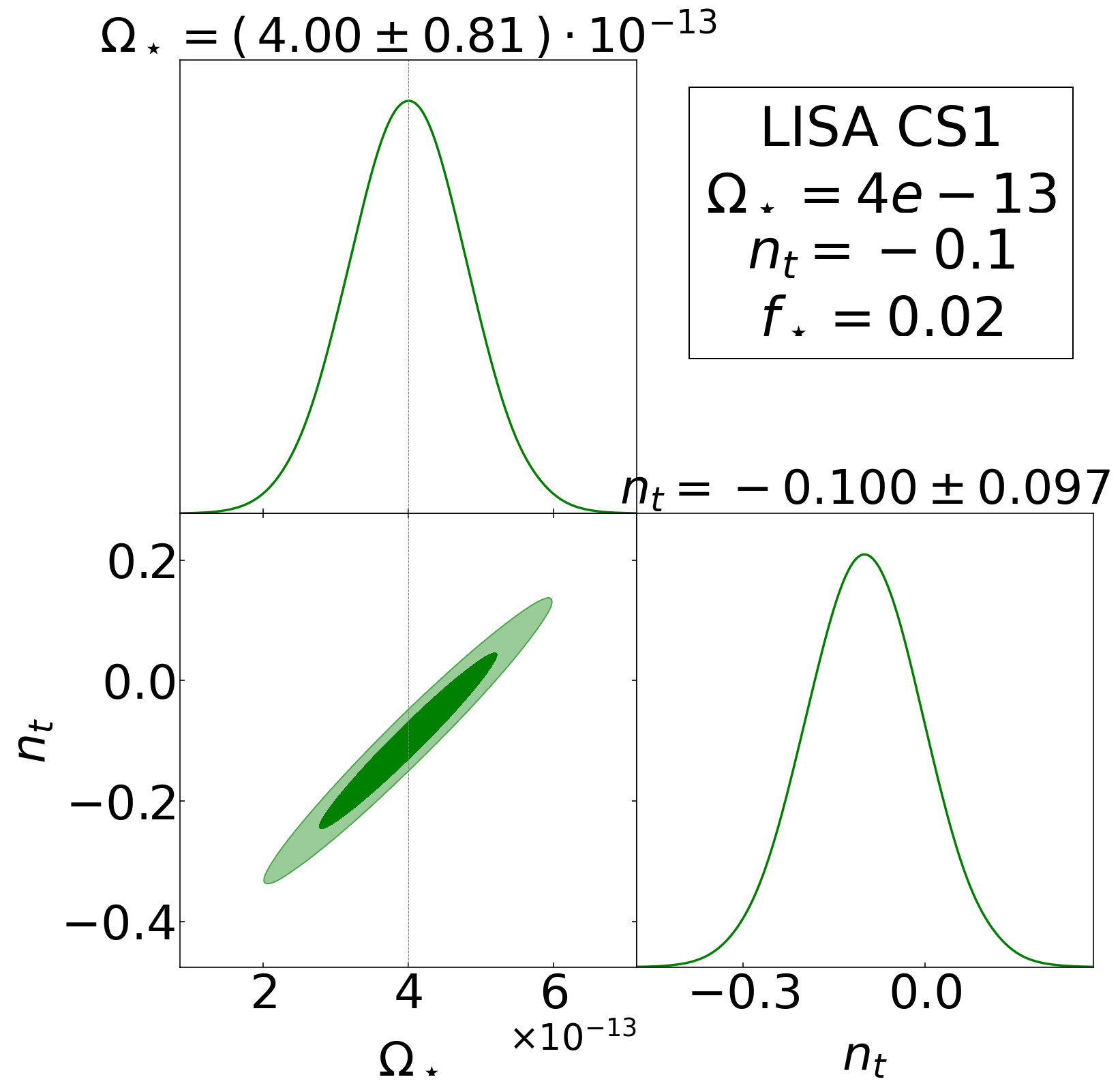}
 \includegraphics[width=0.39\textwidth]{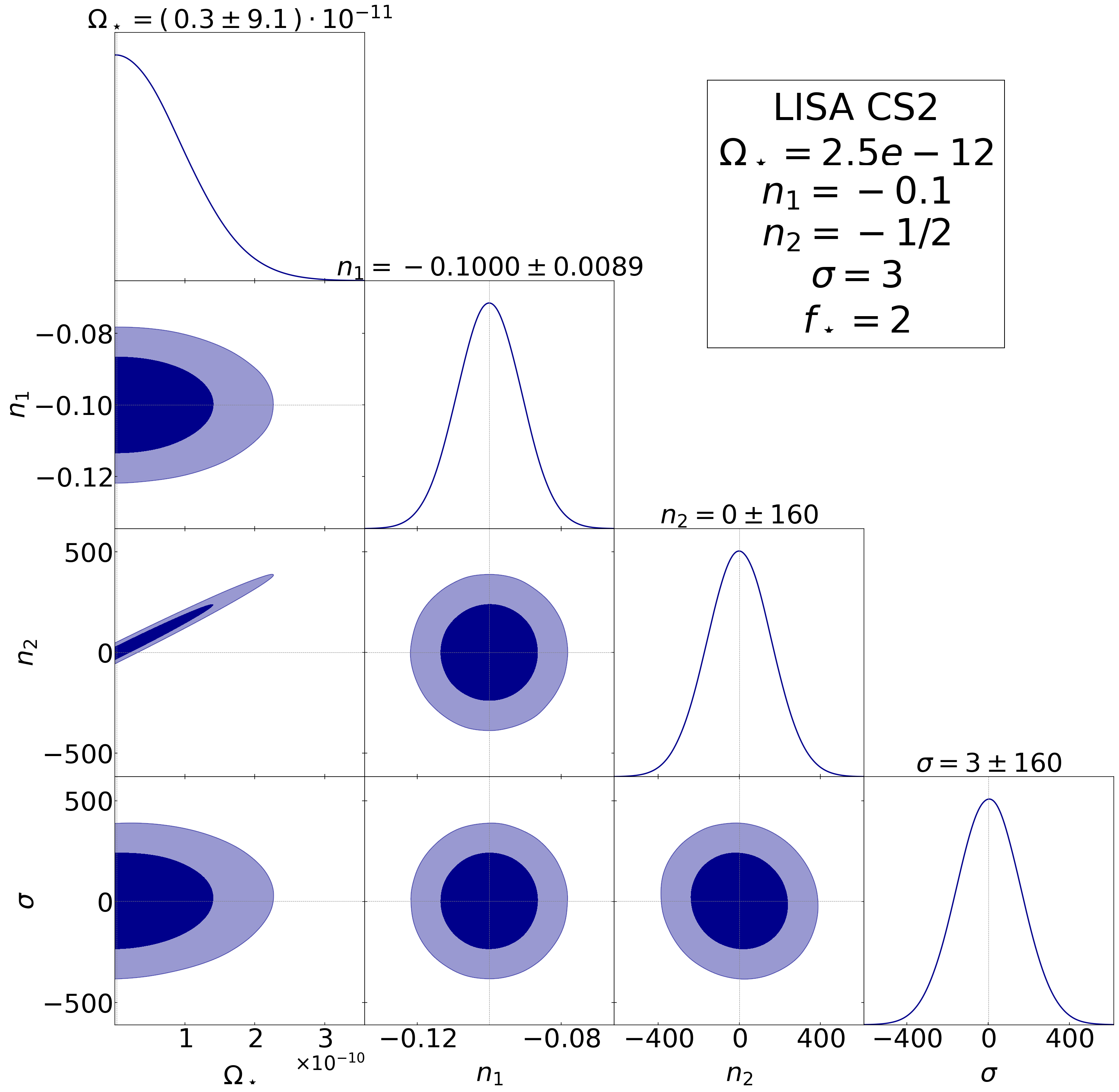}\\
  \includegraphics[width=0.39\textwidth]{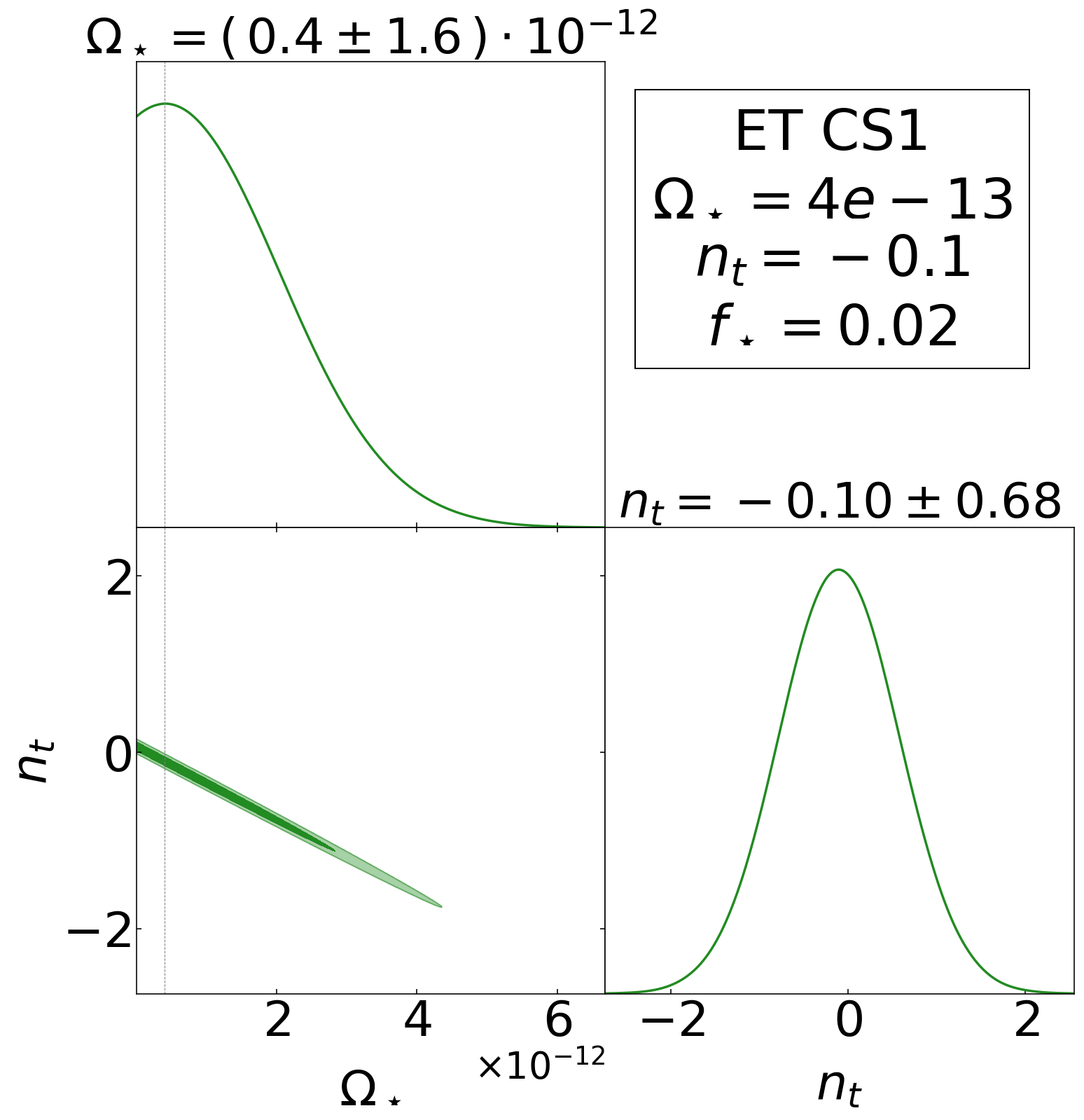}
    \includegraphics[width=0.39\textwidth]{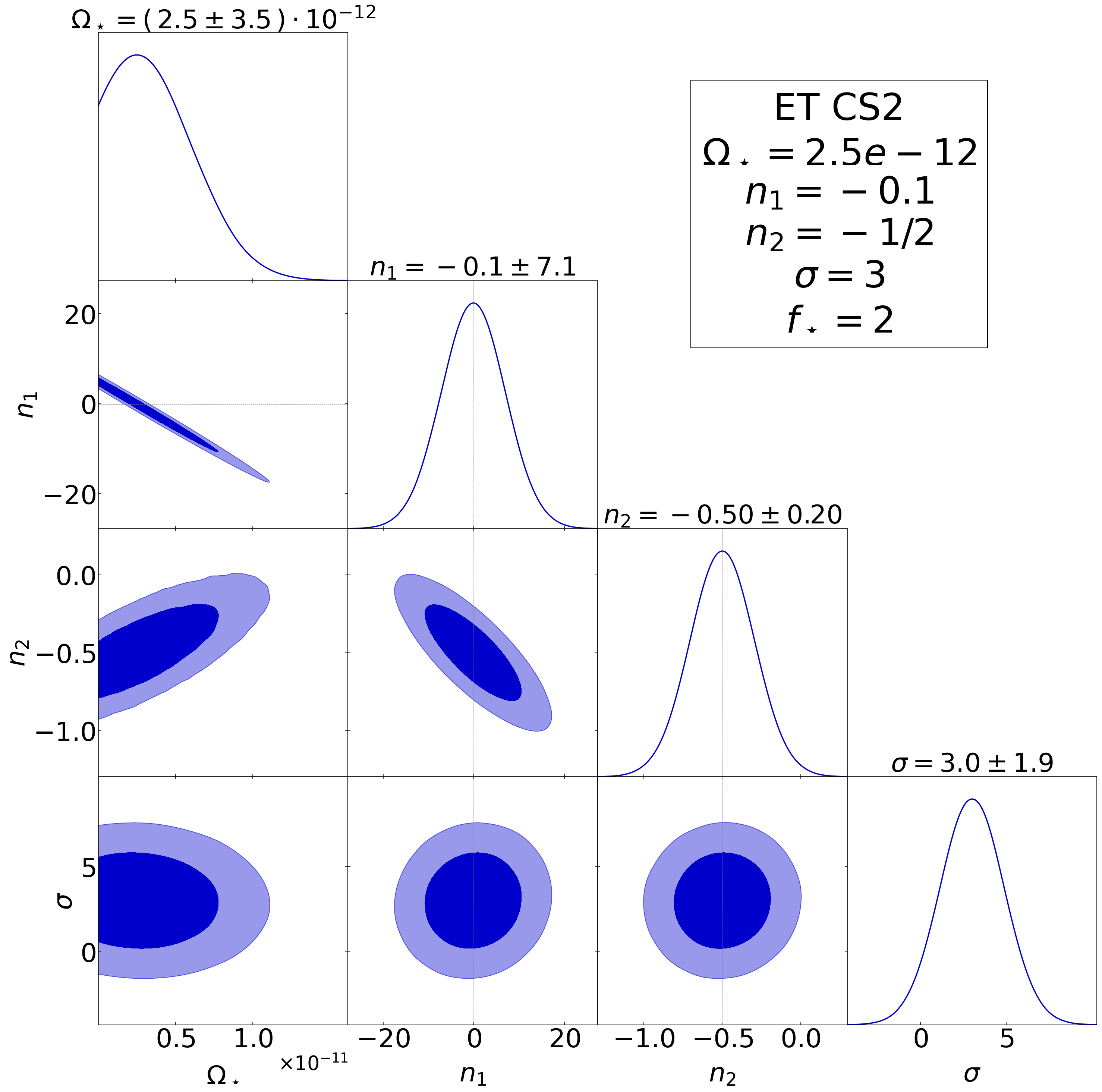}\\
 \includegraphics[width=0.39\textwidth]{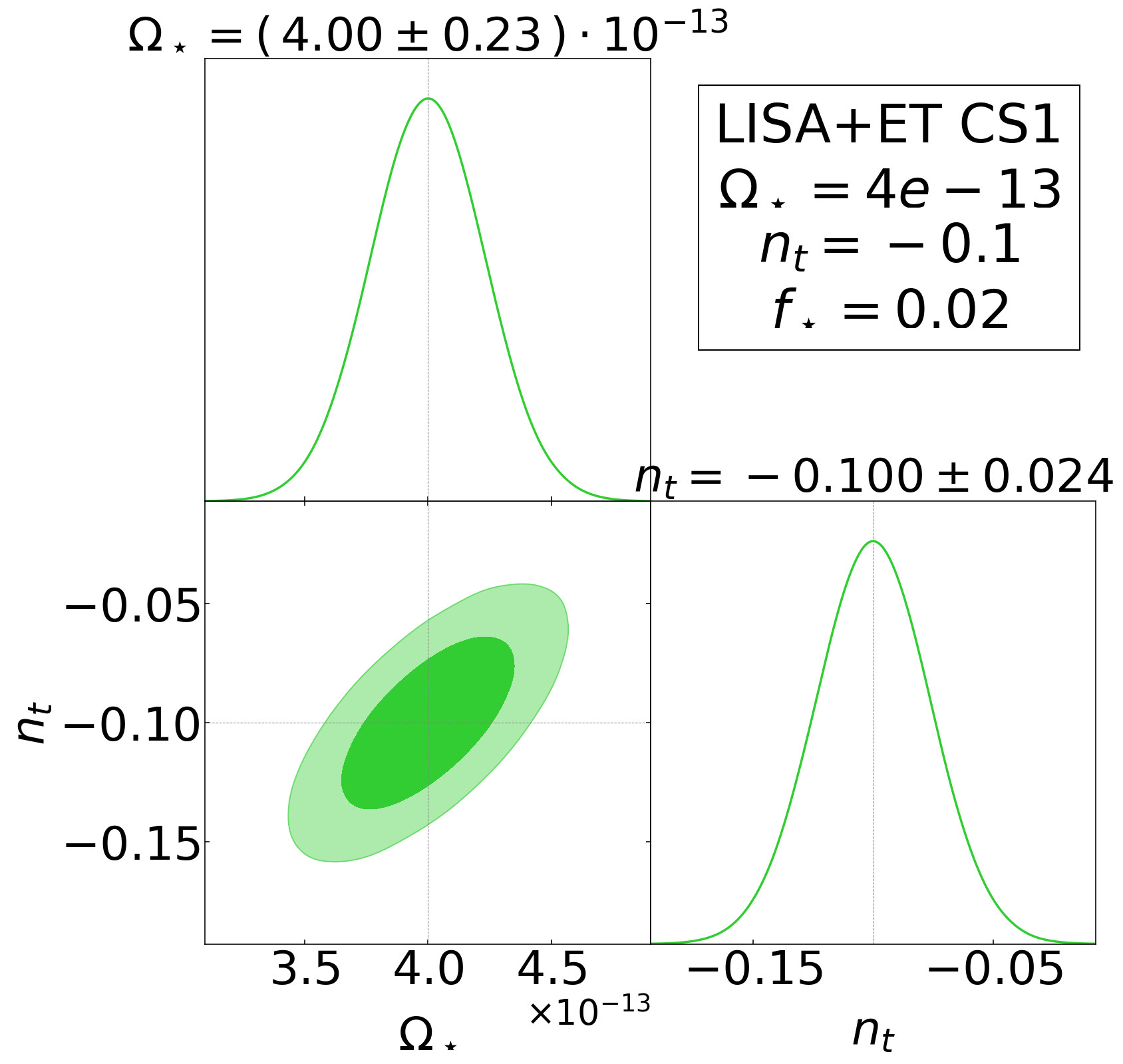}
  \includegraphics[width=0.39\textwidth]{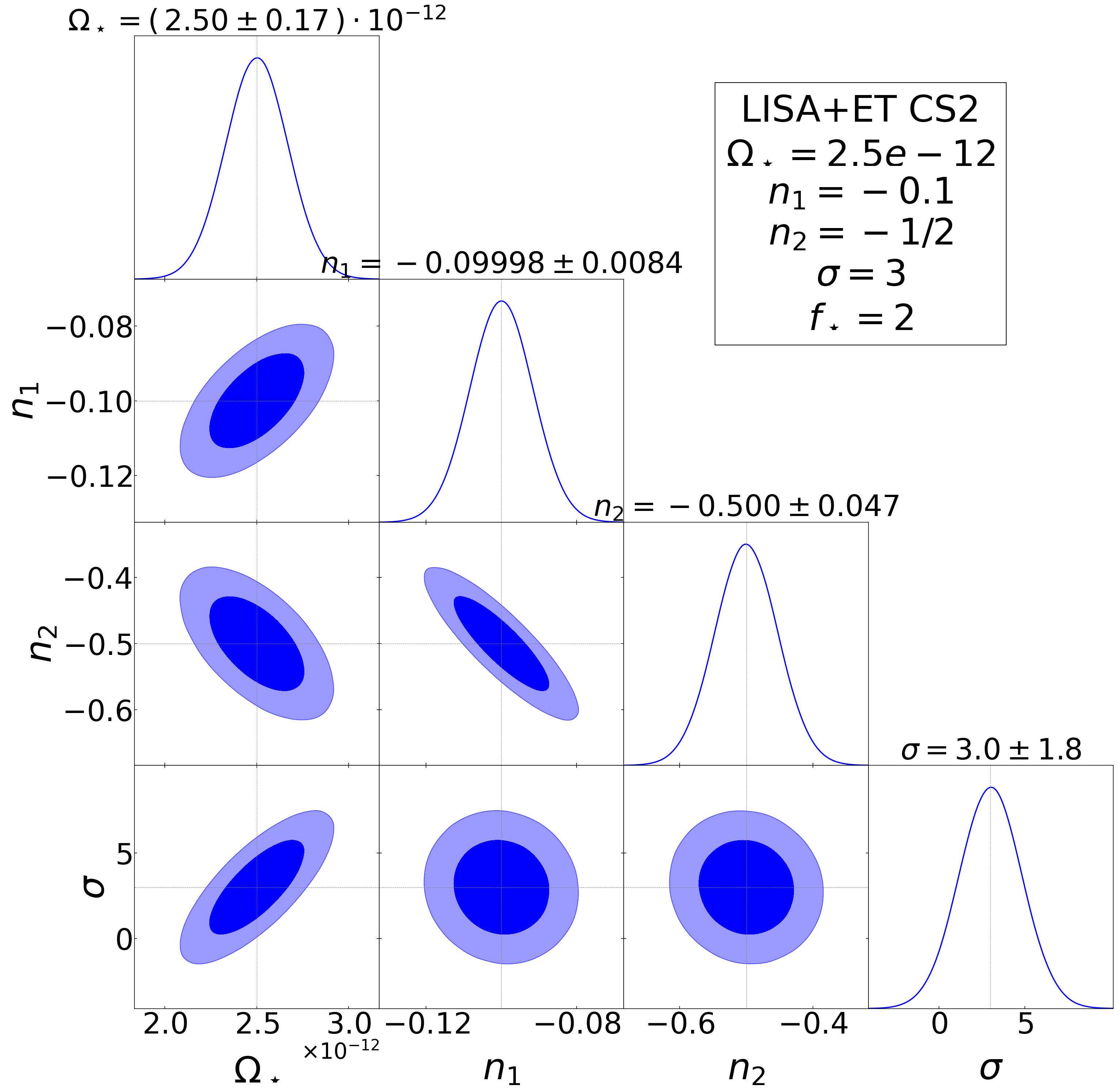}
    \caption{\small  \it Fisher forecasts for the cosmic string benchmark scenarios
    CS1 and CS2, as summarized in Table \ref{tab:valuesCS}. 
    }
\label{fig:Fishcosmicl}
\end{figure}

 \subsubsection*{What can
  we learn about CS from synergies between  LISA and ET?}
 
The two benchmark scenarios CS1 and CS2 -- a single power-law
 and a broken power-law -- are summarized in Table \ref{tab:valuesCS}.
Again, the value of $\Omega_\star$  in the two cases (and $f_\star$
in CS2) are selected
with the aim of demonstrating the advantages offered by
the synergy of the  two instruments. 
 The
corresponding Fisher analysis is collected in the plots in Fig \ref{fig:Fishcosmicl}.
Similar to the case of phase 
transitions,  this plot demonstrates the advantages of 
synergetic measurements with the two
experiments 
for accurately measuring
the parameters of each benchmark scenario. Only the
two instruments together 
can measure with a 10 percent accuracy
the entire set of parameters. Moreover, a detection
in synergy can reduce apparent degeneracies
characterizing the detection with single instruments for the case CS1 (see 
Fig \ref{fig:Fishcosmicl}, first row).

%%%%%%
 %%%%%%
 \subsection{Cosmological inflation}
   \label{sec_ci}
%%%%%%%%%%%%%%%%%%%%%%%%%%%%
\begin{figure}[t!]
    \centering
    \includegraphics[width=0.31\linewidth]{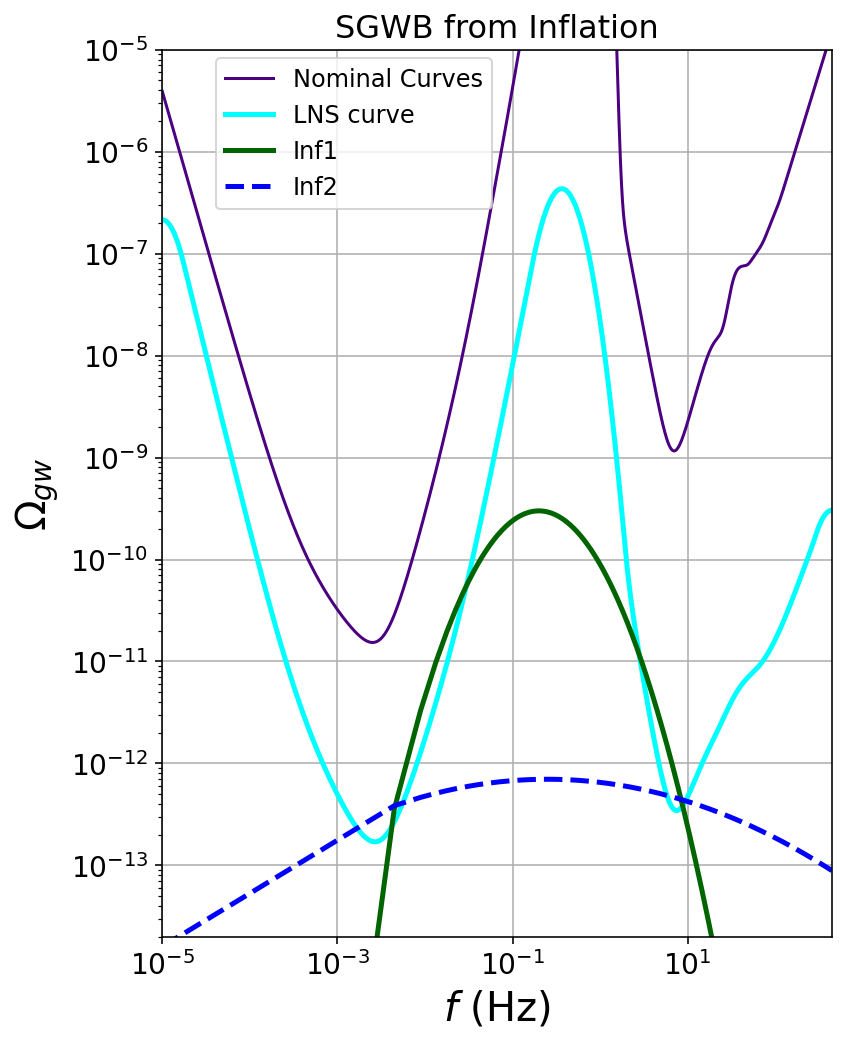}
    \caption{\it 
\small  Examples of SGWB from cosmic inflation, see section \ref{sec_ci}. The purple curves correspond to the LISA and ET
 nominal curves, the turquoise curve corresponds to the integrated sensitivity curve for Log Normal scenarios discussed in Section \ref{sec_sencur}, and the blue and green lines
 represent the benchmark scenarios in Table \ref{tab:valuesinf} }
    \label{fig:inflation1}
\end{figure}

Cosmological inflation is a well studied early universe phenomenon capable of producing
a stochastic background of gravitational waves (see e.g. \cite{Weinberg:2008zzc} for a textbook account).  
While the simplest models of inflation predict a SGWB amplitude too small to be 
detected by LISA, there are several well-motivated scenarios capable to 
raise the amplitude of the spectrum to an observable level
within the band \eqref{int_range}. These scenarios are based on multiple field dynamics
involving vector and axion fields \cite{Barnaby:2010vf,Sorbo:2011rz}, spontaneous breaking of space-time
symmetries \cite{Endlich:2012pz,Cannone:2014uqa,Bartolo:2015qvr}, or secondary effects associated with primordial
black hole production (PBH) \cite{Ananda:2006af,Baumann:2007zm} (for reviews see e.g. \cite{Domenech:2021ztg,Ozsoy:2023ryl}).  Various inflationary sources 
 can provide distinct frequency profiles for $\Omega_{\rm GW}$, which
can be distinguished when detected by GW experiments. In general,
the frequency profile of a SGWB
produced by inflation is much richer in features than SGWB produced
by other phenomena, and it
cannot be described by the broken power law Ansatz of equation \eqref{BPLt}.  It
may include a log-normal profile, multiple peaks,
or shapes characterised by oscillatory
features (see e.g. \cite{Braglia:2024kpo}  for examples, and  
a classification of SGWB templates suitable for describing GW from
different inflationary scenarios). Moreover,
quite interestingly, the SGWB characteristics
depend on the very early cosmological  history
 preceding BBN, which are
imprinted in the SGWB frequency spectrum. Hence,
 before focusing on developing forecasts to detect a
 specific template of inflationary SGWB by means of
 synergies of LISA and ET, we theoretically 
 further motivate how  frequency profiles of SGWB originating  from
 inflation -- more general than equation \eqref{BPLt} -- may provide information on the early universe evolution prior to BBN.

   \medskip
   \noindent
{\textbf{Inflationary first-order tensor perturbations:} The primordial spectral index $n_T$, defined
in terms of log derivative of the power spectrum along the momentum scale, is a crucial quantity for  characterizing the inflationary primordial tensor power spectrum.   Standard single-field slow-roll inflation models predict a red-tilted spectrum, with $n_T$ satisfying the slow roll consistency relation $n_T \approx -r/8$ \cite{Liddle:1993fq}. However, there {exist} several more complex scenarios characterised by  blue-tilted spectra ($n_T > 0$), originating from various  high energy of cosmological models \cite{Brandenberger:2006xi, Baldi:2005gk, Kobayashi:2010cm, Calcagni:2004as, Calcagni:2013lya, Cook:2011hg, Mukohyama:2014gba, Kuroyanagi:2020sfw}. Since the primordial  GW background, after being produced during inflation, exists all throughout cosmic history, the spectrum is a perfect target for the multi-band frequency study  we carry on  in our analysis.

Just like the  GW {spectrum} from cosmic strings, GW from inflation are also ideal targets for probing the period of pre-BBN history, i.e. the 
 universe barotropic parameter $w$ in the post inflationary era. Let us discuss some examples where the the background equation of state deviates from the standard prediction for radiation domination ($1/3$). We can consider 
 models of quintessential inflationary theories  \cite{Peebles:1998qn, Dimopoulos:2001ix, Akrami:2017cir, Bettoni:2021qfs}, or  non-oscillatory inflation models \cite{Ellis:2020krl}. In these cases,  the scalar field (the inflaton or some spectator field) keeps rolling for a long  time even  after inflation ends. As a consequence of this process, the primordial Universe experiences a phase  known as {\em kination} \cite{Joyce:1997fc,Gouttenoire:2021jhk}, during which the scalar kinetic energy fraction becomes the dominant component of energy budget in the Universe. This phase  is  not too long lasting, as the corresponding  energy
 decreases  fast   as $\rho_\phi\propto a^{-6}$ before the onset of the standard radiation domination phase. The corresponding background equation of state  during kination is given as $w=1$, stiffer than the barotropic parameter during RD ($w=1/3$) or {during} matter domination (MD) ($w=0$).    Inflationary tensor perturbations  re-entering the horizon during this phase  receive a boost in their amplitude with respect
 to modes re-entering the horizon during RD.
  See e.g. \cite{Tashiro:2003qp, Bernal:2020ywq,Ghoshal:2022ruy,Berbig:2023yyy,Barman:2023ktz,Cai:2020ovp, Cai:2023ykr}.
  More in general, we can consider   a  stiff era when the barotropic parameter of the Universe {lie} in the range \mbox{$1/3<w<1$}. Such 
 phenomena were investigated in   \cite{Figueroa:2019paj},
finding that in order to get a detectable signal in LISA -- the stiff period in the post-inflationary epochs must be in the range \mbox{$0.46\lesssim w\lesssim 0.56$} with a high inflationary scale $H_{\rm inf} \sim 10^{13}$ GeV and the reheating temperature in the range \mbox{1 MeV$\,\lesssim T_{\rm reh}\lesssim\,$150 MeV} assuming no blue-tilting ($n_T \sim 0$). A realization of this possibility in UV-complete inflationary models was actively studied in details in Ref.~\cite{Dimopoulos:2022mce,Chen:2024roo} along with other interesting predictions. Moreover, {subsequent} cosmological eras can imprint signatures on the resulting GW spectrum \cite{DEramo:2019tit, Bernal:2020ywq,Gouttenoire:2021jhk,Gouttenoire:2021wzu,Dunsky:2021tih,Berbig:2023yyy}. 
 Other well-studied examples of such scenarios include:  a long-lived heavy scalar field generating an early matter era~\cite{McDonald:1989jd,Moroi:1999zb, Visinelli:2009kt,Erickcek:2015jza,  Nelson:2018via,Cirelli:2018iax,Gouttenoire:2019rtn,Allahverdi:2020bys}; a very fast rolling scalar field generating a kination era \cite{Spokoiny:1993kt,Joyce:1996cp,Peebles:1998qn,Poulin:2018dzj,Gouttenoire:2021jhk,Gouttenoire:2021wzu,Co:2021lkc,Ghoshal:2022ruy,Heurtier:2022rhf};  a supercooled phase transition \cite{Guth:1980zk,Witten:1980ez,Creminelli:2001th,Randall:2006py,Konstandin:2011dr,Baratella:2018pxi,Ghoshal:2020vud,Baldes:2020kam,Baldes:2021aph,Dasgupta:2022isg,Ferrer:2023uwz}; an extended particle physics sector undergoes decays and scatterings, or  an broad distribution of Primordial Black Holes (PBHs) evaporating  in the early universe, as studied in detail in Refs.\cite{Barrow:1991dn,Dienes:2022zgd, Dienes:2021woi,Bhaumik:2022pil,Bhaumik:2022zdd}.}

\medskip

\noindent
{\textbf{Scalar-induced GW:} Yet another important and well studied source of cosmological GWs or the so-called scalar induced SGWB (second-order tensor perturbation)\cite{Matarrese:1992rp,Matarrese:1993zf,Matarrese:1997ay,Carbone:2004iv,Ananda:2006af,Baumann:2007zm} particular with boosted interests in very recent times \cite{Alabidi:2012ex,Alabidi:2013wtp,Hwang:2017oxa,Espinosa:2018eve,Kohri:2018awv,Cai:2018dig,Bartolo:2018rku,Inomata:2018epa,Yuan:2019udt,Inomata:2019zqy,Inomata:2019ivs,Chen:2019xse,Yuan:2019wwo,DeLuca:2019ufz,Tomikawa:2019tvi,Gong:2019mui,Inomata:2019yww,Yuan:2019fwv,Domenech:2017ems,Domenech:2019quo,Ota:2020vfn,Cai:2019jah,Cai:2019elf,Cai:2019amo,Bhattacharya:2019bvk} due to the connection with dark matter in the form
of primordial black holes.

{Just as the cosmic strings or the first-order inflationary GW, the induced SGWB is also a very well-recognised tool  to test the thermal history of the universe but  leading to different spectral shapes controlled   by different aspects of microphysics.
 Ref.~\cite{Domenech:2019quo} extended the investigations of the induced SGWB for radiation and early matter dominated universes to more arbitrary barotropic parameters $w>0$ and predicts a multi-band GW frequency spectrum including motivations for Primordial Blackhole domination and its evaporation \cite{Bhaumik:2022pil,Bhaumik:2022zdd}.  In Ref. \cite{Domenech:2019quo} it is shown that for an adiabatic perfect fluid the shape of the peak of the spectrum depends on the value of $w$. In similar lines, Ref.~\cite{Hajkarim:2019nbx} shows the impact in the GW spectrum due to the change in the effective degrees of freedom in thermal history, like those occurring during the QCD and electroweak phase transitions in early universe. In a more general set up, Ref.~\cite{Cai:2019cdl} shows that the infrared side of the GW spectrum has a universal slope given a certain $w$.

{The broad frequency profile of  scalar induced GW is related with  the epochs at which the high density scalar fluctuations re-enter the horizon (and may collapse to form PBH). In this manner scalar-induced GW probe the thermal history of the universe,  see Ref. \cite{Domenech:2020kqm}. We remark the difference between this probe of cosmic history with that of inflationary first-order and cosmic strings lies in the shape of the resultant GW spectrum as well as the microphysics involved. 
 }

\medskip

\noindent
{\textbf{Particle production during and after inflation:} 
 Axion or more general pseudoscalar inflation models with particle production\cite{Niu:2022quw,Niu:2022fki,Adshead:2015pva,Adshead:2016iae, Adshead:2018doq, Adshead:2019lbr, Adshead:2019igv, Freese:1990rb, Silverstein:2008sg, McAllister:2008hb, Kim:2004rp, Berg:2009tg, Dimopoulos:2005ac, Pajer:2013fsa} are characterised by a pseudoscalar inflaton $\chi$ which respects an approximate shift symmetry \cite{Freese:1990rb} and a Chern-Simons coupling of the form $\chi F \tilde{F}$ to a $U(1)$   gauge field. $F$ denotes the field strength of the gauge field and $\tilde{F}$ is its dual. Such Chern-Simons  couplings lead to a tachyonic production of a transverse mode of the gauge fields generating a  boosted primordial GW spectrum  \cite{Anber:2006xt,Anber:2009ua, Cook:2011hg, Barnaby:2010vf, Barnaby:2011qe, Barnaby:2011vw, Meerburg:2012id, Anber:2012du, Linde:2012bt, Cheng:2015oqa, Garcia-Bellido:2016dkw, Domcke:2016bkh, Domcke:2016mbx, Peloso:2016gqs, Domcke:2018eki, Cuissa:2018oiw, Niu:2023bsr}. 
Going beyond  axion inflation set-up, even in models in which an axion or an axion-like particle is not the inflaton a SGWB can be produced~\cite{Machado:2018nqk, Machado:2019xuc,Co:2021rhi,Fonseca:2019ypl,Chatrchyan:2020pzh,Ratzinger:2020oct,Eroncel:2022vjg}.
After the axion starts  rolling it induces a tachyonic instability for one of the dark photon helicities, causing vacuum fluctuations to grow exponentially. This effect generates a time-dependent anisotropic stress in the energy-momentum tensor, which ultimately sources the tensor perturbations. The GW formation ends when the tachyonic band closes at temperature
$
T_* \approx \frac{1.2 \sqrt{m_a M_{\rm P}}}{g_*^{1/4} (\alpha \theta)^{2/3}} 
$
where $\alpha$ is the coupling with the dark photon, $\theta$ is the initial misalignment angle, and $m_a$ is the mass of the axion ~\cite{Machado:2018nqk}. For well motivated and suitable values of axion mass and  coupling we can  get a broad multi-band frequency profile
 detectable by LISA and ET, see eq  \eqref{ans_lnp}.

\bigskip

\noindent
{\textbf{Our Ansatz:} 
After discussing theory motivations aimed at
underlying  how the detection of GW from inflation can help in characterizing the thermal
history of the universe, we now focus on forecasting the detectability of  a specific
SGWB template, with a 
log-normal profile
\begin{equation}
    \Omega_{\rm GW} = \Omega_{\star}  \exp{\left[{-\frac{\ln^2\left({f}/{f_\star}\right)}{2\rho^2}}\right]}.
    \label{ans_lnp}
\end{equation}
The previous equation is
characterised by three free parameters: $\Omega_{\star}$, $f_\star$, $\rho$. These parameters
control the amplitude, position, and sharpness of the peak, respectively. The log-normal Ansatz \eqref{ans_lnp} is  qualitatively different from the BPL  profile of eq \eqref{BPLt}. 
The SGWB profile of eq \eqref{ans_lnp} can be generated by the axion or axion spectator
models of inflation  \cite{Namba:2015gja,Dimastrogiovanni:2016fuu,Thorne:2017jft} described above. Hence, it is theoretically well motivated. We nevertheless  emphasize that other cosmic inflation models lead to different SGWB frequency profiles, hence
they require separate dedicated analyses. See
e.g. the recent work \cite{Braglia:2024kpo} for a classification of possible inflationary SGWB shapes.
Explicit particle physics models leading to eq \eqref{ans_lnp}
 are based on the dynamics of the aforementioned models
 based spectator axion field
 $\chi$, which rolls during a fraction $\Delta N$ of e-folds
 of inflation.  In this epoch,
 the axion excites vector
gauge fields, through the coupling $\chi F \tilde F$.
The dynamics and energy density of
the latter produces a
sizeable SGWB with the
lognormal profile \eqref{ans_lnp}. The SGWB peak
$f_\star$ occurs  at scales
corresponding to modes leaving the horizon during the epoch 
of fastest roll of $\chi$. The height
of the peak is sensitive to the quantity
$\dot \chi$, while its width depends
on $\Delta N$. Hence, each of the
parameters characterizing the ansatz
\eqref{ans_lnp} has a clear physical
interpretation in terms of quantities characterizing 
well-motivated underlying scenarios.

\begin{figure}[H]
\centering
        \includegraphics[width=0.39\textwidth]{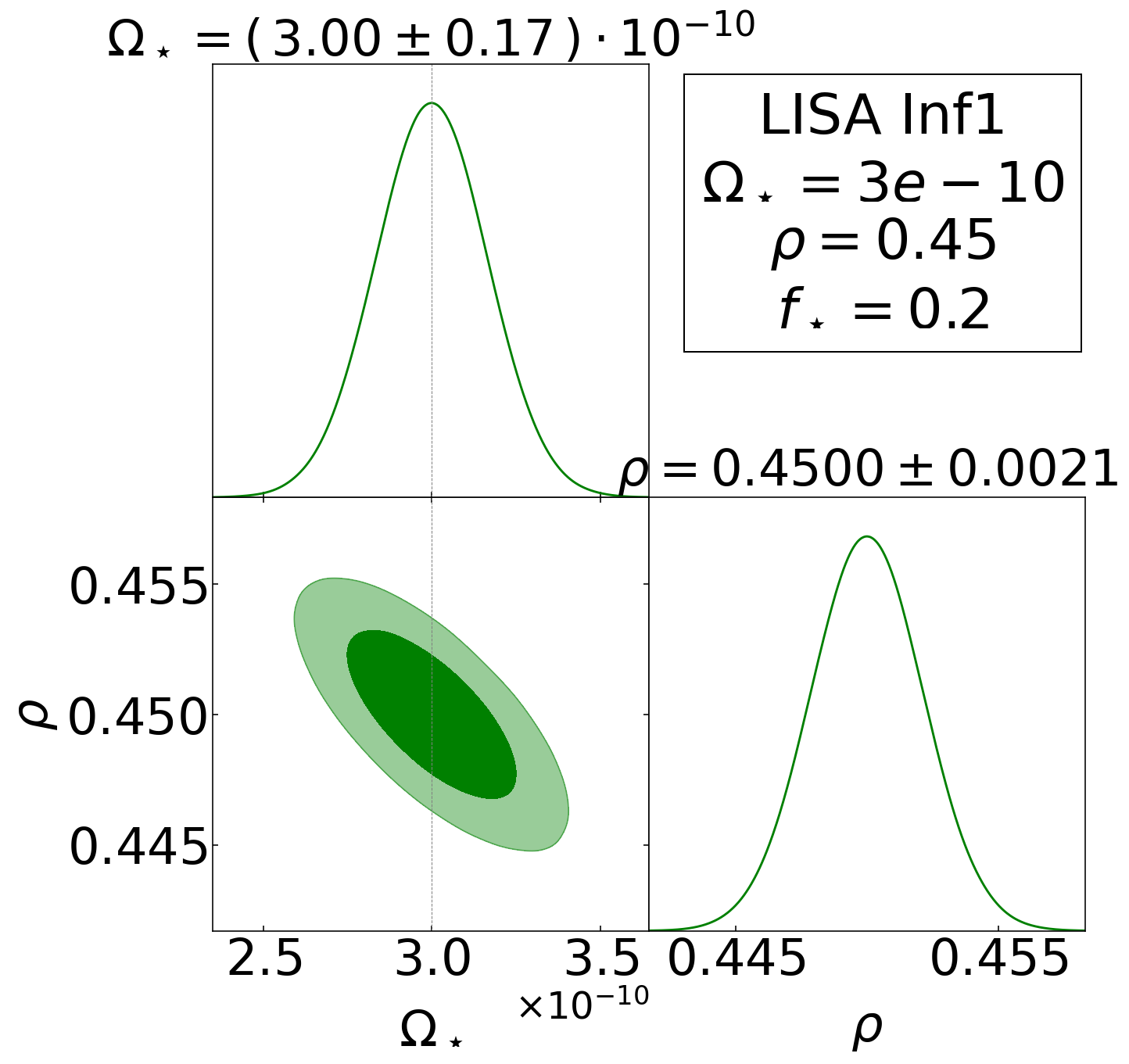}
 \includegraphics[width=0.39\textwidth]{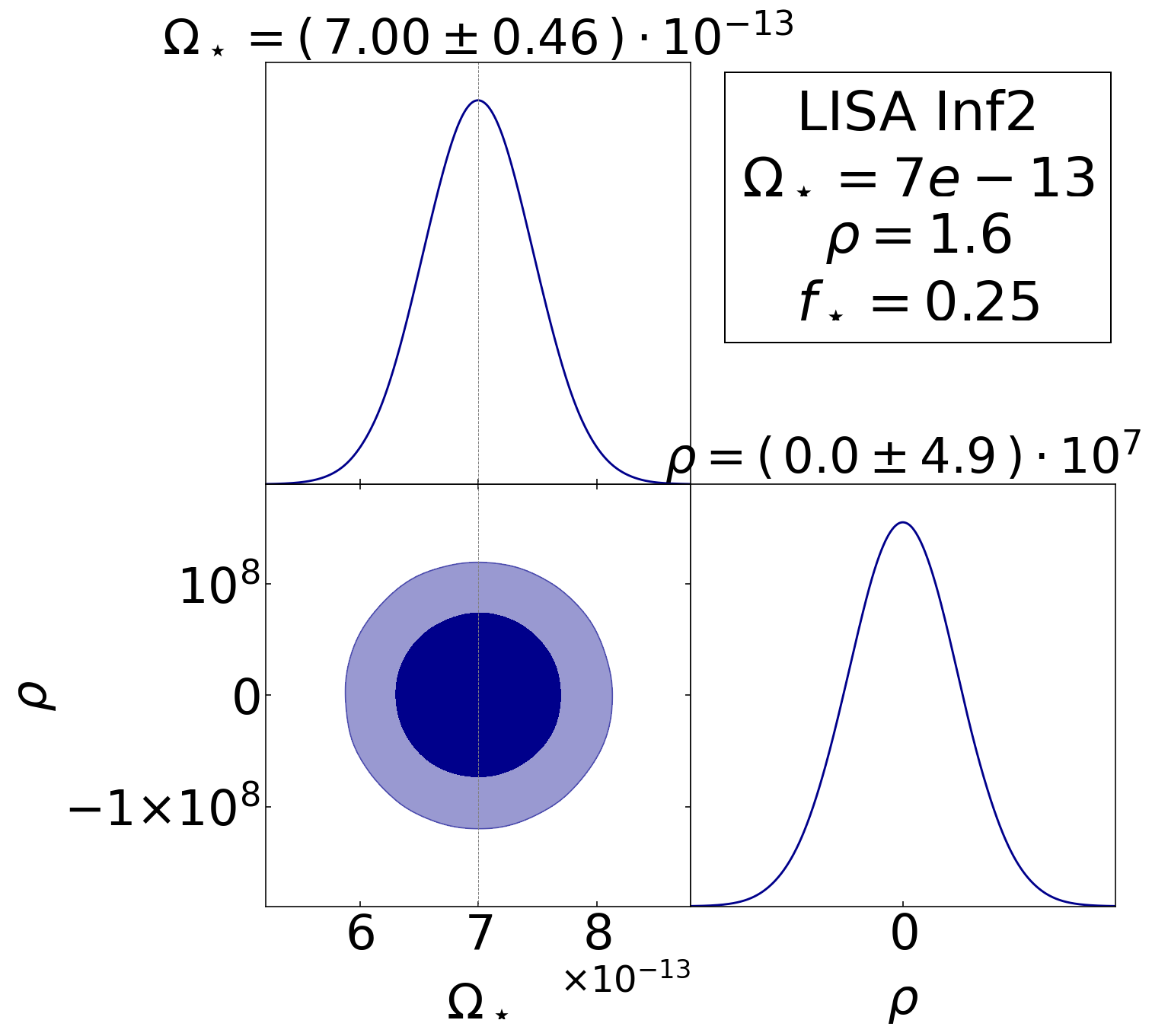}\\
  \includegraphics[width=0.39\textwidth]{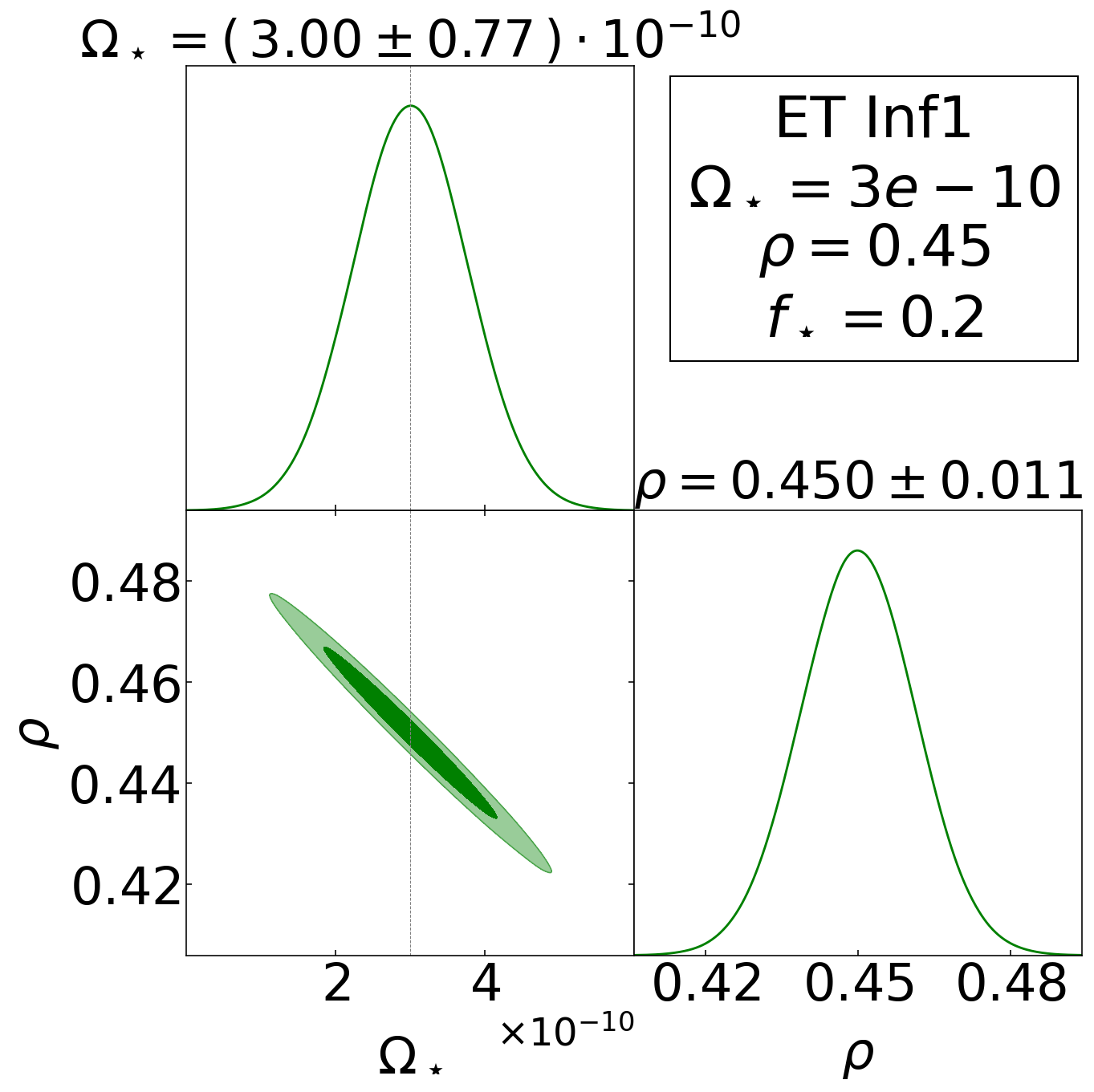}
    \includegraphics[width=0.39\textwidth]{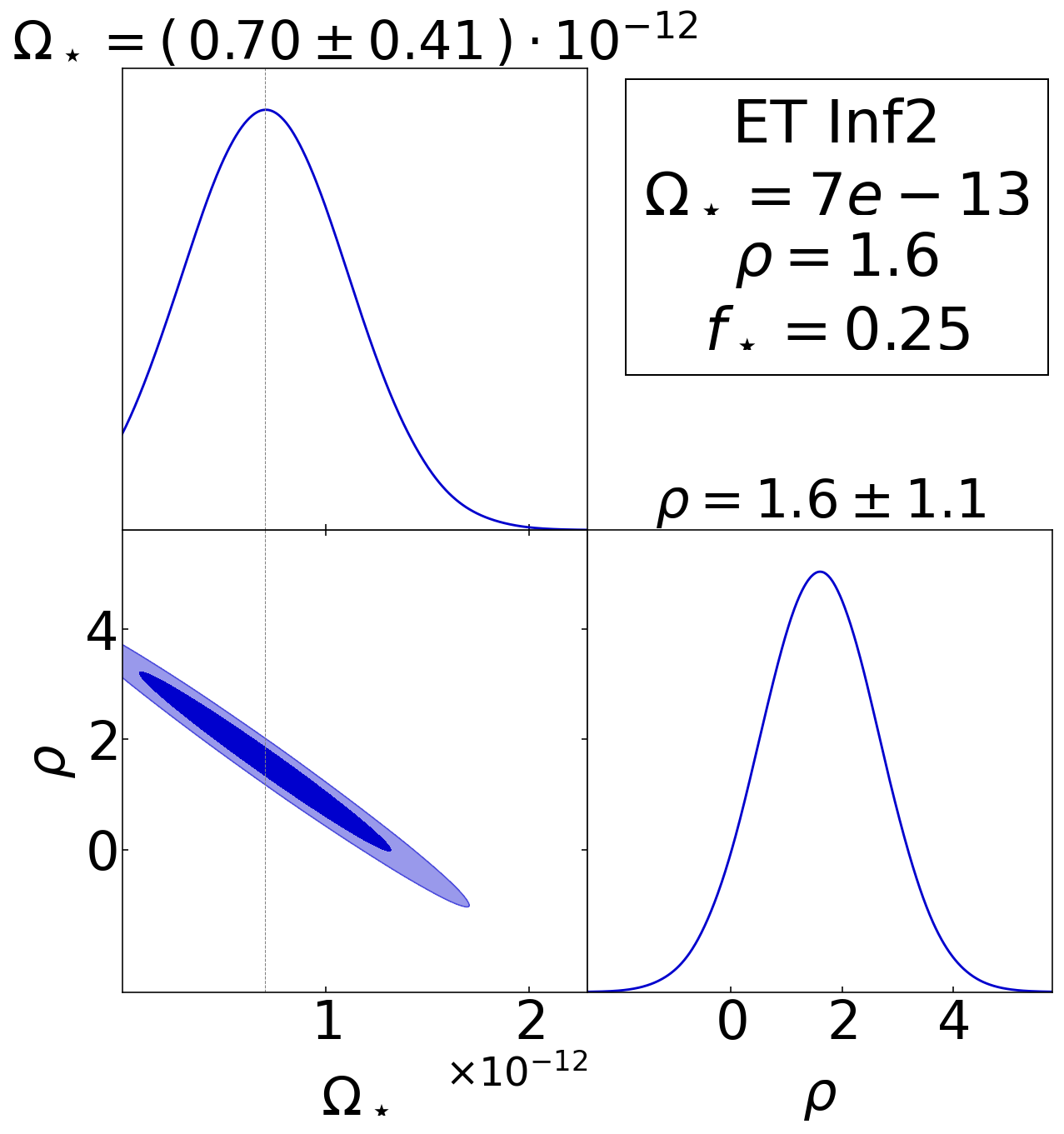}\\
 \includegraphics[width=0.39\textwidth]{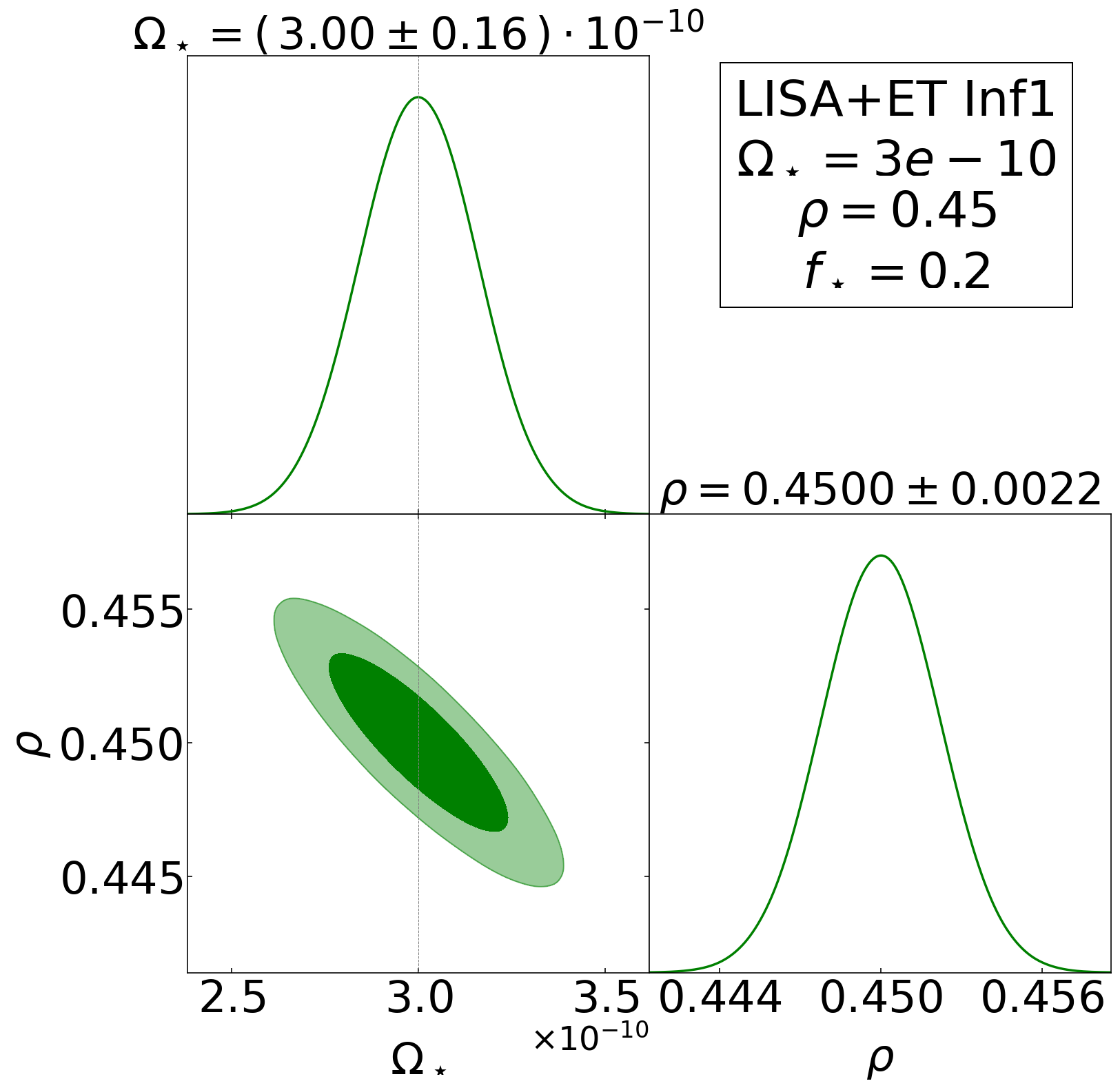}
  \includegraphics[width=0.39\textwidth]{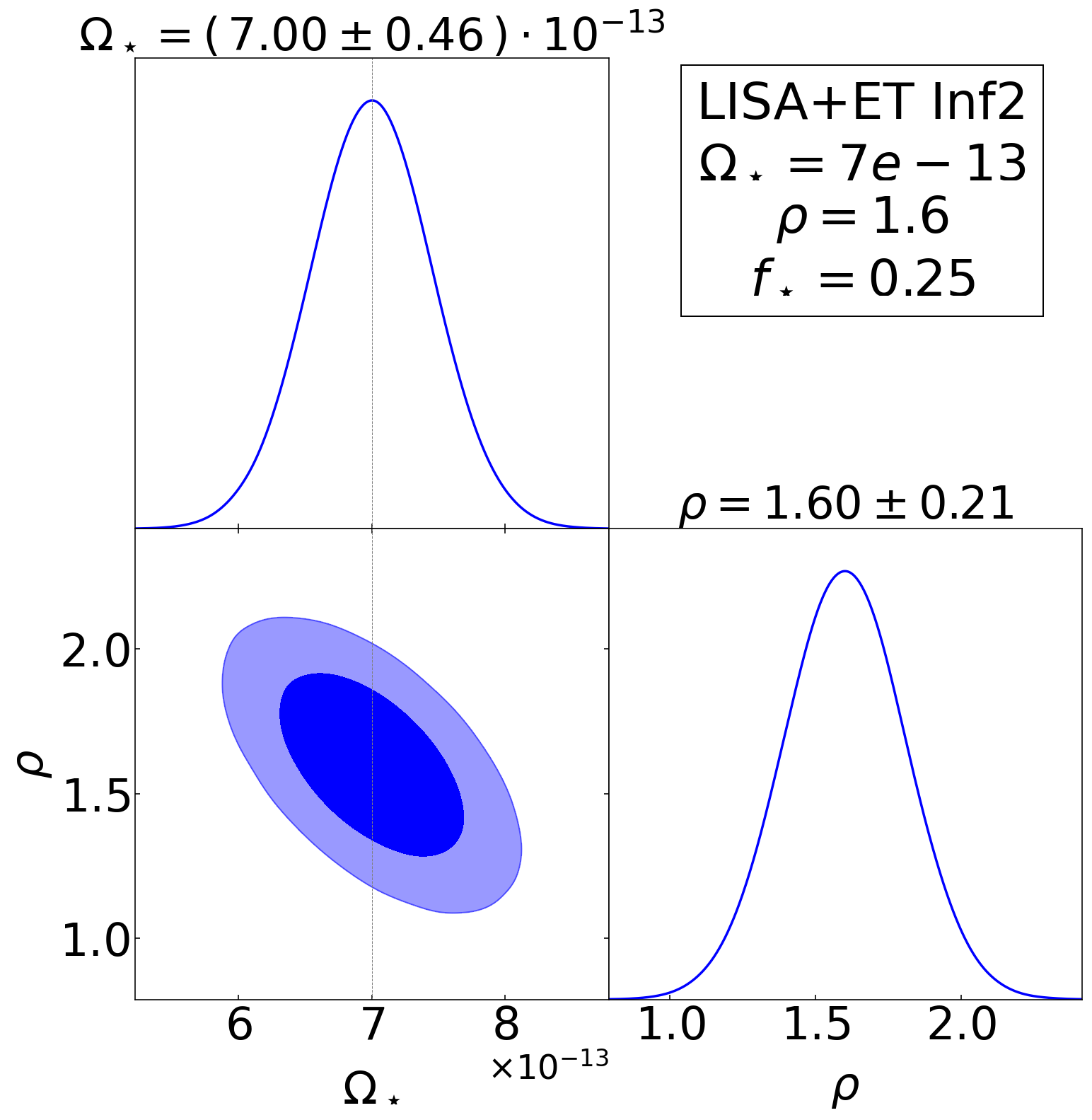}
    \caption{\it \small Fisher forecasts for the cosmic inflation benchmark scenarios
    Inf1 and Inf2, as summarized in Table \ref{tab:valuesinf}.  }
\label{fig:Fishinflation}
\end{figure}

\begin{table}[h!]
\begin{center}
\begin{tabular}{| c | c | c | c | c | c | }
%\hline
\hline
 {\rm }& \cellcolor[gray]{0.9} $\Omega_\star$  &\cellcolor[gray]{0.9}$f_\star$&\cellcolor[gray]{0.9}$\rho$\\
\hline
\cellcolor[gray]{0.9} {\rm Inf1} & $3\times10^{-10}$ & $0.2$ & $0.45$
\\
\hline
\cellcolor[gray]{0.9} {\rm Inf2} & $7\times10^{-13}$ & $0.25$ & $1.6$
\\
\hline
\end{tabular}
\caption{\it \small Benchmark  values  for  cosmological inflation scenarios described
by Ansatz \eqref{ans_lnp}. \label{tab:valuesinf} }
\end{center}	
\end{table}

In Table \ref{tab:valuesinf} we collect two representative scenarios (Inf1 and 2)
and the associated
benchmark values for the parameters corresponding to the log-normal 
  ansatz \eqref{ans_lnp}. We present their 
profiles in the right panel of Fig \ref{fig:inflation1}, where we also
depict in blue the corresponding log-normal sensitivity curve (to be discussed in the next section). The parameters
are selected such that their profiles peak in the middle
between the LISA and ET bands, with different amplitudes and different
peak sharpness. In Fig \ref{fig:Fishinflation} we plot the corresponding Fisher
forecasts.  LISA alone would be able to measure the parameters with good accuracy in scenario Inf1. On the other hand, scenario Inf2 would benefit much from the synergy between the two experiments. 

In conclusion, the synergy between
LISA and ET can help in distinguishing
and characterizing early universe
sources of SGWB. In the next
two sections, we  develop
and expand upon this topic, considering
further concepts and observables
to exploit the potential of making
detections with the two experiments
together.

%%%%%%%%%%%%%%%%
%%%%%%%%%%%%%%%%
%%%%%%%%%%%%%%%%%%%%%%%%%%%
\section{The notion of integrated sensitivity curves}
\label{sec_sencur}
%%%%%%
%%%%%%
Is there a simple, intuitive way to know
whether a given SGWB profile can be detected
by GW experiments? The answer is affirmative, thanks
to the notion of a sensitivity curve. 
In this section we discuss various versions 
of sensitivity curves for LISA, ET, and the two
experiments {operating} together. 

\smallskip

The concept of nominal sensitivity curve offers a visual tool to intuitively understand whether a certain GW source {with its frequency profile} can be detected  by a GW experiment. 
 If  a given GW signal has
 a sufficiently large amplitude to cross the {sensitivity} curve, it 
    will automatically be detected by the particular experiment with a signal-to-noise ratio of greater than  {unity}.
 The frequency profile of the nominal sensitivity curve depends on the noise sources affecting a given experiment, and on the response
of the latter to a GW input.  We already discussed and represented the nominal sensitivity curves
in section \ref{sec_corr_exp} (see Fig \ref{fig:ETLISAnoms}).

By inspecting
 Fig \ref{fig:ETLISAnoms}, we can see 
that the {frequency} regions of maximal sensitivity for LISA and ET are
different (we call them ${\cal B}_{\rm LISA}$ and ${\cal B}_{\rm ET}$), and span the ranges
\bea
\label{ind_bands}
{\cal B}_{\rm LISA}&\simeq&10^{-5}\, \le\,f/{\rm Hz}\le 10^{-1}, 
\\
{\cal B}_{\rm ET}&\simeq&10^{0}\, \le\,f/{\rm Hz}\le 445.
\label{ind_bandsa}
\eea
At   face value,  the  detectors do not   cover  well (due to poor sensitivity) the intermediate region in between 
say $8\times 10^{-2}\le f/{\rm Hz}\le 2$. Also, the minimal nominal sensitivity of LISA to $\Omega_{\rm GW}$ is around one order
of magnitude larger than ET.
We have cut off the upper limit of the frequency range for ET at $445$ Hz, because beyond this frequency the sensitivity of the instrument decreases beyond the previously mentioned BBN limit of $\Omega_{\rm GW}\le 1.7\times 10^{-6}$, and therefore those frequencies are not of interest to us in this work.

However,
in representing the nominal sensitivity curves as discussed above, we do not make use of the crucial fact that the
SGWB signal is {extended over decades of frequency ranges}. {If such a broad frequency profile exists we may integrate over the entire frequency range to allow us obtain
more crucial information on the signal by  having more large SNR.} A broad frequency profile suggests  that we can integrate over the frequency range, allowing us to obtain
further information on  the signal by   collecting more
SNR (see formula \eqref{snr_tog}). We already came across this feature
while discussing Fisher forecasts in section \ref{sec_theory}.
For this reason, \cite{Thrane:2013oya} introduced the  notion of power law sensitivity curve \footnote{{For certain signals like phase transition which carries a peak, peaked integrated sensitivity may fare better than power law one, see Ref. \cite{Schmitz:2020syl} for details. However in this paper we focus to the former case for overall comparison purposes.}} as a useful visual device  to understand  whether a given signal can be detected by a GW experiment.
Following  \cite{Thrane:2013oya}, we start by assuming  that a given signal is described by a power law profile
\begin{equation}
    \Omega_{\rm GW}\,=\,\Omega_\star\,\left( f/f_\star \right)^{n_T},
\end{equation} 
over
the frequency band we are interested in,
with $f_\star$ a given
reference frequency. The spectral tilt $n_T$ is not exactly known though: we assume that it can vary over an interval between two fiducial values.
  For each value of the tilt, we can determine the minimal value of the signal amplitude $\Omega_\star$ the ensures that the corresponding SNR overcomes a certain threshold. We shall consider
  \be
  {\rm SNR}\,=\,5 \hskip0.5cm, \hskip0.5cm { T}\,=\,3\,{\rm years}\,\,.
  \ee
  Then, we  determine the envelope of the resulting curves associated with the various spectral tilts, and draw for each frequency the maximal signal amplitude after scanning over all the tilts. 

\begin{figure}[t!]
	\centering
    \includegraphics[width=0.3\linewidth]{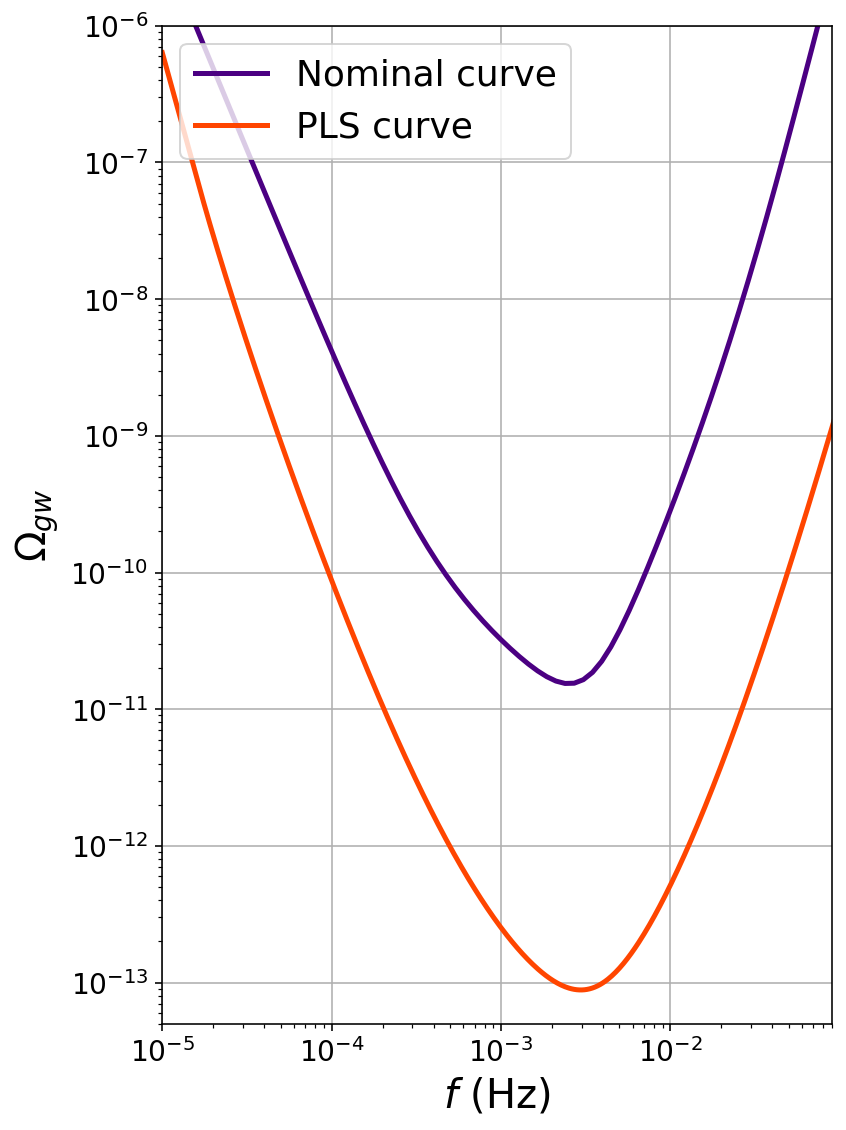}
	\includegraphics[width=0.3\linewidth]{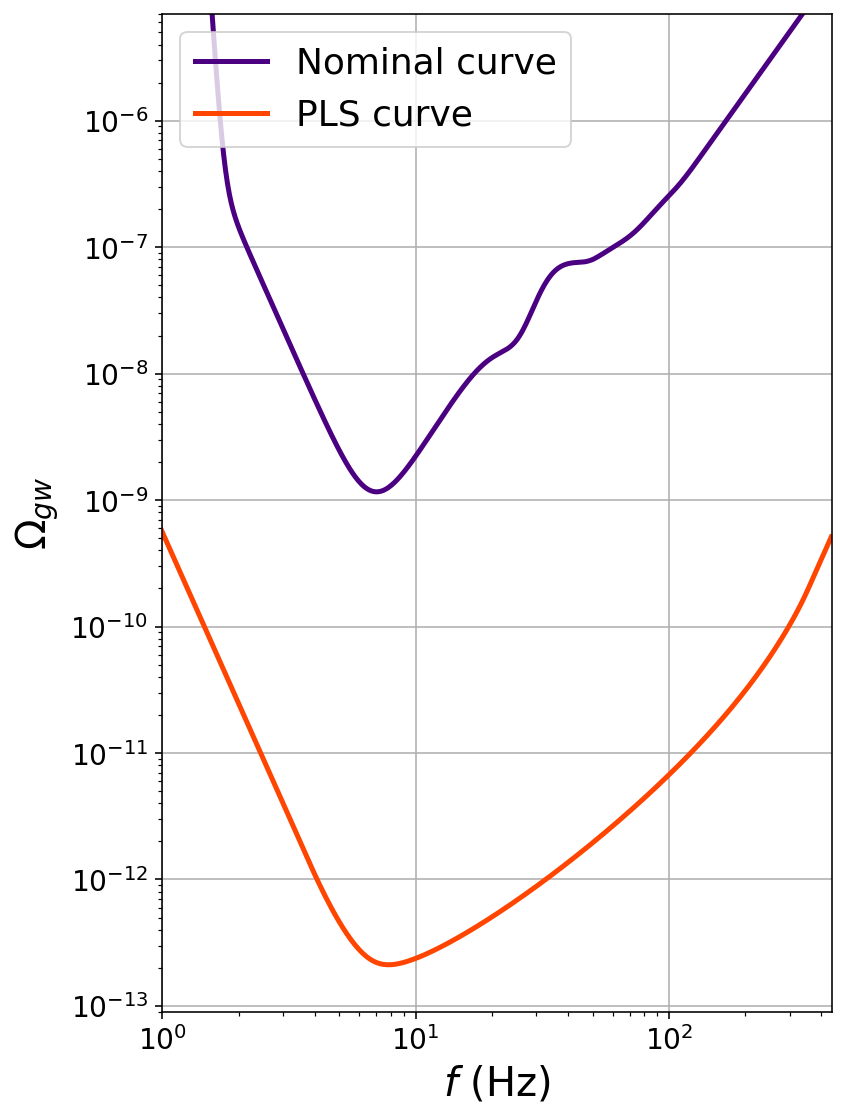}
    \includegraphics[width=0.3\linewidth]{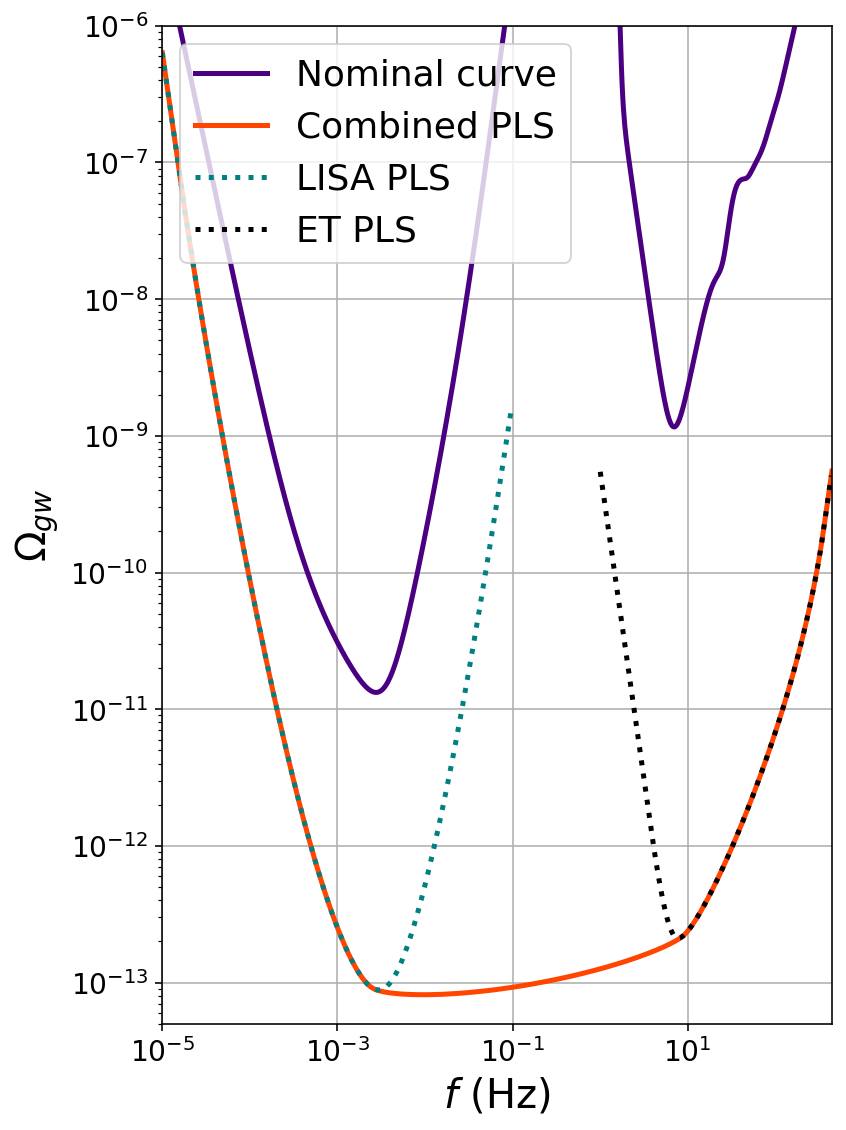}
	 \caption{\small  \it Orange lines: power-law sensitivity (PLS) curves for LISA ({\bf left panel}), ET ({\bf central panel})
	 and the two experiments combined ({\bf right panel}). For the nominal sensitivity curves in purple we
	 use the analytical fits of Appendix \ref{app_fit}.
	   }
	\label{fig:etlispl}
\end{figure}

The result is the so-called power law sensitivity curve (PLS) which we represent in Fig \ref{fig:etlispl}.
We allow $n_T$ to vary between $-9/2<n_T<9/2$, with these numbers
chosen for better visualization purposes. 
 The left and central panels show the standard PLS curves associated with LISA and ET, built following the aforementioned algorithm. Every power law spectrum passing above these curves can be detected by LISA (left panel) or ET (central panel) with ${\rm SNR}>5$. In drawing these curves, we have assumed that the signal is a power law over the entire range of frequencies \eqref{int_range}, but we measure it with
one experiment only. 
 Notice that, differently from the nominal curves, the integration over frequencies associated with the   notion of PLS curves  leads
to  a sensitivity for 
ET  comparable with the  one of LISA.

In plotting the combined   PLS in the right panel of Fig \ref{fig:etlispl}, we
take advantage of the fact that both LISA and ET can measure the same signal independently, hence they can provide a multiband detection
of a given SGWB. 
Again, we assume that the SGWB signal is a power law for the {\it entire} frequency range \eqref{int_range}. 
 We form the total SNR$_{\rm tot}$ using eq \eqref{snr_tog}, and  use this quantity to draw the combined PLS in the right panel of Fig  \ref{fig:etlispl}.
%%%%

\smallskip
The plots in Fig \ref{fig:etlispl} demonstrate that the PLS
curves gain  orders of magnitude in sensitivity with respect to the nominal curves. If a power law SGWB signal passes above the PLS (but below the nominal sensitivity curve) it can nevertheless be detected by the experiment.  Notice that when LISA and ET operate together, the PLS has a low amplitude also within the frequency interval between the sensitivity bands of the individual instruments, where
the system would seem to have low sensitivity (see comment after eq \eqref{ind_bandsa}). In fact, if a power law profile crosses  in the middle of each of two experiment bands, {e.g. $\Omega_{\rm GW} \sim  10^{-10}$ at a frequency $f \sim 10^{-2}$} (above the PLS of the plot of Fig \ref{fig:etlispl}, right panel), it  certainly crosses the sensitivity curve of one or the other experiment, as it grows towards larger or smaller frequencies through the band \eqref{int_range}. This feature, due to our
hypothesis that the signal is a power-law
in the entire band \eqref{int_range},
explains the much improved PLS sensitivity in the intermediate regions in the right panel of Fig \ref{fig:etlispl}, when compared to the PLS sensitivity of the individual experiments.

%%%%%%%%%%%%%%%
%\newpage
\subsection*{Broken power-law and log-normal sensitivity curves}
\label{sub_BPLsen}

\begin{figure}[t!]
	\centering
			\includegraphics[width=0.3\linewidth]{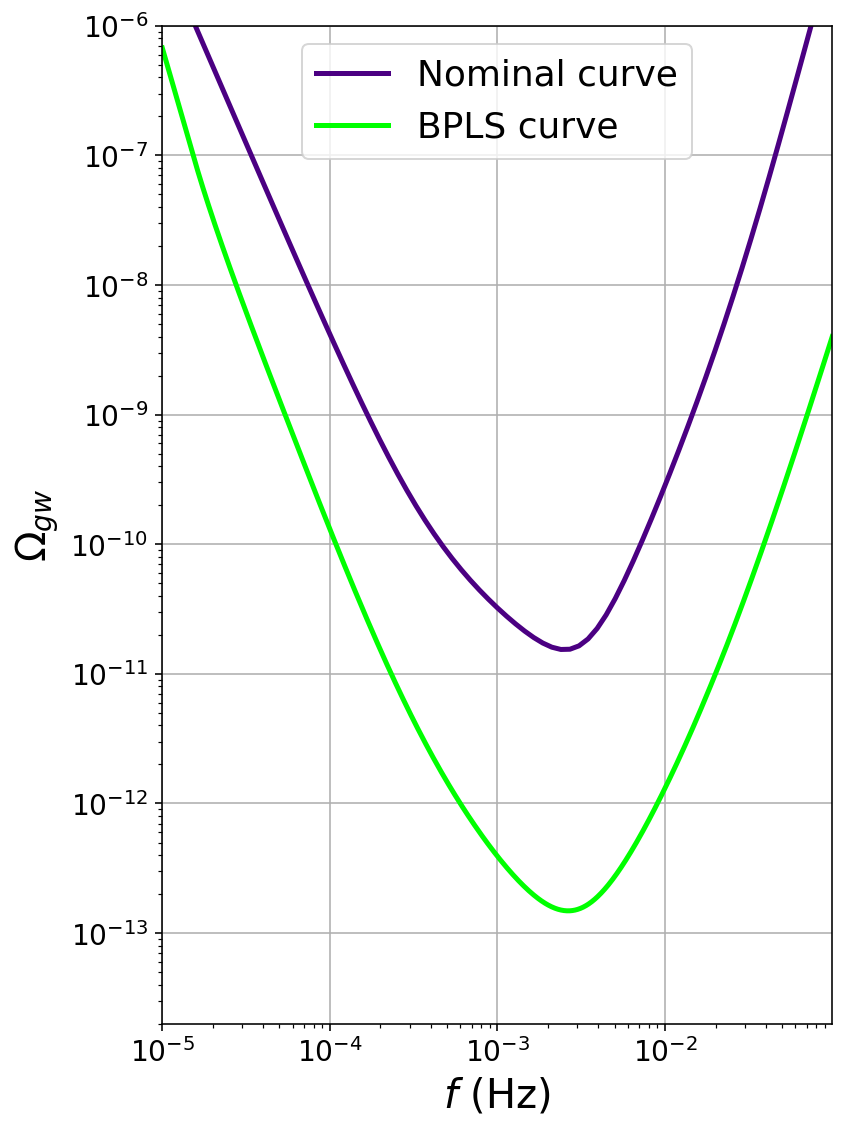}
		\includegraphics[width=0.3\linewidth]{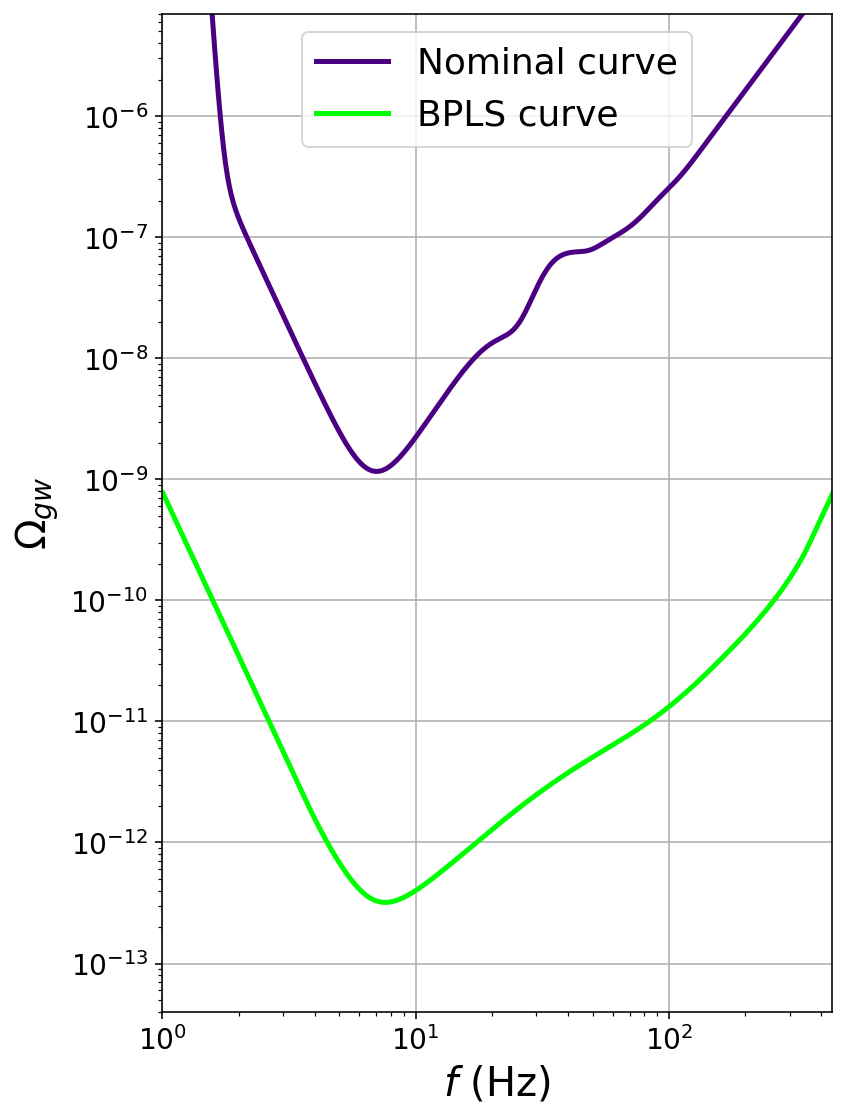}
	\includegraphics[width=0.3\linewidth]{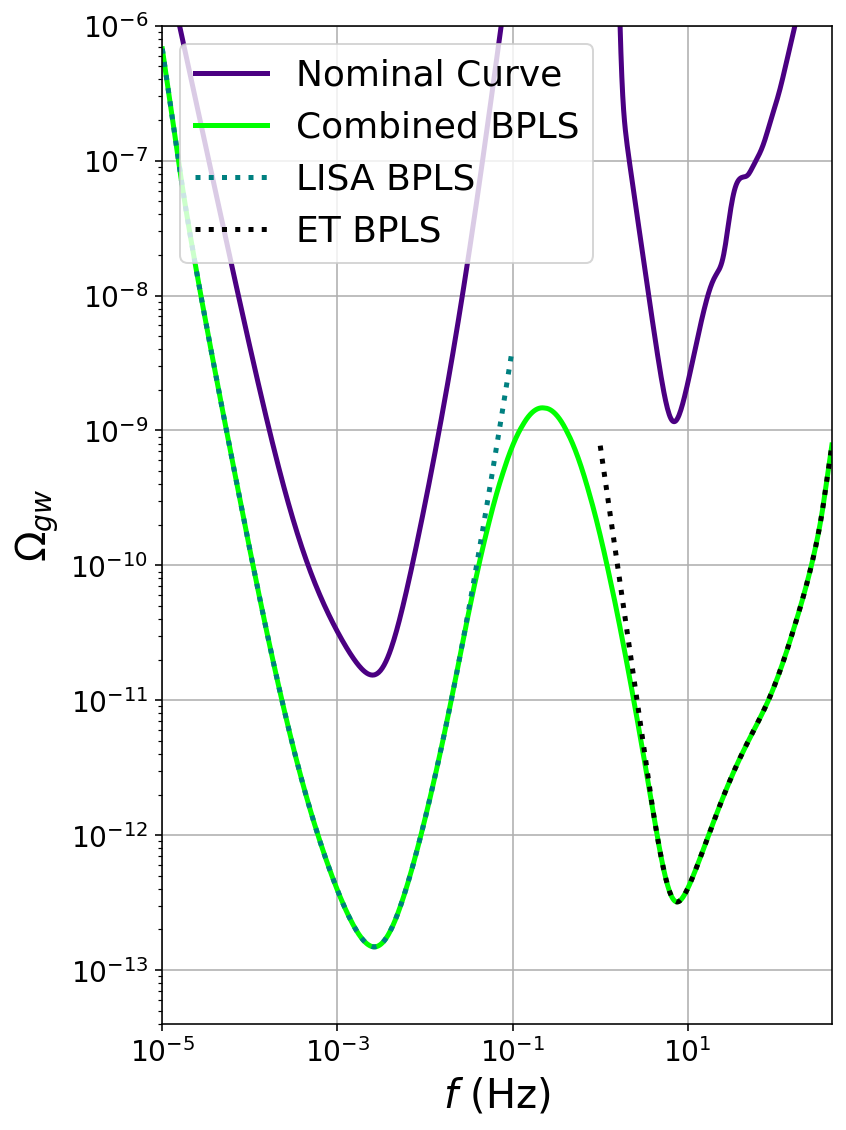}
	\caption{\small  \it{Representation of the broken power-law (BPLS) sensitivity curves,
	as discussed in the main text. 
  }}
	\label{fig:sigmabpls}
\end{figure}
After reviewing the concept of power law sensitivity curves, we shall now
discuss the other families of integrated sensitivity curves. 
As discussed in section \ref{sec_theory}, we are 
interested in SGWB profiles described by a broken
power-law, with the frequency dependence given by the function \eqref{BPLt},
or log-normal profiles associated
with ansatz \eqref{ans_lnp}. For this reason, we
go beyond the concept of the PLS  curve \cite{Thrane:2013oya} analysed above, and we discuss the
concept of BPLS sensitivity curve as introduced in 
\cite{Chowdhury:2022gdc}. (See also
\cite{Schmitz:2020syl} for previous related arguments.) We can ask what is the sensitivity of GW experiments towards detecting a SGWB {with a particular SNR detection threshold}, assuming  a broken
power-law profile within the range \eqref{int_range}.

In this case, there are several parameters we can vary: the spectral tilts $n_{1,2}$ of the growing and decreasing part of the spectrum, the position $f_\star$ of the break, and the parameter $\sigma$ controlling the smoothness of the transition.  
  We independently vary over the spectral tilts $n_{1,2}$
in the interval $-9/2 \le n_{1,2} \le 9/2$, over $1<\sigma<10.2$, as well as over the values
of $f_\star$ in the  ranges of sensitivity of the system.
 We  determine the minimal amplitude  in eq \eqref{BPLt} to ensure we reach an SNR=$5$ for each set of values of the parameters we examine, and we draw the envelope of the corresponding curves.
In the left and central panel of Fig \ref{fig:sigmabpls} we focus on the individual experiments LISA and ET,
varying $f_\star$
respectively within the ${\cal B}_{\rm LISA}$ and ${\cal B}_{\rm ET}$ bands of eq \eqref{ind_bands}.
 On the right panel we consider the two experiments
together, and vary the break
position $f_\star$ over the entire range
\eqref{int_range}.

While for the LISA-only and ET-only cases the BPLS curves result in similar shapes and amplitude as the PLS curves 
in the previous subsection (compare the left and central panels of Fig \ref{fig:etlispl} and Fig \ref{fig:sigmabpls}), the BPLS 
curve for the two experiments
together is very different
compared to the PLS curve
for the two
experiments (compare the
right panels of  Fig \ref{fig:etlispl} and Fig \ref{fig:sigmabpls}). The reason for this difference is 
that  the break of the BPL might occur in between the LISA and ET bands, where the sensitivity is reduced, and BPL spectra with large spectral tilts (in absolute value) might enter only partially within the sensitivity region of an experiment. Nevertheless, the plot suggests that  we can accurately detect BPL SGWB  profiles with a break in the middle of LISA and ET bands, and
with a relatively small amplitude at the break position. This is a property that
we have already explored in the previous section with a Fisher analysis of selected
benchmark models for  phase transition and cosmic string scenarios.

\begin{figure}[t!]
	\centering
    \includegraphics[width=0.3\linewidth]{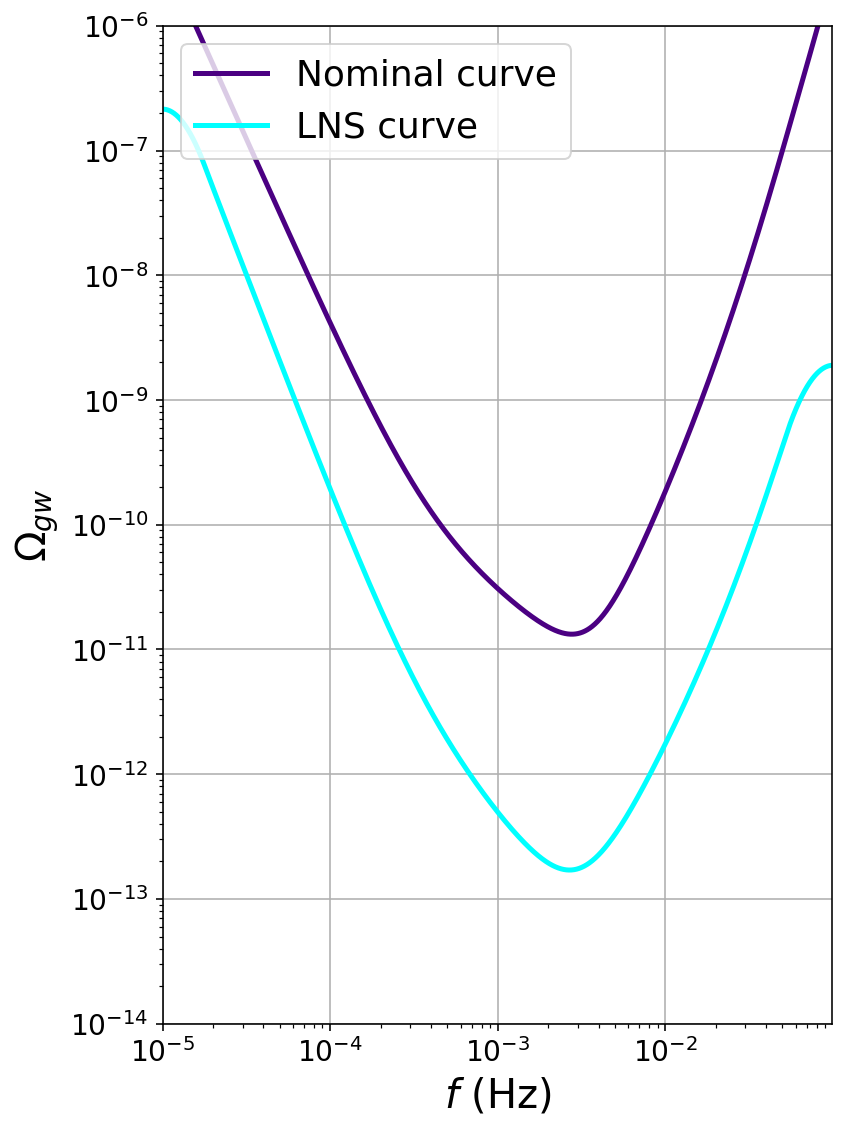}
	       \includegraphics[width=0.3\linewidth]{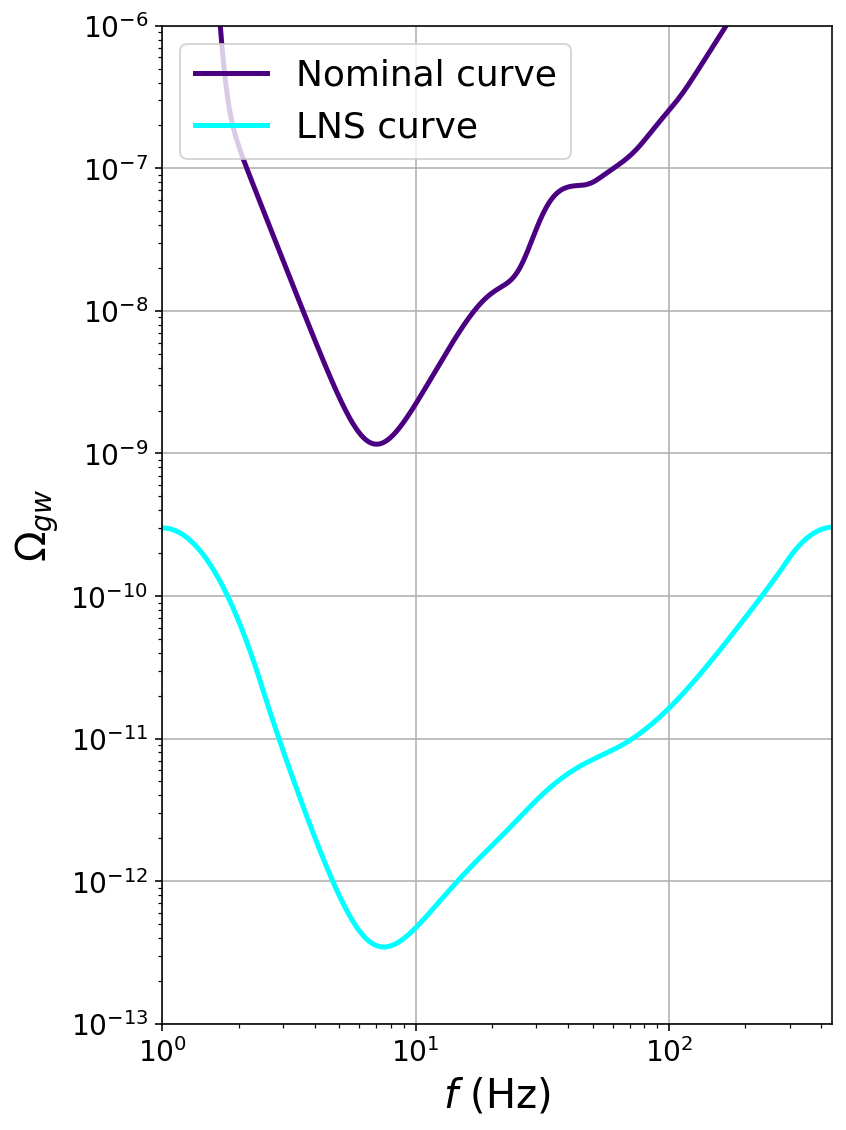}
	       	\includegraphics[width=0.3\linewidth]{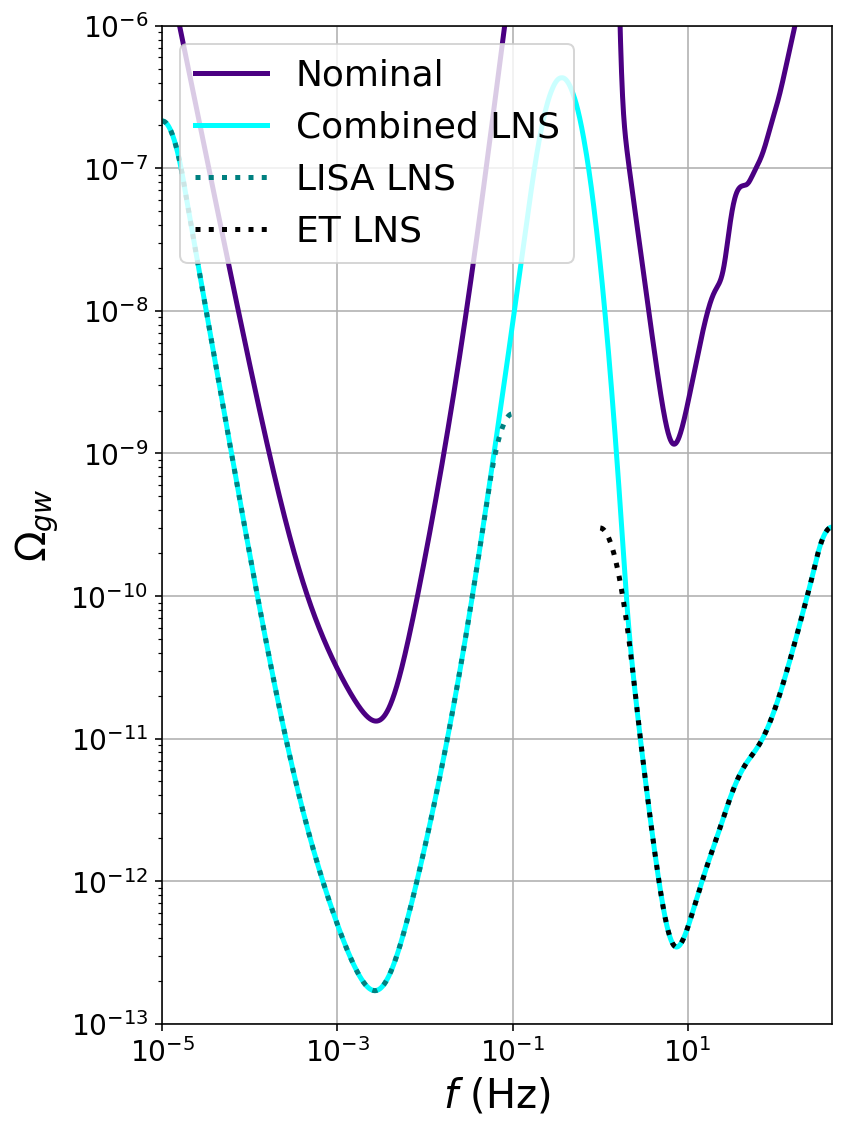}
	\caption{\small  \it{Log-normal
	integrated sensitivity curves, built following the method discussed in the main text.
	 }}
	\label{fig:etlisalogns}
\end{figure}

The method outlined above can also be applied
to other SGWB profiles, such as the log-normal
described by ansatz \eqref{ans_lnp}. In fact, we can build integrated
sensitivity curves varying over the parameters
characterizing eq \eqref{ans_lnp}. We do so
in Fig \ref{fig:etlisalogns}, varying over $0.4<\rho<1$, while $f_\star$ varies
over the sensitivity bands of the experiments as described in the BPLS case. We notice that
in this case the sensitivity of the combined LISA+ET system is not
as good as the BPLS curves of Fig \ref{fig:sigmabpls} in the intermediate band
between LISA and ET maximal sensitivities.
This could be because a lognormal spectrum with a sharp peak in the intermediate band may not enter the sensitivity regions of either of the two experiments at all. Nevertheless, it is important to include both profiles and carefully study the differences in their detection prospects. If we compare Figure 3 with Figure 7, we can see that a GW spectrum detected with a peak in the range~$10^{-9}\lesssim \Omega_{\rm GW} \lesssim 5 \times10^{-7}$ and $0.1~{\rm Hz}\lesssim f\lesssim0.9~{\rm Hz}$ is much more likely to be a broken power-law spectrum than a lognormal one. Once we know the shape of the spectrum, we can extract the parameters in the functional forms of the profiles [cf. eqs. (3.2) and (3.8)] and use them to understand important information regarding the underlying physics, as explained in the sections delineating the two profiles. Since the different parameters correspond to different physical processes, an accurate theoretical modelling of the signal template is important for making forecasts.

\bigskip

As we have learned, 
the concept of integrated sensitivity curves offers an immediate
tool to understand (or guess) the results of more sophisticated
analyses based on Fisher forecasts, regarding
the combined sensitivity of LISA and ET to SGWB signals. It allows
us to visually understand, in a semi-quantitative manner, 
 {\it why} the two experiments operating together are
 more powerful for
  detecting and characterizing a  SGWB. Besides the $\Omega_{\rm GW}$
profile,  we shall now proceed to
discussing another observable which can benefit from joint 
GW detection: the non-Gaussianity of the  SGWB.

\section{The case of a non-Gaussian
signal }
\label{sec_nonG}

In this  section, we wish to  discuss {at a preliminary level} yet another
observable which can benefit from synergies between
LISA and ET: the non-Gaussian features
of the SGWB. Non-Gaussianities arise
whenever non-linearities
are important in the formation
and characterization of the SGWB. It
is a well-studied theme in the context
of the cosmic microwave background. Yet, 
its physics needs to be developed further
for SGWB sources.
  
   We focus here on non-Gaussianities
in the GW 
statistics
produced during inflation (but see   \cite{Kumar:2021ffi} for related
studies in the context of PT). See e.g.\cite{Bartolo:2018qqn}, section 5
for a review. 
It is well-known that intrinsic non-Gaussianities
of the SGWB are difficult to directly measure with interferometers \cite{Bartolo:2018rku,Margalit:2020sxp}.
In fact, Shapiro time-delay effects  ruin  phase correlations
that are essential for characterizing most of the non-Gaussian
shapes which are possible to directly detect through $n$-point function measurements.
 Exceptions are shapes corresponding to soft
limits of correlation functions, such as squeezed \cite{Dimastrogiovanni:2019bfl,Powell:2019kid} 
or collapsed limits \cite{Tasinato:2022xyq} of three or higher
point functions. In this case, momenta characterizing the Fourier modes
entering correlation functions get aligned, and avoid the previously
mentioned dephasing time-delay effects (see \cite{Powell:2019kid,Tasinato:2022xyq}
for extended discussions of aspects
of the physics
involved)~\footnote{Another possibility, which we will not further explore in this context,
  is  to avoid correlating
 the GW signal directly, but to instead form three (or higher)-point functions
of the SGWB {\it anisotropies} \cite{Bartolo:2019oiq,Bartolo:2019yeu}.
 Interestingly, cross-correlations
 between CMB and SGWB anisotropies
 can also be used
 to test inflationary 
 mixed tensor-scalar non-Gaussianities
 \cite{Adshead:2020bji,Tasinato:2023ioq}.}.

\smallskip

Intuitively, the synergy between the LISA and ET detectors -- which operate at well-separated  frequency scales --
 represents 
an invaluable opportunity to probe
 soft limits of GW higher point functions.  Soft limits contain a large
 wealth of physics information (see e.g. \cite{Assassi:2012zq}), which would be interesting to acquire.
 Here we take a small step in this direction, and investigate the response of the LISA-ET system
 to the collapsed limit of the four-point function (the system has vanishing response
 to the squeezed limit of $3$-point function \cite{Seto:2009ju}).
We assume that the GW four-point correlator is described by the following  ansatz:
\bea
\label{colcor}
&&
\langle
h^{\lambda_1}(f_1, \hat n_1)
\,
h^{\lambda_2}(f_2, \hat n_2)
\,
h^{\lambda_3}(f_3, \hat n_3)
\,
h^{\lambda_4}(f_4, \hat n_4)
\rangle_{f_1\ll f_3}
%\,=\,
\nonumber
\\
&&
=\,\frac{\delta^{\lambda_1 \lambda_2}\,\delta^{\lambda_3 \lambda_4} }{2}
\,\delta(f_1-f_2) \delta(f_3-f_4)\,
\delta^{(3)}(\hat n_1+\hat n_2)\,
\delta^{(3)}(\hat n_3+\hat n_4)\,
\delta^{(3)}(\hat n_1-\hat n_3)\,
\,S(f_1, f_3).
\nonumber
\\
\eea
 The above four-point correlator in Fourier space describes a closed quadrilateral with  momenta 
aligned and two-by-two equal in magnitude, enhanced in a soft counter-collinear limit with a frequency $f_1$ much smaller than $f_3$.
Since $f_{\rm LISA}\ll f_{\rm ET}$, such a soft regime can be probed by our set-up. 
We shall not discuss
theoretical motivations and model building perspectives
leading to the ansatz \eqref{colcor}. This will be covered elsewhere, including  further analysis of its consequences for  GW
experiments. 
Instead, we shall
enquire how the LISA-ET system in synergy responds
to the collapsed correlator described
by eq \eqref{colcor}. 
 We 
wish to measure the four-point amplitude $S(f_1, f_3)$ in synergy between LISA and ET. The response
of the system can be obtained by a generalization of the analysis reviewed
in section \ref{sec_corr_exp}. 

The four-point function corresponding to  the measured signal -- the generalization  of eq \eqref{tostwpofu} to higher point correlations -- is 
\bea
&&
\langle 
\Phi_{a_{1\,,b_1 c_1}}(f_1)
\Phi_{a_{2\,,b_2 c_2}}(f_2)
\Phi_{a_{3\,,b_3 c_3}}(f_3)
\Phi_{a_{4\,,b_4 c_4}}(f_4)
\rangle
\nonumber
\\
&=&\frac{\delta(f_1-f_2) \delta(f_3-f_4)}{2}
\,\left[
R^{(4)}(f_1, f_3)\,S(f_1,f_3)+N
\right],
\eea
with $R^{(4)}(f_i)$ being the four-point response function, and $N$ the noise (to avoid cumbersome expressions, we drop
 indexes  labelling the interferometer channels).
 In writing the
 previous formula, we make use of our assumption
\eqref{colcor}
for the GW response function. 
 Geometrically,
the above quantity correlates measurements made at two arms
of LISA and two arms of ET, in the limit $f_1\ll f_3$.  
 $R^{(4)}(f)$ 
 is a generalization
of the two-point response function of eq \eqref{resccor},  formally given by 
the  integral
\bea
R^{(4)}(f_1, f_3)&=&\int
\frac{d^2 \hat n}{4\pi}\,
 %\frac{d^2 \hat n_2}{4\pi}\,
\left[
F_{{a_1}_{b_1 c_1}}^+ (f_1, \hat n)
 F_{{a_2}_{b_2 c_2}}^+ (-f_1, \hat n)
 +
 F_{{a_1}_{b_1 c_1}}^\times (f_1, \hat n)
 F_{{a_2}_{b_2 c_2}}^\times (-f_1, \hat n)
\right]_{\rm LISA}\,
\nonumber
\\
&&
\times 
\left[
F_{{a_3}_{b_3 c_3}}^+ (f_3, \hat n)
 F_{{a_4}_{b_4 c_4}}^+ (-f_3, \hat n)
 +
 F_{{a_3}_{b_3 c_3}}^\times (f_3, \hat n)
 F_{{a_2}_{b_4 c_4}}^\times (-f_3, \hat n)
\right]_{\rm ET}\,.
\label{genfpfa}
\eea

\noindent
As for the case of two-point response functions,
orthogonal channels can be found and orthogonal response
functions can be numerically evaluated. The complication is that
the response function
we are dealing with is a four-dimensional tensor. It has four indexes
corresponding to each interferometer
 channel we correlate in the soft limit  -- two for LISA and two for ET.  
 The result depends on the relative
 positions
 among all arms. As a very first
 step to address the subject, here we shall fix
 the ET arms along the directions $(ab)$
 and $(ac)$ of eq \eqref{dirLIar}, and we
 do not attempt
 to diagonalise the
 channels in the ET sector. We
 instead diagonalise  the
 two remaining indexes corresponding
 to the LISA channels.

We again call the orthogonal
LISA channels as
$A^\ell$, $E^\ell$, $T^\ell$. The computation
of such a four-point response
function is similar to the diagonalization
discussed in section \ref{sec_corr_exp}, although
the results have  different
amplitude and frequency dependence
with respect to the two-point ones. 
The reason is that 
we have to deal with extra angular integrations when  computing the integrals in eq \eqref{genfpfa}. For definiteness, we fix
the positions of the vertexes of the interferometer as in Footnote \ref{foot_position}. It would be interesting
and important to extend our
analysis to more general arm orientations.

\medskip

\noindent
\begin{figure}[t!]
	%\centering
	\noindent\includegraphics[width=0.5\linewidth]{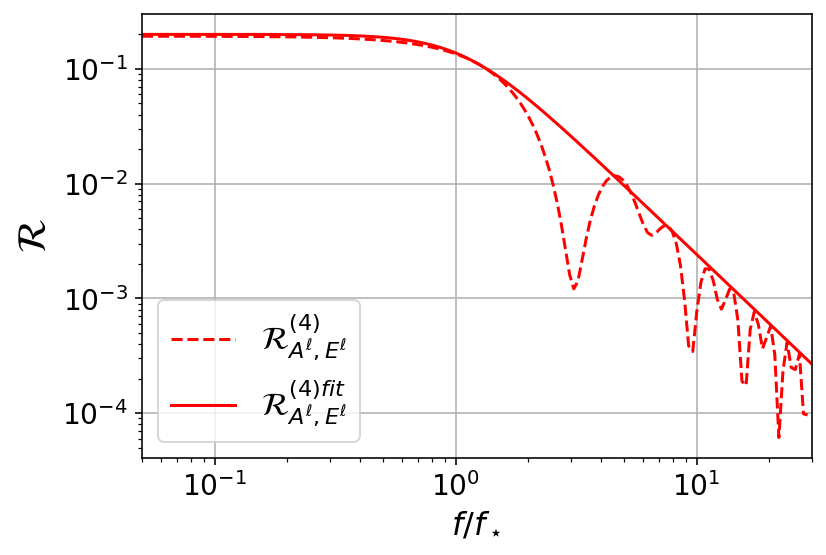}
\hfill
 \includegraphics[width=0.5
\linewidth]{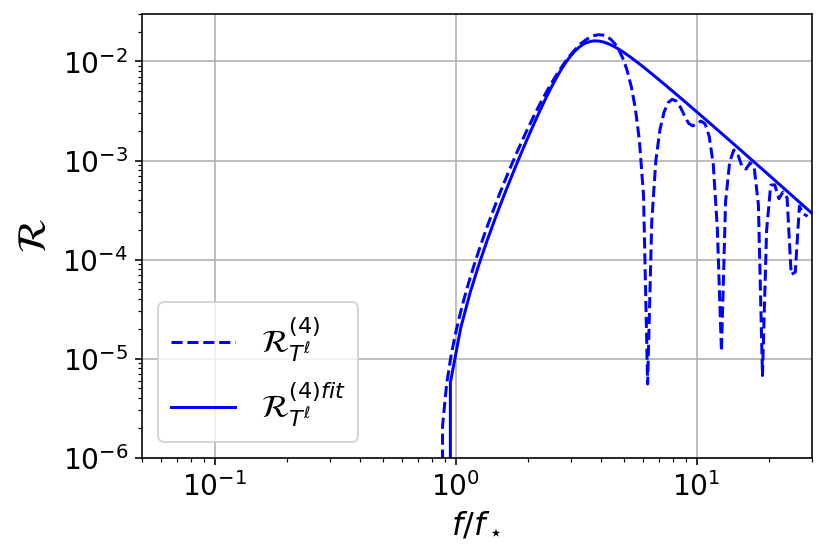}
	\caption{\small \it The response function for the collapsed four-point function, as measurable by the LISA and ET system, for the $A$, $E$ ({\bf left panel}), $T$ ({\bf right panel}) channels in the LISA sector. Dashed lines: numerical results. Continuous lines: the analytical approximations of eqs \eqref{anap4a}, \eqref{anap4t}. }
	\label{fig:resp2}
\end{figure}

%\noindent
In
the small-frequency limit, the four-point response  for the orthogonal channels $A^\ell=E^{\ell}$ and $T^\ell$
channels result
\bea
R^{(4)}_{A^\ell}&=&R_{E^\ell}\,=\,\frac{27}{140}+{\cal O}\left({f}/{f_\star} \right)^2,
\\
R^{(4)}_{T^\ell}&=&-\frac{17}{295680} \left(\frac{f}{f_\star} \right)^4+{\cal O}\left({f}/{f_\star} \right)^6.
\eea
Notice that their amplitudes are different from the two-point cases of eqs \eqref{ra2p} and \eqref{rt2p}, and
the $R_{T^\ell}$ now starts at small frequencies with a $(f/f_\star)^4$ contribution,  instead of a $(f/f_\star)^6$ as in eq \eqref{rt2p}.
The complete frequency dependence of the response functions can be easily obtained numerically: see Fig \ref{fig:resp2}. Suitable
analytical approximations for these two quantities  are give by
\bea
\label{anap4a}
R^{(4)}_{A^\ell,\,E^\ell}(f)&=&\frac{27}{140} \left( 1+\left(
\frac{f}{1.1\,f_\star}
\right)^3 \right)^{-2/3},
\\
\label{anap4t}
R^{(4)}_{T^\ell}(f)&=&\left(\frac{f}{3.3 \,f_\star} \right)^4 \left(0.00315 \left(\frac{f}{ f_\star} \right)  - 0.0104 \right)  
\left(1+\left(\frac{f}{3.3 \,f_\star} \right)^{7.2} \right)^{-1}.
\eea
These formulae provide the starting point for probing soft limits of GW correlation functions by 
considering synergies among detectors operating at different frequencies. This is a topic
with several theoretical and phenomenological ramifications that
 we
plan to develop elsewhere.

\section{Outlook}
\label{sec_concl}

In the next decades the  GW interferometers LISA and ET will hopefully be working  around the same time.  They
 will operate over different frequency ranges, but will have similar integrated sensitivities to the amplitude of the SGWB.  
It is important to embark on the quest to investigate what new physics we may learn from synergies between these two detectors. We take a first small step towards that direction in this paper. Particularly we have focused on cosmological sources of GWs, leading to a SGWB characterised by a large amplitude and a broad frequency spectrum
spanning several decades in frequency. Operating at different frequency scales, LISA and ET together will have the opportunity to detect distinct  features  of GWs produced by the same cosmological source. We quantitatively demonstrated this possibility by discussing various early-universe
examples motivated
by phase transitions, cosmic strings, and inflation, showing that the synergy of the two detectors can  improve our measurements of the parameters characterising a cosmological GW source. Moreover,  the two experiments operating in tandem can be sensitive to features of early universe cosmic expansion before big-bang nucleosynthesis, which affect the SGWB frequency profile. This probe of early universe of the pre-BBN epoch is challenging if not impossible  to test otherwise. Besides considering the GW spectrum, we additionally made a preliminary study of the sensitivity of LISA and ET to  soft limits of higher order GW correlation functions.  Given that these experiments operate over different frequency bands, their synergy constitutes an ideal direct probe of  squeezed  limits of non-Gaussian GW correlators, and of  its rich physical content.

We leave the important discussion of astrophysical SGWB and/or astrophysical noise sources to a future study. It is well known  that astrophysical sources of  SGWB can also lead to a broad spectrum of GWs, typically characterised by  a broken power law profile. Its shape is controlled by the type of sources of GWs  (see e.g. \cite{Regimbau:2011rp} for a review).  
In order to extract the signal and distinguish between a cosmological SGWB from the one generated by the astrophysical foregrounds, it is necessary  to subtract the astrophysical signals expected with sensitivities of BBO and ET or CE windows of frequency ranges \cite{Cutler:2005qq,Regimbau:2016ike}.
As well known a binary white dwarf galactic and extra-galactic astrophysical foreground also present in LISA is the dominant component as shown in Refs. \cite{Farmer:2003pa, Rosado:2011kv, Moore:2014lga}.
This issue is quite well studied for the case of galactic white dwarfs in the LISA band (see e.g. \cite{Babak:2023lro} for a recent analysis).
Therefore it should be possible to be subtracted \cite{Kosenko:1998mv,Adams:2010vc, Adams:2013qma}
in order to disentangle our alluded 
cosmological effects and signal. In the entire analysis in our present work we assume that
such subtractions will be possible. If LISA and ET operate in synergy, and if the cosmological sources lead to a sufficiently broad GW spectrum, it would be  possible to obtain extra information about the signal at  ET frequencies in order to `dig out' the properties of GWs in the LISA band through matched filtering techniques.

Other important simplifications we made are related to the fact that we neglected the relative motion
 between the two detectors, and we made simplifying assumptions about the directions of the interferometer arms.  Also, we considered the noise models to be fixed, and we did not marginalize over the noise parameters.
All these hypotheses will need to be extended in a more complete analysis. 
We leave all these interesting questions to  future studies.

{Ushering in the era of gravitational wave
astronomy with the planned network of GW detectors worldwide aspires to and perhaps will be able to achieve measurement
precisions that are orders of magnitude better with respect to the present day detectors. This new era of
GW detectors,  particularly with LISA and ET will make the dream of testing fundamental BSM microphysics, e.g.
scales of new physics symmetry breaking, scale of primordial cosmic inflation, probing pre-BBN cosmic epochs, a reality forthcoming in a not-so-distant future.}

\subsection*{Acknowledgments}
The work of AMB and GT is partially funded by STFC grant ST/X000648/1.  
DC would like to thank the Indian Institute of Astrophysics and the Department of Science and Technology, Government of India, for support through the INSPIRE Faculty Fellowship
grant no. DST/INSPIRE/04/2023/000110. 
DC would also like to thank the Indian Institute of Science for support through the C. V. Raman postdoctoral fellowship.
DC acknowledges the use of high-performance
computational facilities at the Supercomputer Education and Research Centre (SERC) of the Indian Institute of Science and the NOVA HPC
cluster at the Indian Institute of Astrophysics.
For the purpose of open access, the authors have applied a Creative Commons Attribution licence to any Author Accepted Manuscript version arising.
Research Data Access Statement: No new data were generated for this manuscript.

\begin{appendix}

%%%%%%%%%%%%%%%%%%%%%%%%
\section{Analytical fits to nominal curves}
\label{app_fit}
%%%%%%%%%%%%%%%%%

For LISA we use the analytical fits to the nominal
sensitivity curve of \cite{Flauger:2020qyi}, using $h=0.67$ (see   Fig \ref{fig:ETLISAnoms} right panel).
In this appendix, we report the expressions
for the fit to the  nominal ET-D \footnote{\url{http://www.et-gw.eu/index.php/etsensitivities}}   sensitivity curve for the Einstein telescope,
 which we have presented in the left panel of Fig \ref{fig:ETLISAnoms}. 
 The fit is given by
\begin{equation}
    \Omega_{\rm GW}(f) = 0.88 \times (t_1+t_2)\times t_3t_4t_5t_6\,,
    \label{eq:etd-90}
\end{equation}
with,
\begin{align*}
    t_1 & = \left[9x^{-30} + 5.5\times 10^{-6}x^{-4.5}+28\times10^{-13}x^{3.2}\right]
   \times\left(\frac{1}{2}-\frac{1}{2}\tanh\left(0.06\left(x-42\right)\right)\right), \\
   t_2 & = \left[1\times10^{-13}x^{1.9} + 20\times 10^{-13}x^{2.8}\right]\times\frac{1}{2}\tanh\left(0.06\left(x-42\right)\right),\\
   t_3 & = 1-0.475\exp\left(-\frac{\left(x-25\right)^2}{50}\right), \\
   t_4 & = 1-5\times10^{-4}\exp\left(-\frac{\left(x-20\right)^2}{100}\right),\\
   t_5 & = 1-0.2\exp\left(-\frac{\left(\left(x-47\right)^2\right)^{0.85}}{100}\right), \\
   t_6 & = 1-0.12\exp\left(-\frac{\left(\left(x-50\right)^2\right)^{0.7}}{100}\right)-0.2\exp\left(-\frac{\left(x-45\right)^2}{250}\right)+0.15\exp\left(-\frac{\left(x-85\right)^2}{400}\right),
\end{align*}
where $x = {f}/{1 \,{\rm Hz}}$.

\noindent
We determined this fit by trial and error. 
Note that the nominal sensitivity curve provided for the Einstein Telescope, which we are fitting with the above function, is for a pair of interferometers with an opening angle of $90^{\circ}$~\cite{Hild:2010id}. In order to obtain the fit for the ET-D triangular configuration with an opening angle of $60^{\circ}$, we have to multiply the expression~(\ref{eq:etd-90}) with a factor of $0.816^2$.
\end{appendix}

%\newpage
%\clearpage

{\small
\addcontentsline{toc}{section}{References}
\bibliographystyle{utphys}

\bibliography{LISA_ET_refsV2}
}

\end{document}